\documentclass[conference]{IEEEtran}

\usepackage{array, calc, color, multicol}
\usepackage{mathtools}

\DeclarePairedDelimiter\floor{\lfloor}{\rfloor}
\usepackage{graphics, epstopdf, graphicx, epsfig, psfrag, epsf, subfigure}
\usepackage{amssymb, amsfonts, amsmath, times}
\usepackage{relsize}
\usepackage{multicol,balance, verbatim}
\usepackage{textcomp, mathrsfs, stmaryrd, setspace}
\usepackage{amssymb}
\usepackage{soul}
\usepackage[dvipsnames]{xcolor}
\usepackage{cite}
\usepackage{url}
\usepackage{amsmath,stackengine}
\stackMath

\newcommand{\ie}{{\em i.e., }}
\newcommand{\Ie}{{\em I.e., }}

\newtheorem{theorem}{Theorem}
\newtheorem{lemma}[theorem]{Lemma}

\newtheorem{example}{Example}

\DeclareGraphicsExtensions{.eps, .pdf, .png, .jpg}   

\newcommand{\Nset}{\mathcal{N}}

\newcommand{\yas}[1]{\textcolor{magenta}{#1}}

\usepackage{lettrine}
\usepackage[ruled,vlined]{algorithm2e}

\IEEEoverridecommandlockouts

\begin{document}


\title{PolyDot Coded Privacy Preserving Multi-Party Computation at the Edge
}

\author{\IEEEauthorblockN{Elahe Vedadi}
\IEEEauthorblockA{
\textit{University of Illinois at Chicago}\\
evedad2@uic.edu}
\and
\IEEEauthorblockN{ Yasaman Keshtkarjahromi}
\IEEEauthorblockA{
\textit{Seagate Technology}\\
yasaman.keshtkarjahromi@seagate.com}
\and
\IEEEauthorblockN{Hulya Seferoglu}
\IEEEauthorblockA{
\textit{University of Illinois at Chicago}\\
hulya@uic.edu}
}

\maketitle

{$\hphantom{a}$}\vspace{-10pt}{}

\begin{abstract}
We investigate the problem of privacy preserving distributed matrix multiplication in edge networks using multi-party computation (MPC). Coded multi-party computation (CMPC) is an emerging approach to reduce the required number of workers in MPC by employing coded computation. Existing CMPC approaches usually combine coded computation algorithms designed for efficient matrix multiplication with MPC. We show that this approach is not efficient. We design a novel CMPC algorithm; PolyDot coded MPC (PolyDot-CMPC) by using a recently proposed coded computation algorithm; PolyDot codes. We exploit ``garbage terms'' that naturally arise when polynomials are constructed in the design of PolyDot-CMPC to reduce the number of workers needed for privacy-preserving computation. We show that entangled polynomial codes, which are consistently better than PolyDot codes in coded computation setup, are not necessarily better than PolyDot-CMPC in MPC setting.

 \end{abstract}
 

\section{\label{sec:introduction}Introduction}

Privacy-preserving distributed computing in edge networks is crucial for Internet of Things (IoT) applications including smart homes, self-driving cars, wearables, etc. Multi-party computation (MPC), which is a privacy-preserving distributed computing framework \cite{scalableMPC}, is a promising approach. The main goal of MPC is to calculate a function of data stored in multiple parties such as end devices and edge servers in edge computing systems. In this paper, we focus on BGW \cite{BGW}, an information theoretic MPC solution due to its lower computing load as well as quantum safe nature \cite{10.1007/3-540-48405-1_4} rather than cryptographic solutions \cite{Yao}, \cite{GMW}. Despite its potential, BGW should adapt to the limited resources of edge networks. 

Coded-MPC (CMPC) \cite{PolynomCMPC, 8613446} aims to improve BGW and make it adaptive to limited edge resources by employing coded computation \cite{SpeedUp-journal, Tradeoff-journal}. Coded computation advocates splitting computationally intensive tasks into smaller ones, coding these sub tasks using error correcting codes, and distributively processing coded tasks in parallel at workers (end devices or edge servers in our setup). This idea turns out to address the straggling workers problem \cite{SpeedUp-journal, Tradeoff-journal}. CMPC uses the coded computation idea in MPC setup to reduce the required number of workers, which is limited in edge systems.

Existing CMPC approaches \cite{PolynomCMPC, 8613446} usually combine coded computation algorithms designed for efficient matrix multiplication with MPC. In this paper, we show that this approach is not efficient with regard to reducing the required number of workers as it does not consider an important relationship between coded computation and MPC. Actually, the required number of workers (or efficiency of a code) is directly related with the powers of the created polynomials in coded computation. For example, the efficiency of polynomial codes reduces if there are gaps in the powers of the polynomials in coded computation. On the other hand, our key observation shows that such gaps help to reduce the required number of workers in CMPC setup. 
In particular, when there are gaps among powers of the coded terms, multiplication of the coded terms may create additional terms that we name ``garbage terms'', which can be used to reduce the required number of workers. The next example illustrates our key observation. 


\begin{example} \label{ex:codedMPCMatDot} \emph{MatDot-Coded MPC.}\footnote{Although our PolyDot-CMPC mechanism uses PolyDot codes, we use MatDot codes in this example to explain the ``garbage terms'' in a simple way.}
{Let us assume that there are two end devices; source 1 and source 2} that own matrices $A$ and $B$, respectively. Our objective is to compute $Y = A^TB$, which is a computationally exhaustive task for large  $A$ and $B$ matrices, while preserving privacy. To achieve this goal, end users need the help of edge servers (workers). Assume that matrices $A$ and $B$ are divided into two parts row-wise such that: $A^T = [A_1 \text{ } \text{ } A_2]$ and $B^T = [B_1 \text{ }  \text{ } B_2]$, where $Y = A^TB$ is constructed as $A^TB=A_1B_1+A_2B_2$. 

When the number of colluding workers  is $z=2$, {source 1 and source 2} construct polynomials $F_A(x)=A_1 + A_2x+ \bar{A}_3x^2 + \bar{A}_4x^3$ and $F_B(x)=B_1x + B_2 + \bar{B}_3x^2 + \bar{B}_4x^3$. The first two terms, namely, coded terms in these polynomials are determined by MatDot codes \cite{PolyDotMatDot}, and the second two terms, \ie secret terms, are designed by our proposed PolyDot-CMPC method, which we explain later in the paper. We note that the degree of the secret terms starts from two. The reason is that the multiplication of the coded terms becomes $(A_1 B_1 + A_2 B_2)x + A_1B_2 + A_2 B_1 x^2$, where the only term we need to recover $Y = A^TB$ is $(A_1 B_1 + A_2 B_2)x$. Other terms, namely, $A_1 B_2$ and $A_2 B_1 x^2$, are called \emph{garbage terms}.

After $F_A(\alpha_n)$ and $F_B(\alpha_n)$ are sent from {source 1 and source 2} to workers, worker $n$ determines $H(\alpha_n) = F_A(\alpha_n)F_B(\alpha_n)$, where {${H}(x)=A_1B_2+(A_1 B_1+A_2 B_2)x+\sum_{i=2}^{6} {H}_i x^i$}. Next, each worker $n$ computes the multiplication of $r_n$ with $H(\alpha_n)$ and creates the polynomial $G_n(x)$ as $G_n(x)=r_n H(\alpha_n)+R^{(n)}_0x+R^{(n)}_1x^2$, where the selection of $r_n$'s, $R^{(n)}_0$'s, and $R^{(n)}_1$'s will be explained later in the paper. 
Then, worker $n$ sends  $G_n(\alpha_{n'})$ to worker ${n'}$. After all data exchanges, worker ${n'}$, knowing  $G_n(\alpha_{n'})$, calculates their sum and sends $I(\alpha_{n'})=\sum_{n=1}^{7} G_n(\alpha_{n'})$ to the master (one of the edge devices that would like to get the calculated value of $Y=A^TB$), where $I(x) = A_1 B_1 + A_2 B_2 +
\sum_{n=1}^{7} R^{(n)}_0x+\sum_{n=1}^{7} R^{(n)}_1x^2$. In the last phase, the master reconstructs $I(x)$ once it receives $I(\alpha_n)$ from $1+z=3$ workers. After reconstructing $I(x)$ and determining all coefficients, $Y=A^TB = A_1 B_1 + A_2 B_2$ is calculated in a privacy-preserving manner. The number of terms with non-zero coefficients in polynomial $H(x)$ is equal to $7$. Thus, $7$ workers are required for privacy-preserving computation. We note that, for the same number of colluding workers and matrix partitions, polynomial coded MPC \cite{PolynomCMPC}, which divides matrices $A$ and $B$ into two column-wise partitions, requires 11 workers.\footnote{This example is a special case of both PolyDot-CMPC and Entangled-CMPC \cite{8613446}, when matrices $A$ and $B$ are  partitioned row-wise, but the idea of garbage terms is not discussed in \cite{8613446}.}
\hfill $\Box$
\end{example}

The above example demonstrates the importance of the garbage terms for the efficiency of CMPC algorithms. Based on this observation and exploiting the garbage terms, we design PolyDot-CMPC. We show that PolyDot-CMPC reduces the required number of workers for several colluding workers as compared to  entangled polynomial coded MPC (Entangled-CMPC) \cite{8613446}. This result is surprising as entangled polynomial codes are consistently better than PolyDot codes in coded computation setup \cite{YuFundamentalLimits2018}. We also compare PolyDot-CMPC with baselines; SSMM \cite{Zhu2021ImprovedCF}, and GCSA-NA \cite{9333639}. We show that  PolyDot-CMPC performs better than SSMM \cite{Zhu2021ImprovedCF} and GCSA-NA \cite{9333639} for a range of colluding workers. 

The structure of the rest of this paper is as follows.  Section~\ref{sec:system} presents our system model. Section~\ref{sec:attack} outlines the attack model we consider in this work.  Section~\ref{sec:PolyDotMPC} presents our PolyDot-CMPC algorithm as well as its performance analysis as compared to baselines. Section~\ref{sec:results} provides simulation results of PolyDot-CMPC. Section~\ref{sec:conc} concludes the paper.

\section{System Model} \label{sec:system}
We consider an MPC system containing $E$ sources, $N$ workers, and a master node, where all of them are edge devices with limited resources. There exists no connection among source nodes, but there are connections between sources and workers. All workers are connected to each other, and there exists a connection between the master node and each worker. Private data $\chi_{e}$ is stored at source node $e$. The goal is to compute $Y = f(\chi_{1}, \ldots, \chi_{E})$ in a privacy-preserving manner. The function $f(.)$ stands for any polynomial function, but we focus on the multiplication of two square matrices (which can be easily extended to general matrices). In particular, we consider $\chi_1 = A$ and $\chi_2 = B$, and calculate $Y=f(A, B) = A^TB$. 



Given the above system model, we use the following notation in the rest of this paper. Considering two arbitrary sets $\mathbf{I}$ and $\mathbf{J}$, with integer elements $i, j \in \mathbb{Z}$, we have; (i) $\mathbf{I} + \mathbf{J} = \{ i + j: i \in \mathbf{I},\; j \in \mathbf{J}\} $; (ii) $\mathbf{I} + j = \{i+j: i \in \mathbf{I}\}$; and
(iii) $|\mathbf{I}|$ stands for the cardinality of $\mathbf{I}$. We define $\Omega_{i}^{j}$ as $\Omega_{i}^{j} = \{i, \ldots, j\}$. We show the divisibility with $k|m$, \ie $m$ is divisible by $k$. {Considering a polynomial $f(x) = \sum_{i=0}^n a_ix^i$, $\mathbf{P}(f(x))$ is defined as the set of powers of the terms in $f(x)$ with non-zero coefficients, \ie $\mathbf{P}(f(x)) = \{i \in \mathbb{Z}: 0 \leq i \leq n,\; a_i \neq 0\}$. }
Finally, if a matrix $A$ is divided into $s$ row-wise and $t$ column-wise partitions, it is represented as
\begin{align}\label{eq:blockwise-part}
A = \left[ {\begin{array}{ccc}
   A_{0,0}&
   \ldots & A_{0,t-1}\\   
   \vdots&\ddots&\vdots \\ A_{s-1,0} &
   \ldots&A_{s-1,t-1}
  \end{array} } \right].
  \end{align}
\section{Attack Model}\label{sec:attack}

A semi-honest system model is considered in this paper where all parties (master, workers, and sources) are honest and follow the exact protocol defined by PolyDot-CMPC, but they are eavesdropping and potentially spying about private data. We design PolyDot-CMPC such that it is information theoretically secure against $z$ colluding workers, where $z$ is less than half of the total number of workers, \ie $z < N/2$. More specifically, we provide privacy requirements from source, worker and master nodes' perspective next.

\textit{Sources:} The private data of each source node, should be kept private from all other sources. Our system model satisfies this condition since, source nodes do not communicate. Also, the worker nodes and the master node do not send data to any of the source nodes.

\textit{Workers:} There should not be any privacy violation when workers receive data from sources, communicate with other workers and the master. Such privacy requirement should be satisfied if no more than $z$ workers collude. More formally, the following condition should be satisfied; $\tilde{H}(\chi_{1},\ldots,\chi_{E}|\underset{n \in \Nset_c}{\bigcup} (\{G_{n'}(\alpha_n), n'\in \Omega_{1}^{N}\}, {\underset{e \in \Omega_{1}^{E}}{\cup}}F_e(\alpha_n))) =\tilde{H}(\chi_{1}, \ldots, \chi_{E})$, where $\tilde{H}$ is the Shannon entropy, $\alpha_n$ is from finite field and known by all workers, $G_{n'}(\alpha_n)$ is the data that worker $n$ gets from worker $n'$, $F_e(\alpha_n)$ is the data that worker $n$ gets from source $e$, and $\Nset_c$ is a subset of $\{0, \ldots, N\}$ 
with cardinality less than or equal to $z$. 

\textit{Master:} Everything, except the final result $Y$, should be kept private from the master node. In particular, the following condition should be satisfied; $\tilde{H}(\chi_{1},\ldots, \chi_{E}|Y,\underset{n \in \Omega_{1}^{N}}{\bigcup}I(\alpha_n)) = \tilde{H}(\chi_{1},\ldots, \chi_{E}|Y)$, where $I(\alpha_n)$ is the data received from worker $n$ by the master node. 

\section{PolyDot Coded MPC (PolyDot-CMPC)} \label{sec:PolyDotMPC} 

In this section, we present our PolyDot coded MPC framework (PolyDot-CMPC) that employs PolyDot coding \cite{PolyDotMatDot} to create coded terms. 
Our design is based on leveraging the garbage terms that are not required for computing $Y=A^TB$ and reusing them in the secret terms.


\subsection{PolyDot-CMPC}

\textbf{Sources.} 
{Source 1 and source 2} 
divide matrices $A \in \mathbb{F}^{m \times m}$ and $B \in \mathbb{F}^{m \times m}$ into $s$ row-wise and $t$ column-wise partitions as in (\ref{eq:blockwise-part}), where $s,t \in \mathbb{N}$, and $s|m$ and $t|m$ hold. {Using the splitted matrices $A_{i,j} \in A^T$ and $B_{k,l} \in B$, where $i,l \in \Omega_{0}^{t-1}$, $j,k \in \Omega_{0}^{s-1}$}, they generate polynomials $F_A(x)$ and $F_B(x)$, which consist of coded and secret terms, \ie $F_{i'}(x)=C_{i'}(x)+S_{i'}(x),\; i' \in \{A,B\}$, where $C_{i'}(x)$'s are the coded terms defined by PolyDot codes \cite{PolyDotMatDot}, and $S_{i'}(x)$'s are the secret terms that we construct. Next, we discuss the construction of $S_{i'}(x)$, hence   $F_A(x)$ and $F_B(x)$ in detail. 

Let $\mathbf{P}(C_A(x))$ and $\mathbf{P}(C_B(x))$ be sets of the powers of the polynomials $C_A(x)$ and $C_B(x)$ with  coefficients larger than zero.  $\mathbf{P}(C_A(x))$ and $\mathbf{P}(C_B(x))$ are expressed as
\begin{align}\label{eq:polydot-p(CA)-th}
    \mathbf{P}(C_{A}(x)) = & \{i+tj \in \mathbb{N}: i \in \Omega_{0}^{t-1},\; j \in \Omega_{0}^{s-1}\} \nonumber \\
    = & \{0,\ldots,ts-1\},
\end{align}
\begin{align}\label{eq:polydot-p(CB)-th}
    \mathbf{P}(C_B(x)) = & \{t(s-1-k)+l\theta' \in \mathbb{N}:  k \in \Omega_{0}^{s-1},\; l \in \Omega_{0}^{t-1}\} \nonumber \\
    = & \{tq'+l\theta' \in \mathbb{N}: q' \in \Omega_{0}^{s-1},\; l \in \Omega_{0}^{t-1}\},
\end{align}
where $s,t \in \mathbb{N}$, and $\theta' = t(2s-1)$. 

As seen from (\ref{eq:polydot-p(CA)-th}) and (\ref{eq:polydot-p(CB)-th}), $\mathbf{P}(C_A(x)C_B(x))$ is the set of the powers of the polynomial $C_A(x)C_B(x)$ with  coefficients larger than zero,  and is expressed as $\mathbf{P}(C_A(x)C_B(x))= \{i+t(s-1+j-k)+tl(2s-1) \in \mathbb{N}: i,l \in \Omega_{0}^{t-1},\; j,k \in \Omega_{0}^{s-1} \}$. Furthermore, we know from \cite{PolyDotMatDot} that $Y_{i,l}=\sum_{j=0}^{s-1}A_{i,j}B_{j,l}$,  which are the coefficients of $x^{i+t(s-1)+tl(2s-1)}$ in $C_A(x)C_B(x)$, are the elements of the final result $Y=A^TB$. Therefore, we define $\{i+t(s-1)+tl(2s-1) \in \mathbb{N}: i,l \in \Omega_{0}^{t-1}\}$ as the set of important powers of $C_A(x)C_B(x)$. We define the secret terms $S_A(x)$ and $S_B(x)$ so that the important powers of $C_A(x)C_B(x)$ do not have common terms with $\mathbf{P}(C_A(x)S_B(x))$, $\mathbf{P}(S_A(x)C_B(x))$, and $\mathbf{P}(S_A(x)S_B(x))$. The reason is that $Y_{i,l}$'s should not have any overlap with the other components  for successful recovery of $Y$. The following conditions should hold  to guarantee this requirement. 
\begin{align}\label{eq:non_eq-polydot-thrm4}
    & \text{C1: } i+t(s-1)+tl(2s-1) \not\in \mathbf{P}(S_{A}(x))+\mathbf{P}(C_B(x)), \nonumber \\
   & \text{C2: } i+t(s-1)+tl(2s-1) \not\in \mathbf{P}(S_{A}(x))+\mathbf{P}(S_B(x)), \nonumber \\
    & \text{C3: } i+t(s-1)+tl(2s-1) \not\in \mathbf{P}(S_B(x))+\mathbf{P}(C_{A}(x)),
\end{align} where $i, l \in \Omega_{0}^{t-1}$ and $s,t \in \mathbb{N}$. We determine $\mathbf{P}(S_A(x))$ and $\mathbf{P}(S_B(x))$ according to the following set of rules; (i) determine all elements of $\mathbf{P}(S_{A}(x))$, starting from the minimum possible element, satisfying C1 in (\ref{eq:non_eq-polydot-thrm4}), (ii) fix $\mathbf{P}(S_{A}(x))$ in C2 of (\ref{eq:non_eq-polydot-thrm4}), and find all elements of the subset of $\mathbf{P}(S_B(x))$, starting from the minimum possible element, that satisfies C2; we call this subset as $\mathbf{P'}(S_B(x))$, (iii) determine all elements of the subset of $\mathbf{P}(S_B(x))$, starting from the minimum possible element, that satisfies C3 in (\ref{eq:non_eq-polydot-thrm4}); we call this subset as $\mathbf{P''}(S_B(x))$, and (iv) find the intersection of $\mathbf{P'}(S_B(x))$ and $\mathbf{P''}(S_B(x))$ to form $\mathbf{P}(S_B(x))$.
In our PolyDot-CMPC mechanism, we define the polynomials $F_A(x)$ and $F_B(x)$, based on the above strategy as formalized in Theorem \ref{th:S_i(x)andC_i(x)-strategy-polydot}.
\begin{theorem}\label{th:S_i(x)andC_i(x)-strategy-polydot}
With the following design of $F_A(x)$ and $F_B(x)$ in PolyDot-CMPC, the conditions in (\ref{eq:non_eq-polydot-thrm4}) are satisfied.
{\begin{align}\label{eq:FA1andFA2PolyDotCMPC}
   F_A(x)= \bigg\{ \begin{array}{cc}
   F_{A_1}(x)  &  z > ts-t \text{ and } s,t \neq 1\\
   F_{A_2}(x) & z \leq ts-t \text{ or } t=1 \text{ or } s=1\\
\end{array}\end{align}} 
\begin{align}\label{eq:FAPolyDotCMPC}
     F_{A_1}(x) = & \underbrace{{\sum_{i=0}^{t-1}\sum_{j=0}^{s-1} A_{i,j}x^{i+tj}}}_{\triangleq C_A(x)} + \underbrace{{\sum_{w=0}^{t(s-1)-1}\sum_{l=0}^{p-1}\bar{A}_{(w+\theta'l)}x^{ts+\theta'l+w}}}_{\triangleq S_{A_1}(x)} \nonumber \\
     & \underbrace{{+ \sum_{u=0}^{z-1-pt(s-1)}\bar{A}_{(u+t(s-1)+\theta'(p-1))}x^{ts+\theta'p+u}}}_{\triangleq S_{A_1}(x)},
\end{align}
\begin{align}\label{eq:FA2PolyDotCMPC}
F_{A_2}(x) = & \underbrace{\sum_{i=0}^{t-1}\sum_{j=0}^{s-1} A_{i,j}x^{i+tj}}_{\triangleq C_A(x)} + \underbrace{\sum_{u=0}^{z-1}\bar{A}_{u}x^{ts+\theta' p+u}}_{\triangleq S_{A_2}(x)},
\end{align}
{\begin{align}\label{eq:FBPolyDotCMPC}
   F_B(x)= \bigg\{ \begin{array}{cc}
   F_{B_1}(x)  &  z>\tau \text{ or } t=1 \text{ or } s=1\\
   F_{B_2}(x) & \frac{\tau+1}{2}< z \leq \tau \text{ and } s,t \neq 1\\
   F_{B_3}(x) & z \leq \frac{\tau+1}{2} \text{ and } s,t \neq 1
\end{array}\end{align} 
\begin{align}\label{eq:FB1PolyDotCMPC}
F_{B_1}(x)= \underbrace{\sum\limits_{k=0}^{s-1}\sum\limits_{l=0}^{t-1} B_{k,l}x^{t(s-1-k)+\theta'l}}_{\triangleq C_B(x)}+ \underbrace{\sum\limits_{r=0}^{z-1} \bar{B}_rx^{ts+\theta'(t-1)+r}}_{\triangleq S_{B_1}(x)},
\end{align} 
\begin{align}\label{eq:FB2PolyDotCMPC}
F_{B_2}(x) = & \underbrace{\sum\limits_{k=0}^{s-1}\sum\limits_{l=0}^{t-1} B_{k,l}x^{t(s-1-k)+\theta'l}}_{\triangleq C_B(x)}\nonumber\\+ &\underbrace{\sum\limits_{d=0}^{\tau-z}\sum\limits_{l'=0}^{p'-1} \bar{B}_{(d+\theta'l')}x^{ts+\theta'l'+d}}_{\triangleq S_{B_2}(x)} \nonumber\\+ &\underbrace{\sum\limits_{v=0}^{z-1-p'(\tau-z+1)}\bar{B}_{(v+\tau-z+1+\theta'(p'-1))}x^{ts+\theta'p'+v}}_{\triangleq S_{B_2}(x)},
\end{align}}
\begin{align}\label{eq:FB3PolyDotCMPC}
       F_{B_3}(x) = & \underbrace{\sum_{k=0}^{s-1}\sum_{l=0}^{t-1} B_{k,l}x^{t(s-1-k)+\theta'l}}_{\triangleq C_B(x)} + \underbrace{\sum_{v=0}^{z-1}\bar{B}_{v}x^{ts+v}}_{\triangleq S_{B_3}(x)}
\end{align}
where $p = \min \{ \floor{\frac{z-1}{ts-t}},t-1\}$, {$\tau=\theta'-ts-t$, $p'=\min \{\floor{\frac{z-1}{\tau-z+1}},t-1\}$}. {Moreover,} $\bar{A}_{(w+\theta'l)}$, $\bar{A}_{(u+t(s-1)+\theta'(p-1))}$, and $\bar{A}_{u}$, are selected independently and uniformly at random in $\mathbb{F}^{\frac{m}{t} \times \frac{m}{s}}$, and $\bar{B}_r$, {$\bar{B}_{d+\theta'l'}$, $\bar{B}_{(v+\tau-z+1+\theta'(p'-1))}$}, and $\bar{B}_{v}$ are chosen independently and uniformly at random in $\mathbb{F}^{\frac{m}{s} \times \frac{m}{t}}$. 
\end{theorem}
{\em Proof:}
The proof is provided in Appendix A. \hfill $\Box$

The degrees of secret terms in Theorem~\ref{th:S_i(x)andC_i(x)-strategy-polydot} are selected by exploiting the ``garbage terms'', which are all the terms coming from the multiplication of $C_{A}(x)$ and $C_B(x)$, except for the terms with indices $i+t(s-1)+\theta'l,\; i, l \in \Omega_{0}^{t-1}$, as these terms will be used to recover $Y=A^TB$. 

\textbf{Workers.}
Worker $n$ receives $F_{A}(\alpha_n)$ and $F_B(\alpha_n)$ from {source 1 and source 2}, 
and computes  $H(\alpha_n) = F_{A}(\alpha_n)F_B(\alpha_n)$. Then, worker $n$ calculates $G_n(x)$ as \begin{align} \label{eq:FnPolyDotCMPC}
    G_n(x) = \sum_{i=0}^{t-1}\sum_{l=0}^{t-1}r_n^{(i,l)}{H}(\alpha_n)x^{i+tl}+ \sum_{w=0}^{z-1} R^{(n)}_wx^{t^2+w},
\end{align} where $R^{(n)}_w$'s are selected independently and uniformly at random from $\mathbb{F}^{\frac{m}{t} \times \frac{m}{t}}$, and $r_n^{(i,l)}$'s are obtained satisfying $\sum_{j=0}^{s-1} A_{ij}B_{jl} = \sum_{n=1}^{N} r_n^{(i,l)} {H}(\alpha_n)$ using the Lagrange interpolation rule, and known by all workers.

Next, worker $n$ shares $G_n(\alpha_{n'})$ with other workers ${n'}$. After all the communications among workers, each worker ${n'}$ has access to all $G_{n}(\alpha_{n'})$'s. Worker ${n'}$ computes the summation of all $G_{n}(\alpha_{n'})$'s, and sends this result, \ie $I(\alpha_{n'})$, to the master node, where $ I(x) = \sum_{n=1}^{N} G_n(x)$.

\textbf{Master.} The master node can reconstruct the polynomial $I(x)$ by receiving $\deg(I(x))+1=t^2+z$ results from workers, and it directly gives the desired output $Y = A^TB$. The reason is that the coefficients of the first $t^2$ terms of $I(x)$ are exactly equal to the elements of {the} final result $Y=A^TB$.


\begin{theorem} \label{th:N_PolyDot}
{The required number of workers for multiplication of two massive matrices $A$ and $B$ employing PolyDot-CMPC, in a privacy preserving manner while there exist $z$ colluding workers in the system and due to the resource limitations each worker is capable of working on at most $\frac{1}{st}$ fraction of each input matrix, is expressed as follows}

\begin{align} \label{eq:N-PolyDot-DMPC}
N_{\text{PolyDot-CMPC}} =
\begin{cases}
   \psi_1, & ts<z \;\text{ or } t=1 
   \\
   \psi_2, & ts-t < z \leq ts \text{ and } t,s \neq 1\\
   \psi_3, & ts-2t < z \leq ts-t \text{ and } t,s \neq 1\\
   \psi_4, &  \upsilon' < z \leq ts-2t \text{ and } t,s \neq 1\\
   \psi_5, & z \leq \upsilon' \text{ and } t,s \neq 1\\
   \psi_6, & s=1 \text{ and } t \ge z \text{ and } t \neq 1
\end{cases}
\end{align}
where $\psi_1 = (p+2)ts+\theta'(t-1)+2z-1$, $\psi_2 = 2ts+\theta'(t-1)+3z-1$, $\psi_3 = 2ts+\theta'(t-1)+2z-1$, $\psi_4 = (t+1)ts+(t-1)(z+t-1)+2z-1$, $\psi_5 = \theta't+z$, and $\psi_6=t^2+2t+tz-1$,  $s|m$, and $t|m$ are satisfied, $p=\min\{\floor{\frac{z-1}{\theta'-ts}},t-1\}$, $\theta'= 2ts-t$ and $\upsilon'=\max\{ts-2t-s+2,\frac{ts-2t+1}{2}\}$. 
\end{theorem}
{\em Proof:}
The proof is provided in Appendix B. \hfill $\Box$

\subsection{PolyDot-CMPC in Perspective}\label{sec:polydotcmpc-perspective}
This section provides a theoretical analysis for the number of workers required by PolyDot-CMPC as compared to the baselines; 
Entangled-CMPC \cite{8613446}, SSMM \cite{Zhu2021ImprovedCF} and GCSA-NA \cite{9333639}\footnote{GCSA-NA is constructed for batch matrix multiplication. However, by considering the number of batches as one, it becomes an appropriate baseline to compare PolyDot-CMPC.
}.
\begin{lemma}\label{lemma: regions where N_polydot<N_entangled}
PolyDot-CMPC {is more efficient} than Entangled-CMPC with regards to requiring smaller number of workers in the following regions:
\begin{enumerate}
    \item  $z>ts,\; p<\frac{t-1}{s},\; t\neq 1$
    \item $ts-s<z\leq ts,\; t-1>s,\; s,t \neq 1$
    \item $(t-1)^2<z<t(t-1),\; s=t-1,\; s,t \neq 1$
    \item $ts-t-\min\{0,1-\frac{2s-5}{t-3}\}<z\leq ts-s,\; t>3,\; s \neq 1$
    \item $s=2,\; t=3,\;z=4$
    \item $t=2,\; s=2,\; z=1,2$ 
    \item $\max\{st-t-s-\frac{2}{t-2}, ts-2t\} < z \leq ts-t,\; t>2,\; t\ge s,\; s \neq 1$
   \item $t<s\leq 2t,\; ts-s<z\leq ts-t,\; s,t \neq 1$ 
     \item $t=2,\; 3\leq s \leq 4,\; 2(s-2)<z\leq 2(s-1)$ 
    \item $st-2t < z \leq ts-s,\; t>2,\; t< s\leq 2t$
    \item $s>2t,\; ts-2t<z\leq ts-t,\; s,t \neq 1$ 
    \item $2t\ge s,\; \max\{ts-2t-s+2, \frac{ts-2t+1}{2}\} < z \leq \min\{st-2t, 2ts-t^2+t-2s+1\},\; s,t \neq 1$ 
    \item $s>2t,\; ts-s<z\leq ts-2t,\; t\neq 1,2$ 
    \item $4<s<z<2s-4,\; t=2$ 
    \item $ts-2t-s+2<z<ts-s,\; 2t<s,\; s,t \neq 1$
    \item $st-2s-t-\frac{1}{t-1}<z\leq \max\{ts-2t-s+2, \frac{ts-2t+1}{2}\},\; s,t \neq 1$. 
\end{enumerate}
In all other regions for the values of the system parameters $s, t$, and $z$, PolyDot-CMPC requires the same or larger number of workers.
\end{lemma}
{\em Proof:}
The proof is provided in Appendix C.A. \hfill $\Box$
\begin{lemma}\label{lemma: regions where N_polydot<N_ssmm}
PolyDot-CMPC performs better than SSMM in terms of requiring smaller number of  workers in the following two regions:
\begin{enumerate}
    \item $z > \max\{ts,ts-t+\frac{pts}{t-1}\}, t\neq 1$
    \item $\frac{t-1}{t-2}(st-t)<z \leq ts$.
\end{enumerate}
In all other regions for the values of the system parameters $s, t$, and $z$, PolyDot-CMPC requires the same or larger number of workers.
\end{lemma}
{\em Proof:}
The proof is provided in Appendix C.B. \hfill $\Box$

\begin{lemma}\label{lemma: regions where N_polydot<N_gcsana}
PolyDot-CMPC performs better than GCSA-NA in terms of requiring smaller number of  workers  in the following regions:
\begin{enumerate}
    \item $z>ts, p<\frac{t-1}{s}, t\neq 1$
    \item $s<t, ts-t<z \leq \min\{ts, t(t-1)-1\}$
    \item $z \leq ts-t$
    \item $s=1, t>z, t\neq 2$.
\end{enumerate}
In all other regions for the values of the system parameters $s, t$, and $z$, PolyDot-CMPC requires the same or larger number of workers.
\end{lemma}
{\em Proof:}
The proof is provided in Appendix C.C. \hfill $\Box$


\begin{figure}[t!]
\centering
\subfigure{ 
\scalebox{.17}{ \includegraphics{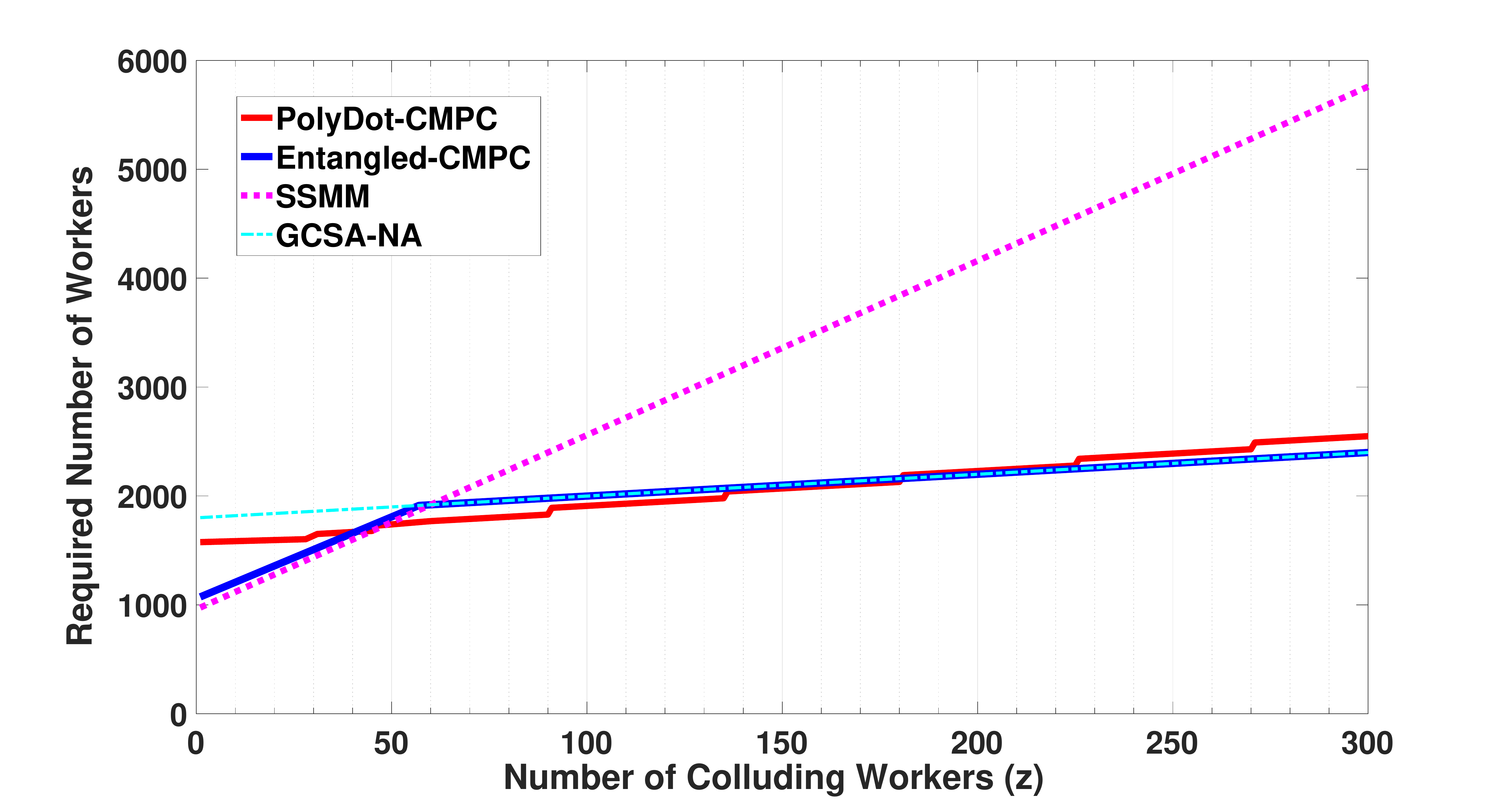}}}
\vspace{-10pt}
\caption{Required number of workers versus number of colluding workers. The parameters are set to $s=4,\; t=15$ and $1 \leq z \leq 300$.}
\label{fig:NvsZ-comp}
\vspace{-5pt}
\end{figure}

\section{Performance Evaluation} \label{sec:results} 

{In this section, the performance of PolyDot-CMPC is evaluated via simulations and compared with the baseline methods, (i) Entangled-CMPC \cite{8613446}, (ii) SSMM \cite{Zhu2021ImprovedCF}, and (iii) GCSA-NA \cite{9333639}. In this setup we have $m=36000$, \ie both $A$ and $B$ are square matrices with the size of $36000 \times 36000$. } 


Fig.~\ref{fig:NvsZ-comp}, shows the number of workers required for computing $Y=A^TB$ versus the number of colluding workers. This figure, is an example for the  analysis provided in Section \ref{sec:polydotcmpc-perspective} for specific values of $s=4,\; t=15$, and $1 \leq z \leq 300$. For small number of colluding workers, \ie $1 \leq z \leq 48$, SSMM \cite{Zhu2021ImprovedCF} performs the best as it requires minimum number of workers. PolyDot-CMPC performs better than all the baselines when $49 \leq z \leq 180$. On the other hand, GCSA-NA \cite{9333639} and Entangled-CMPC \cite{8613446} have similar performance and perform better than the other mechanisms when $181 \leq z \leq 300$. These results confirm Lemmas \ref{lemma: regions where N_polydot<N_entangled}, \ref{lemma: regions where N_polydot<N_ssmm}, and \ref{lemma: regions where N_polydot<N_gcsana} as PolyDot-CMPC performs better than the baselines for a range of colluding workers.



\begin{figure}[t!]
\centering
\subfigure{ 
\scalebox{.17}{ \includegraphics{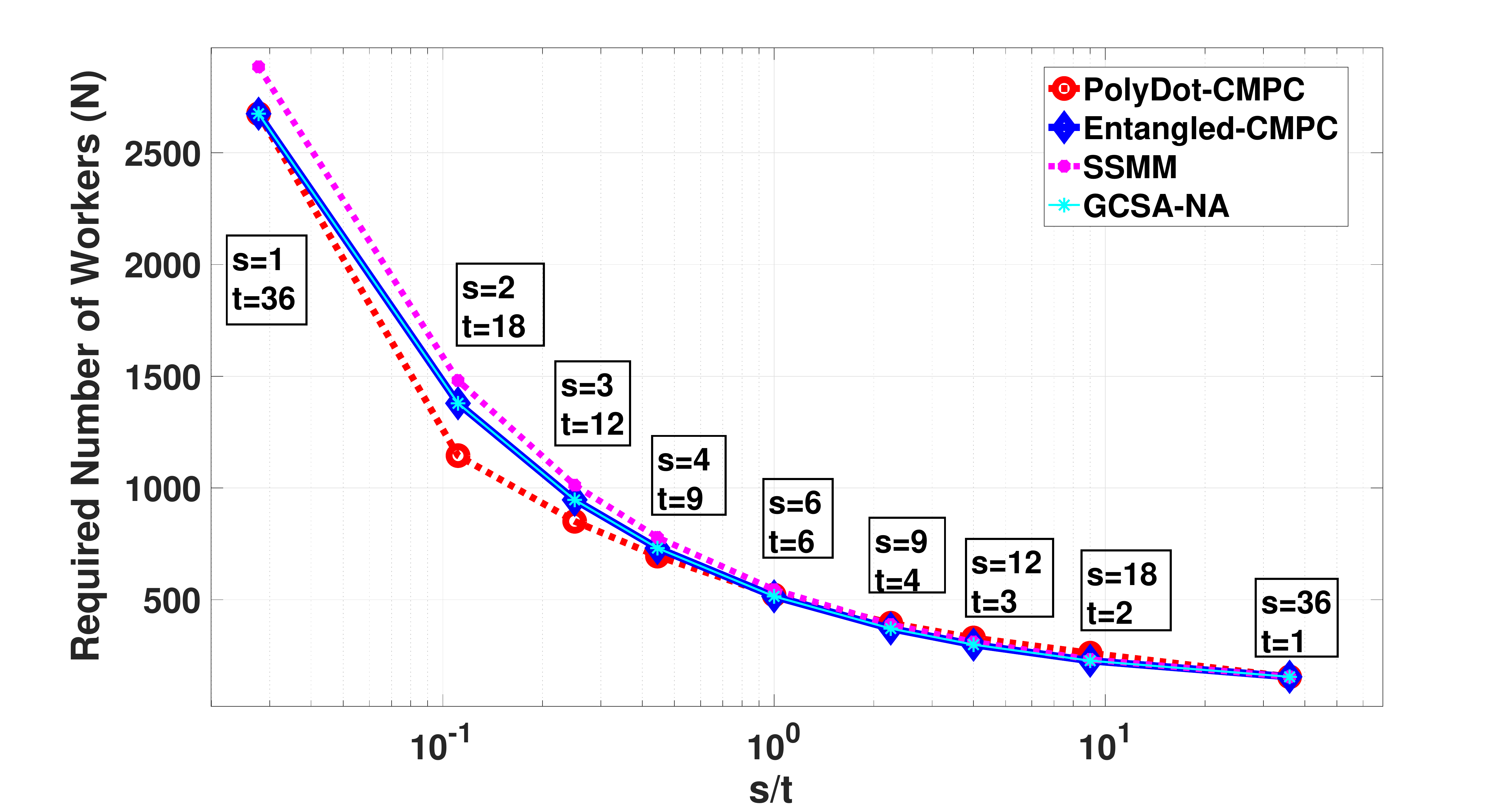}}} 
\vspace{-10pt}
\caption{Required number of workers versus $s/t$ for fixed $z=42$ and $st=36$.} 
\label{fig:N-PolyDot}
\vspace{-5pt}
\end{figure}

Fig.~\ref{fig:N-PolyDot} illustrates the required number of workers versus $s/t$, the number of row partitions divided by the number of column partitions, for fixed $z=42$ and $st=36$. 
As seen, PolyDot-CMPC performs better than the other baseline methods concerning the required number of workers for $(s,t) \in \{(2,18),(3,12),(4,9)\}$, since in this scenario we have $42=z>ts=36$, and for these values of $s,t$, we have $p$ equal to $2$, $1$ and $1$, respectively. Thus, conditions 1 in Lemmas \ref{lemma: regions where N_polydot<N_entangled}, \ref{lemma: regions where N_polydot<N_ssmm}, and \ref{lemma: regions where N_polydot<N_gcsana} are satisfied. However, for  $(s,t) \in \{(1,36),(6,6),(9,4),(12,3),(18,2),(36,1)\}$, these conditions are no longer satisfied. Also, we can see that the required number of workers for all methods is directly related to the number of column partitions, $t$.

\section{Conclusion} \label{sec:conc}
We have {studied} the problem of privacy preserving matrix multiplication in edge networks using MPC. {We have proposed a new coded privacy-preserving computation mechanism;  PolyDot-CMPC, which is designed by employing PolyDot codes}. We have used ``garbage terms'' that naturally arise when polynomials are constructed in the design of PolyDot-CMPC to reduce the number of workers needed for privacy-preserving computation.  We have analyzed and simulated PolyDot-CMPC, and demonstrated that the garbage terms are important in the design and {efficiency} of CMPC algorithms. 


\bibliographystyle{IEEEtran}

\bibliography{refs}

\begin{thebibliography}{10}
\providecommand{\url}[1]{#1}
\csname url@samestyle\endcsname
\providecommand{\newblock}{\relax}
\providecommand{\bibinfo}[2]{#2}
\providecommand{\BIBentrySTDinterwordspacing}{\spaceskip=0pt\relax}
\providecommand{\BIBentryALTinterwordstretchfactor}{4}
\providecommand{\BIBentryALTinterwordspacing}{\spaceskip=\fontdimen2\font plus
\BIBentryALTinterwordstretchfactor\fontdimen3\font minus
  \fontdimen4\font\relax}
\providecommand{\BIBforeignlanguage}[2]{{%
\expandafter\ifx\csname l@#1\endcsname\relax
\typeout{** WARNING: IEEEtran.bst: No hyphenation pattern has been}%
\typeout{** loaded for the language `#1'. Using the pattern for}%
\typeout{** the default language instead.}%
\else
\language=\csname l@#1\endcsname
\fi
#2}}
\providecommand{\BIBdecl}{\relax}
\BIBdecl

\bibitem{scalableMPC}
J.~Saia and M.~Zamani, ``Recent results in scalable multi-party computation,''
  in \emph{SOFSEM 2015: Theory and Practice of Computer Science}, G.~F.
  Italiano, T.~Margaria-Steffen, J.~Pokorn{\'y}, J.-J. Quisquater, and
  R.~Wattenhofer, Eds.\hskip 1em plus 0.5em minus 0.4em\relax Berlin,
  Heidelberg: Springer Berlin Heidelberg, 2015, pp. 24--44.

\bibitem{BGW}
M.~Ben-Or, S.~Goldwasser, and A.~Wigderson, ``Completeness theorems for
  non-cryptographic fault-tolerant distributed computation,'' in
  \emph{Providing Sound Foundations for Cryptography: On the Work of Shafi
  Goldwasser and Silvio Micali}, 2019, pp. 351--371.

\bibitem{10.1007/3-540-48405-1_4}
U.~Maurer, ``Information-theoretic cryptography,'' in \emph{Advances in
  Cryptology --- CRYPTO' 99}, M.~Wiener, Ed.\hskip 1em plus 0.5em minus
  0.4em\relax Berlin, Heidelberg: Springer Berlin Heidelberg, 1999, pp. 47--65.

\bibitem{Yao}
A.~C.-C. Yao, ``How to generate and exchange secrets,'' in \emph{27th Annual
  Symposium on Foundations of Computer Science (sfcs 1986)}, 1986, pp.
  162--167.

\bibitem{GMW}
S.~M. O.~Goldreich and A.~Wigderson, ``How to play any mental game,'' in
  \emph{Proc. of the 19th STOC}, 1987, pp. 218--229.

\bibitem{PolynomCMPC}
H.~Akbari-Nodehi and M.~A. Maddah-Ali, ``Secure coded multi-party computation
  for massive matrix operations,'' \emph{IEEE Transactions on Information
  Theory}, vol.~67, no.~4, pp. 2379--2398, 2021.

\bibitem{8613446}
H.~A. Nodehi, S.~R.~H. Najarkolaei, and M.~A. Maddah-Ali, ``Entangled
  polynomial coding in limited-sharing multi-party computation,'' in \emph{2018
  IEEE Information Theory Workshop (ITW)}, 2018, pp. 1--5.

\bibitem{SpeedUp-journal}
K.~{Lee}, M.~{Lam}, R.~{Pedarsani}, D.~{Papailiopoulos}, and K.~{Ramchandran},
  ``Speeding up distributed machine learning using codes,'' \emph{IEEE
  Transactions on Information Theory}, vol.~64, no.~3, March 2018.

\bibitem{Tradeoff-journal}
S.~{Li}, M.~A. {Maddah-Ali}, Q.~{Yu}, and A.~S. {Avestimehr}, ``A fundamental
  tradeoff between computation and communication in distributed computing,''
  \emph{IEEE Transactions on Information Theory}, vol.~64, no.~1, pp. 109--128,
  Jan 2018.

\bibitem{PolyDotMatDot}
M.~Fahim, H.~Jeong, F.~Haddadpour, S.~Dutta, V.~Cadambe, and P.~Grover, ``On
  the optimal recovery threshold of coded matrix multiplication,'' in
  \emph{2017 55th Annual Allerton Conference on Communication, Control, and
  Computing (Allerton)}.\hskip 1em plus 0.5em minus 0.4em\relax IEEE, 2017, pp.
  1264--1270.

\bibitem{YuFundamentalLimits2018}
Q.~Yu, M.~A. Maddah-Ali, and A.~S. Avestimehr, ``Straggler mitigation in
  distributed matrix multiplication: Fundamental limits and optimal coding,''
  \emph{IEEE Transactions on Information Theory}, vol.~66, no.~3, pp.
  1920--1933, 2020.

\bibitem{Zhu2021ImprovedCF}
J.~Zhu, Q.~Yan, and X.~Tang, ``Improved constructions for secure multi-party
  batch matrix multiplication,'' \emph{IEEE Transactions on Communications},
  vol.~69, pp. 7673--7690, 2021.

\bibitem{9333639}
Z.~Chen, Z.~Jia, Z.~Wang, and S.~A. Jafar, ``Gcsa codes with noise alignment
  for secure coded multi-party batch matrix multiplication,'' \emph{IEEE
  Journal on Selected Areas in Information Theory}, vol.~2, no.~1, pp.
  306--316, 2021.

\end{thebibliography}

\section*{Appendix A: Proof of Theorem \ref{th:S_i(x)andC_i(x)-strategy-polydot}} 
We first 
determine $\mathbf{P}(S_A(x))$ and $\mathbf{P}(S_B(x))$ and then derive $F_{A}(x)$ and $F_{B}(x)$, accordingly.

Based on our strategy for determining $\mathbf{P}(S_A(x))$ and $\mathbf{P}(S_B(x))$, we: (i) first find all elements of $\mathbf{P}(S_{A}(x))$, starting from the minimum possible element, satisfying C1 in (\ref{eq:non_eq-polydot-thrm4}), (ii) then fix $\mathbf{P}(S_{A}(x))$, containing the $z$ smallest elements, in C2 of (\ref{eq:non_eq-polydot-thrm4}), and find all elements of the subset of $\mathbf{P}(S_B(x))$, starting from the minimum possible element, that satisfies C2; we call this subset as $\mathbf{P'}(S_B(x))$, (iii) find all elements of the subset of $\mathbf{P}(S_B(x))$, starting from the minimum possible element, that satisfies C3 in (\ref{eq:non_eq-polydot-thrm4}); we call this subset as $\mathbf{P''}(S_B(x))$, and (iv) finally, find the intersection of $\mathbf{P'}(S_B(x))$ and $\mathbf{P''}(S_B(x))$ to form $\mathbf{P}(S_B(x))$. Next, we explain these steps in details. 

\emph{(i) Find all elements of $\mathbf{P}(S_A(x))$ satisfying C1 in (\ref{eq:non_eq-polydot-thrm4}).} 

For this step, using (\ref{eq:polydot-p(CB)-th}) and C1 in (\ref{eq:non_eq-polydot-thrm4}), we have: 
\begin{align}\label{non_eq1-polydot'}
    &i+t(s-1)+tl(2s-1) \not\in \{t(s-1)-tq+tl'(2s-1)\} \nonumber \\
    & +\mathbf{P}(S_A(x)),  0 \leq q \leq s-1,\; 0 \leq i, l, l' \leq t-1,\; s, \nonumber \\
    & t \in \mathbb{N}
\end{align}
which is equivalent to:
\begin{align}\label{eq:non_eq1-polydot_2}
    \beta+\theta' l'' \not\in \mathbf{P}(S_A(x)),
\end{align}
 for $l''=(l-l')$, $\theta' = t(2s-1)$ and $\beta = i+tq$. From (\ref{non_eq1-polydot'}), the range of the variables $\beta$ and $l''$ are derived as $\beta \in \{0,\ldots,ts-1\}$ and $l'' \in \{-(t-1),\ldots,(t-1)\}$. However, knowing the fact that all powers in $\mathbf{P}(S_A(x))$ are from $\mathbb{N}$, we consider only $l'' \in \{0,\ldots,t-1\}$.\footnote{The reason is that for the largest value of $\beta$, \ie $\beta = ts-1$ and largest value of $l'' \in \{-(t-1),\ldots,-1\}$, \ie $l''=-1$, $\beta+\theta'(l'')$ is equal to $ts-1+(2ts-t)(-1) = t(1-s)-1$, which is negative for $s,t \in \mathbb{N}$. Therefore, for all $l'' \in \{-(t-1),\ldots,-1\}$ in (\ref{eq:non_eq1-polydot_2}), $\beta+\theta'l''$ is negative.} 
Considering different values of $l''$ from the interval $l''\in\{0,...,t-1\}$ in (\ref{eq:non_eq1-polydot_2}), we have: 
\begin{align}
   &\mathbf{P}(S_A(x)) \notin \{0,\ldots,ts-1\}, \nonumber \\
   &\mathbf{P}(S_A(x)) \notin \{\theta',\ldots,ts-1+\theta'\}, \nonumber \\
   &\mathbf{P}(S_A(x)) \notin \{2\theta',\ldots,ts-1+2\theta'\}, \nonumber \\
& \ldots\nonumber \\
   & \mathbf{P}(S_A(x)) \notin \{(t-1)\theta',\ldots,ts-1+(t-1)\theta'\}.
\end{align}
Using the complement of the above intervals, the intervals that $\mathbf{P}(S_A(x))$ can be selected from, is derived as follows: 
\begin{align}\label{eq:P(RA)_set_representation}
    \mathbf{P}(S_A(x)) \in & \{ts,\ldots,\theta'-1\} \cup \{ts+\theta',\ldots,2\theta'-1\} \cup \ldots \nonumber \\
    & \cup \{ts+(t-1)\theta',\ldots,+\infty\}, s,t > 1
\end{align}
\begin{align}\label{eq:P(RA)_set_representation_s=1}
    \mathbf{P}(S_A(x)) \in 
    \{t^2,\ldots,+\infty\}, s=1
\end{align}
\begin{align}\label{eq:P(RA)_set_representation_t=1}
    \mathbf{P}(S_A(x)) \in 
    \{s,\ldots,+\infty\}, t=1
\end{align}
Note that the required number of powers with non-zero coefficients for the secret term $S_A(x)$ is $z$, \ie
\begin{equation}
    |\mathbf{P}(S_A(x))| = z.
\end{equation}
Since our goal is to make the degree of polynomial $F_A(x)$ as small as possible, we choose the $z$ smallest powers from the sets in (\ref{eq:P(RA)_set_representation}) to form $\mathbf{P}(S_A(x))$. 
Note that in (\ref{eq:P(RA)_set_representation}), there are $t-1$ finite sets and one infinite set, where each finite set contains $\theta'-ts$ elements. 
Therefore, based on the value of $z$, we use the first interval and as many remaining intervals as required for $z > \theta'-ts$, and the first interval only for $ z \leq \theta'-ts$. 

\begin{lemma}\label{lem:P(SA)-z large}
If $z > \theta'-ts$ and $s,t \neq 1$, the subsets of all powers of polynomial $S_A(x)$ with non-zero coefficients is defined as the following:
\begin{align}\label{eq:finite_P(RA)_set_representation-z large}
    \mathbf{P}(S_A(x)) = &\Big(\bigcup\limits_{l=0}^{p-1} \{ts+\theta' l,\ldots,(l+1)\theta'-1\}\Big) \nonumber\\
    & \cup \{ts+p\theta',\ldots,ts+p\theta'+z-1-p(\theta'-ts)\}\\
    =&\{ts+\theta'l+w, l\in\Omega_0^{p-1}, w\in \Omega_0^{t(s-1)-1}\} \nonumber \\
    &\cup \{ts+\theta'p+u, u\in\Omega_0^{z-1-pt(s-1)}\}\label{eq:psa12}.
\end{align}
\end{lemma}
{\em Proof:}
For the case of $z > \theta'-ts$ and $s,t \neq 1$, the number of elements in the first interval of (\ref{eq:P(RA)_set_representation}), which is equal to $\theta'-ts$, is not sufficient for selecting $z$ powers. Therefore, more than one interval is used; we show the number of selected intervals with $p+1$, where $p \ge 1$ is defined as $p=\min\{\floor{\frac{z-1}{\theta'-ts}},t-1\}$. With this definition, the first $p$ selected intervals are selected in full, in other words, in total we select $p(\theta'-ts)$ elements to form the first $p$ intervals in (\ref{eq:finite_P(RA)_set_representation-z large}). The remaining $z-p(\theta'-ts)$ elements are selected from the $(p+1)^\text{st}$ interval of (\ref{eq:P(RA)_set_representation}) as shown as the last interval of (\ref{eq:finite_P(RA)_set_representation-z large}). (\ref{eq:psa12}) can be derived from (\ref{eq:finite_P(RA)_set_representation-z large}) by replacing $\theta'$ with its equivalence, $2ts-t$.
\hfill $\Box$

\begin{lemma}\label{lem:P(SA)-z small}
If $z \leq \theta'-ts$ and $s,t \neq 1$, the subsets 
of all powers of polynomial $S_A(x)$ with non-zero coefficients is defined as the following:
\begin{align}\label{eq:finiteP(SA)-polydot-second-scenario}
\mathbf{P}(S_A(x)) = \{ts,\dots,ts+z-1\} \nonumber\\
= \{ts+u, u\in \Omega_0^{z-1}\}.
\end{align}
\end{lemma}
{\em Proof:}
In this scenario for $z \leq \theta'-ts$, the first interval of (\ref{eq:P(RA)_set_representation}) is sufficient to select all $z$ elements of $\mathbf{P}(S_A(x))$, therefore, $z$ elements are selected from the first interval of (\ref{eq:P(RA)_set_representation}), as shown in (\ref{eq:finiteP(SA)-polydot-second-scenario}). 
\hfill $\Box$

\begin{lemma}\label{lem:P(SA)-s=1t=1}
If $s=1$, the subsets of all powers of polynomial $S_A(x)$ with non-zero coefficients is defined as the following:
\begin{align}\label{eq:psafors=1}
    \mathbf{P}(S_A(x)) = \{t^2,\dots,t^2+z-1\}
    \nonumber\\
= \{t^2+u, u\in \Omega_0^{z-1}\},
\end{align}
and if $t=1$, it is defined as:
\begin{align}\label{eq:psafort=1}
    \mathbf{P}(S_A(x)) = \{s,\dots,s+z-1\}
    \nonumber\\
= \{s+u, u\in \Omega_0^{z-1}\}.
\end{align}
\end{lemma}
{\em Proof:}
If $s=1$, $z$ smallest elements are selected from (\ref{eq:P(RA)_set_representation_s=1}), as shown in (\ref{eq:psafors=1}) and if $t=1$, $z$ smallest elements are selected from (\ref{eq:P(RA)_set_representation_t=1}), as shown in (\ref{eq:psafort=1}). 
\hfill $\Box$

\emph{(ii) Fix $\mathbf{P}(S_{A}(x))$ in C2 of (\ref{eq:non_eq-polydot-thrm4}), and find the subset of $\mathbf{P}(S_B(x))$ that satisfies C2; we call this subset as $\mathbf{P'}(S_B(x))$.}

In this step, we consider the four cases of $s=1$, $t=1$, $z>\theta'-ts,\;s,t \neq 1$, and $z \leq \theta'-ts,\; s,t \neq 1$ and derive $\mathbf{P'}(S_B(x))$ as summarized in Lemmas \ref{lem:P'(SB)-s=1}, \ref{lem:P'(SB)-t=1}, \ref{lem:P'(SB)-z large} and \ref{lem:P'(SB)-z small}, respectively.

\begin{lemma}\label{lem:P'(SB)-s=1}
If $s=1$, $\mathbf{P'}(S_B(x))$ is defined as the following:
\begin{align}\label{eq:P'(RB)_set_s=1}
    \mathbf{P'}(S_B(x)) =  \{0,\ldots,+\infty\}.
\end{align}
\end{lemma}
{\em Proof:}
In this scenario, we use (\ref{eq:psafors=1}) defined for $\mathbf{P}(S_A(x))$. By replacing $\mathbf{P}(S_A(x))$ in C2 we have the following: 
\begin{align}\label{eq:non_eq2-polydot''-s=1}
   & \text{C2: } i+tl \not\in \{t^2,\dots,t^2+z-1\}+\mathbf{P'}(S_B(x)),
\end{align}
which can be equivalently written as:
\begin{align}\label{eq:non_eq3-polydot''-s=1}
   & \text{C2: } \{0,\dots,t^2-1\} \not\in \{t^2,\dots,t^2+z-1\}+\mathbf{P'}(S_B(x)).
\end{align}
From the above equation, any non-negative elements for $\mathbf{P'}(S_B(x))$ satisfies this constraint.  
This completes the proof.\hfill $\Box$

\begin{lemma}\label{lem:P'(SB)-t=1}
If $t=1$, $\mathbf{P'}(S_B(x))$ is defined as the following:
\begin{align}\label{eq:P'(RB)_set_t=1}
    \mathbf{P'}(S_B(x)) =  \{0,\ldots,+\infty\}.
\end{align}
\end{lemma}
{\em Proof:}
In this scenario, we use (\ref{eq:psafort=1}) defined for $\mathbf{P}(S_A(x))$. By replacing $\mathbf{P}(S_A(x))$ in C2 we have the following: 
\begin{align}\label{eq:non_eq2-polydot''-t=1}
   & \text{C2: } s-1 \not\in \{s,\dots,s+z-1\}+\mathbf{P'}(S_B(x)).
\end{align}
From the above equation, any non-negative elements for $\mathbf{P'}(S_B(x))$ satisfies this constraint. 
This completes the proof.\hfill $\Box$

   \begin{lemma}\label{lem:P'(SB)-z large}
If $z > \theta'-ts$ and $s,t \neq 1$, $\mathbf{P'}(S_B(x))$ is defined as the following:
\begin{align}\label{eq:P'(RB)_set_representation-zlarge}
    \mathbf{P'}(S_B(x)) = 
    &\Big(\bigcup\limits_{l'=0}^{t-2} \{\theta' l',\ldots,(l'+1)\theta'-ts\}\Big) \\ 
    &\cup \{(t-1)\theta',\ldots,+\infty\}.
\end{align}
\end{lemma}
{\em Proof:}
In this scenario, we use (\ref{eq:finite_P(RA)_set_representation-z large}) defined for $\mathbf{P}(S_A(x))$ when $z > \theta'-ts$, which can be equivalently written as:
\begin{align}
 \mathbf{P}(S_A(x)) = \Bigg\{ \begin{array}{cc}
   ts+\theta' l''+w,  &  l''\in \Omega_{0}^{p-1},\; w \in \Omega_{0}^{\theta'-ts-1} \\
   ts+\theta' l''+u,  &  l''= p,\; u\in \Omega_{0}^{z-1-p(\theta'-ts)}
\end{array}
\end{align}
and then replace $\mathbf{P}(S_A(x))$ in C2 using the above equation:
\begin{align}\label{eq:non_eq2-polydot''}
   & \text{C2: } i+t(s-1)+\theta' l \not\in \nonumber \\
   & \Bigg\{ \begin{array}{cc}
   ts+\theta' l''+w+\mathbf{P'}(S_B(x)),  &  l''\in \Omega_{0}^{p-1},\; w \in \Omega_{0}^{\theta'-ts-1}  \\
   ts+\theta' l''+u+\mathbf{P'}(S_B(x)),  &  l''=p,\; u\in \Omega_{0}^{z-1-p(\theta'-ts)}
\end{array}
\end{align}
Equivalently:
\begin{align}\label{eq:non_eq2-polydot'''-}
  & \mathbf{P'}(S_B(x)) \not \in \nonumber \\
  & \Bigg\{ \begin{array}{cc}
   i-t-w+\theta'(l-l''), &  l''\in \Omega_{0}^{p-1},\; w \in \Omega_{0}^{\theta'-ts-1}  \\
   i-t-u+\theta'(l-p),  &  u \in \Omega_{0}^{z-1-p(\theta'-ts)} 
\end{array}
\end{align}
By simplifying the above equation, we have:
\begin{align}\label{eq:non_eq2prime-polydot'''}
  & \mathbf{P'}(S_B(x)) \not \in \nonumber \\
  & \Bigg\{ \begin{array}{cc}
   \hat{i}-w+\theta'\hat{l}, &  
   \hat{i}\in \Omega_{-t}^{-1},\;
   \hat{l}\in \Omega_{-(p-1)}^{t-1},\; w \in \Omega_{0}^{\theta'-ts-1}  \\
   \hat{i}-u+\theta'\Tilde{l},  & 
   \hat{i}\in \Omega_{-t}^{-1},\; \tilde{l}\in \Omega_{-p}^{t-1-p},\; u \in \Omega_{0}^{z-1-p(\theta'-ts)} 
\end{array}
\end{align}
Knowing the fact that all powers in $\mathbf{P'}(S_B(x))$ are in $ \mathbb{N}$, we consider only $\hat{l}, \tilde{l} \ge 1$ as $\hat{l}, \tilde{l} < 1$ results in negative powers of $\mathbf{P'}(S_B(x))$\footnote{The reason is that $i'-w$ and $i'-u$ are always negative. If $\hat{l}, \tilde{l}$ are also negative or equal to zero, $i'-w+\theta'\hat{l}$ and $i'-u+\theta'\tilde{l}$ are negative.}. 
This results in:
\begin{align}\label{eq:non_eq2-polydot'''}
& \mathbf{P'}(S_B(x)) \not \in  \nonumber \\
& \Bigg\{ \begin{array}{cc}
   \mathbf{V}_1\\
   \mathbf{V}_2,
\end{array} \\
& =\Bigg\{ \begin{array}{cc}
   \hat{i}-w+\theta'\hat{l}, &  
   \hat{i}\in \Omega_{-t}^{-1},\;
   \hat{l}\in \Omega_1^{t-1},\; w \in \Omega_{0}^{\theta'-ts-1}\\
   \hat{i}-u+\theta'\Tilde{l},  & 
   \hat{i}\in \Omega_{-t}^{-1},\; \tilde{l}\in \Omega_1^{t-1-p},\; u \in \Omega_{0}^{z-1-p(\theta'-ts)}
\end{array}
\end{align} 
\begin{lemma} \label{lemma:D1_D2}
$\mathbf{V}_2$ defined in (\ref{eq:non_eq2-polydot'''}) is a subset of $\mathbf{V}_1$: $V_2 \subset V_1$. 
\end{lemma}
{\em Proof:}
To prove this lemma, we consider two cases of\footnote{Note that from the definition of $p=\min\{\floor{\frac{z-1}{\theta'-ts}},t-1\}$, $p$ is less than or equal to $t-1$.} (i) $p=t-1$ and (ii) $p<t-1$. For the first case of $p=t-1$, $V_2$ is an empty set as the upper bound of $\tilde{l}$, \ie $t-1-p$, becomes less than its lower bound, \ie $1$. Thus $\mathbf{V}_2 \subset \mathbf{V}_1$ for $p=t-1$. In the following, we consider the second case of $p<t-1$ and prove that $V_2 \subset V_1$. 
\begin{align}\label{eq:p}
    & p = \min\{\floor{\frac{z-1}{\theta'-ts}},t-1\}, \;\;\; p<t-1 \nonumber \\
    \Rightarrow & p = \floor{\frac{z-1}{\theta'-ts}} \nonumber \\
    \Rightarrow & p+1 > \frac{z-1}{\theta'-ts} \nonumber \\
    \Rightarrow & \theta'-ts > z-1-p(\theta'-ts) \nonumber \\
    \Rightarrow & \theta'-ts \geq z-p(\theta'-ts).
\end{align}

Using (\ref{eq:p}), $u \subset w$ in (\ref{eq:non_eq2-polydot'''}). In addition, $\Tilde{l} \subset \hat{l}$, as $p\ge 0$. Therefore, $\mathbf{V}_2$ is a subset of $\mathbf{V}_1$ for the second case of $p<t-1$, as well.  
This completes the proof. \hfill $\Box$

Using Lemma \ref{lemma:D1_D2}, we can reduce (\ref{eq:non_eq2-polydot'''}) to:
\begin{align}
    \mathbf{P'}(S_B(x)) \not\in \hat{i}-w+\theta' \hat{l}, \hat{i}\in \Omega_{-t}^{-1},\;
   \hat{l}\in \Omega_1^{t-1},\; w \in \Omega_{0}^{\theta'-ts-1}
\end{align}
By replacing $\theta'$ with its equivalence $t(2s-1)$, the range of variation for $\hat{i}-w$ is $\hat{i}-w \in \{-ts+1, \ldots, -1\}$. Therefore, by considering different values of $\hat{l}$, the above equation is expanded as:
\begin{align}
%
   & \mathbf{P'}(S_B(x)) \not\in \{\theta'-ts+1,\ldots,\theta'-1\},  \nonumber \\
   & \mathbf{P'}(S_B(x)) \not\in \{2\theta'-ts+1,\ldots,2\theta'-1\}, \nonumber \\ 
& \ldots \nonumber \\ 
   & \mathbf{P'}(S_B(x)) \not\in \{(t-1)\theta'-ts+1,\ldots,(t-1)\theta'-1\}.
\end{align}
Using the complement of the above intervals, the intervals that $\mathbf{P'}(S_B(x))$ can be selected from, is derived as follows:
\begin{align}
    & \mathbf{P'}(S_B(x)) \in \{0,\ldots,\theta'-ts\} \cup \{\theta',\ldots,2\theta'-ts\} \cup \ldots \cup \nonumber \\
    & \{(t-1)\theta',\ldots,+\infty\}.
\end{align}
This completes the proof of Lemma \ref{lem:P'(SB)-z large}.
\hfill $\Box$\\
  \begin{lemma}\label{lem:P'(SB)-z small}
If $z \leq \theta'-ts$ and $s,t \neq 1$, $\mathbf{P'}(S_B(x))$ is defined as the following:
\begin{align}\label{eq:P'(RB)_set_representation-z small}
    \mathbf{P'}(S_B(x)) = 
    &\Big(\bigcup\limits_{l'=0}^{t-2} \{\theta' l',\ldots,(l'+1)\theta'-z-t\}\Big) \\ 
    &\cup \{(t-1)\theta',\ldots,+\infty\}.
\end{align} 
\end{lemma}
{\em Proof:}
To determine $\mathbf{P'}(S_B(x))$, we need to find a subset of $\mathbf{P}(S_B(x))$ that satisfies C2. By replacing $\mathbf{P}(S_A(x))$ from Lemma \ref{lem:P(SA)-z small} in C2, we have:
\begin{align}\label{eq:satisfyC2-polydot-secondscenario}
   & \text{C2: } i+t(s-1)+\theta' l \not\in ts+r+\mathbf{P'}(S_B(x)).
   \end{align}
   Equivalently:
   \begin{align}\label{eq:satisfyC2-polydot-secondscenario'}
       \mathbf{P}(S_B(x)) \not\in i-r-t+\theta' l,
   \end{align}
   where $i,l \in \{0,\dots,t-1\}$ and $r \in \{0,\dots,z-1\}$. By expanding the above equation we have:
\begin{align}
   & \mathbf{P'}(S_B(x)) \not\in \{-z-t+1,\ldots,-1\}, \nonumber \\
   & \mathbf{P'}(S_B(x)) \not\in \{\theta'-z-t+1,\ldots,\theta'-1\},  \nonumber \\
   & \mathbf{P'}(S_B(x)) \not\in \{2\theta'-z-t+1,\ldots,2\theta'-1\}, \nonumber \\ 
& \ldots \nonumber \\ 
   & \mathbf{P'}(S_B(x)) \not\in \{(t-1)\theta'-z-t+1,\ldots,(t-1)\theta'-1\}. 
\end{align}
Using the complement of the above intervals, the intervals that $\mathbf{P'}(S_B(x))$ can be selected from, is derived as follows:
\begin{align}
    & \mathbf{P'}(S_B(x)) = \{0,\ldots,\theta'-z-t\} \cup \{\theta',\ldots,2\theta'-z-t\} \cup \nonumber \\ &\ldots \cup
     \{(t-1)\theta',\ldots,+\infty\}.
\end{align} 
This completes the proof.
\hfill $\Box$\\ \\ \\
\emph{(iii) Find the subset of $\mathbf{P}(S_B(x))$ that satisfies C3 in (\ref{eq:non_eq-polydot-thrm4}); we call this subset as $\mathbf{P''}(S_B(x))$}.

In this step, we consider the three cases of $s=1$, $t=1$ and $s,t \geq 2$, and derive $\mathbf{P''}(S_B(x))$ as summarized in Lemmas \ref{lem:p''SB-s=1}, \ref{lem:p''SB-t=1} and \ref{lem:p''SB-s,t not 1}.

\begin{lemma}\label{lem:p''SB-s=1}
If $s=1$, $\mathbf{P''}(S_B(x))$ is defined as the following:
\begin{align}\label{eq:P''(SB)_set_s=1}
    \mathbf{P''}(S_B(x))=\{t^2,\dots,+\infty\}.
\end{align}    
\end{lemma}
{\em Proof:}
By replacing $\mathbf{P}(C_A(x))$ from (\ref{eq:polydot-p(CA)-th}) in C3, we have
\begin{align}\label{eq:C3-polydot-eq1-s=1}
     \text{C3: }& i+tl \not\in  
     \{0,\ldots,t-1\}+\mathbf{P''}(S_B(x)),
\end{align}
which can be equivalently written as:
\begin{align}
     & \{0,\dots,t^2-1\} \not\in  
     \{0,\ldots,t-1\}+\mathbf{P''}(S_B(x)),
     \nonumber\\
     \Rightarrow & \{-t+1,\dots,t^2-1\} \not\in  
     \mathbf{P''}(S_B(x))\label{eq:C3-polydot-eq3-s=1}
\end{align}
From the above equation, the elements of $\mathbf{P''}(S_B(x))$ can be selected from any positive integer greater than $t^2-1$. This completes the proof. \hfill $\Box$

\begin{lemma}\label{lem:p''SB-t=1}
If $t=1$, $\mathbf{P''}(S_B(x))$ is defined as the following:
\begin{align}\label{eq:P''(SB)_set_t=1}
    \mathbf{P''}(S_B(x))=\{s,\dots,+\infty\}.
\end{align}    
\end{lemma}
{\em Proof:}
By replacing $\mathbf{P}(C_A(x))$ from (\ref{eq:polydot-p(CA)-th}) in C3, we have
\begin{align}\label{eq:C3-polydot-eq1-t=1}
     \text{C3: }& s-1 \not\in  
     \{0,\ldots,s-1\}+\mathbf{P''}(S_B(x)),
\end{align}
which can be equivalently written as:
\begin{align}\label{eq:C3-polydot-eq2-t=1}
     & \{0,\dots,s-1\} \not\in \mathbf{P''}(S_B(x)).
\end{align}
From the above equation, the elements of $\mathbf{P''}(S_B(x))$ can be selected from any positive integer greater than $s$
This completes the proof. \hfill $\Box$

\begin{lemma}\label{lem:p''SB-s,t not 1}
For any $s,t \geq 2$ and $z \in \mathbb{N}$, $\mathbf{P''}(S_B(x))$ is defined as the following:
\begin{align}\label{eq:P''(RB)_set_representation}
    \mathbf{P''}(S_B(x))= 
    &\Big(\bigcup\limits_{l''=0}^{t-2} \{ts+\theta' l'',\ldots,(l''+1)\theta'-t\}\Big) \nonumber\\
    & \cup \{ts+(t-1)\theta',\ldots,+\infty\}.
\end{align}.
\end{lemma}

{\em Proof:}
By replacing $\mathbf{P}(C_A(x))$ from (\ref{eq:polydot-p(CA)-th}) in C3, we have
\begin{align}\label{eq:non_eq3-polydot'}
     & i+t(s-1)+\theta' l \not\in  
     \{0,\ldots,ts-1\}+\mathbf{P''}(S_B(x)).
\end{align}
Equivalently,
\begin{align}
    \mathbf{P''}(S_B(x)) \not\in 
    \{-t+1,\ldots,ts-1\}+\theta' l.
\end{align}
By expanding the above equation for different values of $l$, we have:
\begin{align}
   & \mathbf{P''}(S_B(x)) \not\in \{-t+1,\ldots,ts-1\},  \nonumber \\
& \mathbf{P''}(S_B(x)) \not\in \{-t+1+\theta',\ldots,ts-1+\theta'\},   \nonumber \\
  &  \mathbf{P''}(S_B(x)) \not\in \{-t+1+2\theta',\ldots,ts-1+2\theta'\},   \nonumber \\
& \ldots   \nonumber \\
  &  \mathbf{P''}(S_B(x)) \not\in \{-t+1+(t-1)\theta',\ldots,ts-1+(t-1)\theta'\}. \nonumber  
\end{align}
We define $\mathbf{P''}(S_B(x))$ as the complement of the above intervals:
\begin{align}\label{eq:P''(RB)_set_representation-1}
    \mathbf{P''}(S_B(x))= 
    &\Big(\bigcup\limits_{l''=0}^{t-2} \{ts+\theta' l'',\ldots,(l''+1)\theta'-t\}\Big) \nonumber\\
    & \cup \{ts+(t-1)\theta',\ldots,+\infty\}.
\end{align}
\hfill $\Box$

\emph{(iv) 
Find the intersection of $\mathbf{P'}(S_B(x))$ and $\mathbf{P''}(S_B(x))$ to form $\mathbf{P}(S_B(x))$.} 

In this step, we consider four regions for the range of variable $z$, (a) $z > \theta'-ts, s,t \neq 1$, (b) $\theta'-ts-t < z \leq \theta'-ts, s,t \neq 1$, (c) $\frac{\theta'-ts-t+1}{2} < z \leq \theta'-ts-t, s,t \neq 1$, and (d) $z \leq \frac{\theta'-ts-t+1}{2}, s,t \neq 1$, as well as the special cases of (e) $s=1$ and (f) $t=1$, and calculate $\mathbf{P}(S_B(x)$ for each case, as summarized in Lemmas \ref{lem:P(SB)-z large}, \ref{lem:P(SB)-z medium}, \ref{lem:P(SB)-z small}, \ref{lem:P(SB)-z very small}, \ref{lem:P(SB)-s=1}, and \ref{lem:P(SB)-t=1}, respectively.

\begin{lemma}\label{lem:P(SB)-z large}
If $z > \theta'-ts$ and $s,t \neq 1$, the subsets of all powers of polynomials $S_B(x)$ with non-zero coefficients is defined as the following
\begin{align}\label{eq:P(RB)_set_representation-z large}
      \mathbf{P}(S_B(x)) 
        = 
        \{ts+(t-1)\theta'+r, 
        \; &0 \leq r \leq z-1,\; \nonumber \\
    & \theta'=t(2s-1) 
    \}.
\end{align}
\end{lemma}
{\em Proof:}
For this region, we use $\mathbf{P'}(S_B(x))$ defined in Lemma \ref{lem:P'(SB)-z large} and 
$\mathbf{P''}(S_B(x))$ defined in (\ref{eq:P''(RB)_set_representation}):
\begin{align}\label{P'(SB) set representation-M'}
   & \mathbf{P'}(S_B(x)) = \mathbf{M}_1' \cup \mathbf{M}_2', \nonumber \\
    & \mathbf{P''}(S_B(x)) = \mathbf{M}_1'' \cup \mathbf{M}_2'', 
\end{align} where,

\begin{align}\label{eq:def-M1'_M1''_M2'_M2''}
    & \mathbf{M}_1'= \bigcup\limits_{l'=0}^{t-2} \{\theta' l',\ldots,(l'+1)\theta'-ts\}, \nonumber \\
    & \mathbf{M}_1'' = \bigcup\limits_{l''=0}^{t-2} \{ts+\theta' l'',\ldots,(l''+1)\theta'-t\}, \nonumber \\
    & \mathbf{M}_2' = \{(t-1)\theta',\ldots,+\infty\}, \nonumber \\
   & \mathbf{M}_2'' = \{ts+(t-1)\theta',\ldots,+\infty\}.
\end{align}

The intersection of $\mathbf{P'}(S_B(x))$ and $\mathbf{P''}(S_B(x))$ is calculated as:

\begin{align}\label{P(SB)=P'(SB)capP''(SB)}
      \mathbf{P'}(S_B(x)) \cap \mathbf{P''}(S_B(x))
      = &\big(\mathbf{M}_1' \cup \mathbf{M}_2' \big) \cap \big(\mathbf{M}_1'' \cup \mathbf{M}_2'' \big) \nonumber \\
      = & \big(\mathbf{M}_1' \cap \mathbf{M}_1''\big) \cup \big(\mathbf{M}_2' \cap \mathbf{M}_1''\big) \cup \nonumber \\
      & \big(\mathbf{M}_1' \cap \mathbf{M}_2''\big) \cup \big(\mathbf{M}_2' \cap \mathbf{M}_2''\big).
\end{align}
In the following, we calculate $\big(\mathbf{M}_1' \cap \mathbf{M}_1''\big)$, $\big(\mathbf{M}_2' \cap \mathbf{M}_1''\big)$, $\big(\mathbf{M}_1' \cap \mathbf{M}_2''\big)$, and $\big(\mathbf{M}_2' \cap \mathbf{M}_2''\big)$, separately.

\begin{itemize}
    \item Calculating $\big(\mathbf{M}_1' \cap \mathbf{M}_1''\big)$

To calculate $\big(\mathbf{M}_1' \cap \mathbf{M}_1''\big)$, we consider each subset of $\mathbf{M}_1'$, \ie $\{\theta' l',\ldots,(l'+1)\theta'-ts\}$ and show that this subset does not have any overlap with any of the subsets of $\mathbf{M}_1''$, \ie $\{ts+\theta' l'',\ldots,(l''+1)\theta'-t\},  0\leq l''< t-2$; This results in $\big(\mathbf{M}_1' \cap \mathbf{M}_1''\big)=\emptyset$. For this purpose, (i) first we consider the subsets of $\mathbf{M}_1''$, for which $l''< l'$ and show that $\{\theta' l',\ldots,(l'+1)\theta'-ts\}$ falls to the right side of all intervals $\{ts+\theta' l'',\ldots,(l''+1)\theta'-t\}, 0\leq l''< l'$, and (ii) second we consider the subsets of $\mathbf{M}_1''$, for which $l''\ge l'$ and show that $\{\theta' l',\ldots,(l'+1)\theta'-ts\}$ falls to the left  side of all intervals $\{ts+\theta' l'',\ldots,(l''+1)\theta'-t\}, l' \leq l'' \leq t-2 $. 

\begin{figure*}[ht]
		\centering
		\scalebox{1.1}{
		\includegraphics[width=15cm]{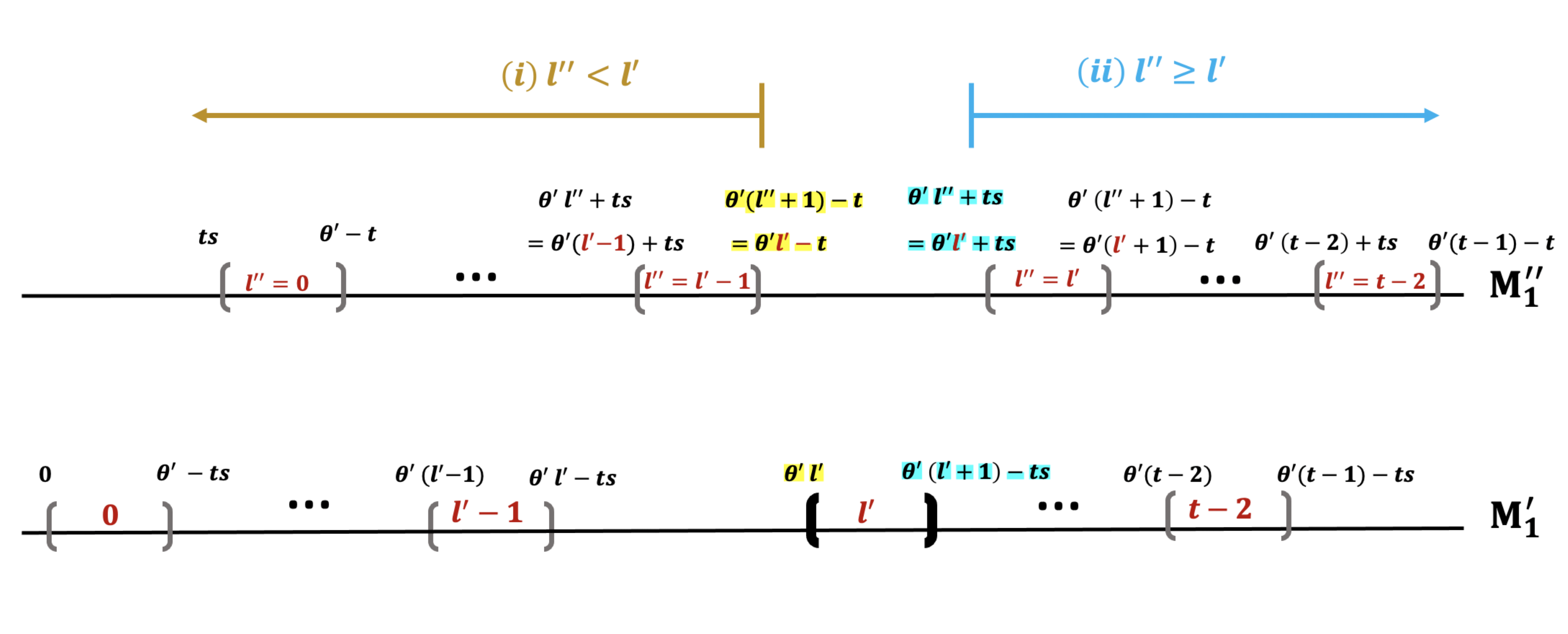}}
	\caption{An illustration showing that $\mathbf{M}_1' \cap \mathbf{M}_1''=\emptyset$ holds in Lemma \ref{lem:P(SB)-z large}. 
	}
\label{Fig_union-M_i'andM_j''=empty}
\vspace{-5pt}
\end{figure*}

(i) 
$l''<l'$: In this case, the largest element of all subsets of $\mathbf{M}_1''$, \ie $\theta' (l''+1)-t$ is less than the smallest element of $\{\theta' l',\ldots,(l'+1)\theta'-ts\}$, as shown in Fig. \ref{Fig_union-M_i'andM_j''=empty}. The reason is that:
    \begin{align}\label{eq:M'1capM''1-l''<l'}
      l'' <l' \Rightarrow & l''+1 \leq l', \nonumber \\
      \Rightarrow &  \theta'(l''+1) \leq \theta' l', \nonumber \\
      \Rightarrow & \theta'(l''+1)-t < \theta' l'.
\end{align}
    
(ii) 
$l'' \ge l'$. In this case, the smallest element of all subsets of $\mathbf{M}_1''$, \ie $\theta' l''+ts$, is greater than the largest element of $\{\theta' l',\ldots,(l'+1)\theta'-ts\}$, as shown in Fig. \ref{Fig_union-M_i'andM_j''=empty}. The reason is that:
    \begin{align}
          l' \leq l''\Rightarrow & \theta' l' \leq \theta' l'', \nonumber \\
          \Rightarrow & \theta' l' -t < \theta' l'', \nonumber \\
          \Rightarrow & \theta' l' -t+ts < \theta' l''+ts, \nonumber \\ 
          \Rightarrow & \theta' l' -t+2ts-ts < \theta' l''+ts, \nonumber \\
          \Rightarrow & \theta' l' +\theta'-ts < \theta' l''+ts, \nonumber \\
          \Rightarrow & \theta' (l'+1)-ts < \theta' l''+ts.
   \end{align}
From (i) and (ii) discussed in the above, we conclude that:
\begin{equation}\label{eq:m'1capm''1} \mathbf{M}_1' \cap \mathbf{M}_1''=\emptyset \end{equation}\\

\item Calculating 
$\big(\mathbf{M}_2' \cap \mathbf{M}_1''\big)$

The largest element of $\mathbf{M}_1''$, $(t-1)\theta' -t$, is always less than $(t-1)\theta'$, which is the smallest element of $\mathbf{M}_2'$. This results in: \begin{equation}\label{eq:m'2capm''1} \mathbf{M}_2' \cap \mathbf{M}_1'' = \emptyset\end{equation}\\ \\
\item Calculating 
$\big(\mathbf{M}_1' \cap \mathbf{M}_2''\big)$

The largest element of $\mathbf{M}_1'$, \ie $(t-1)\theta' -ts$ is always less than $(t-1)\theta'+ts$, which is the smallest element of $\mathbf{M}_2''$. This results in: \begin{equation}\label{eq:m'1capm''2}\mathbf{M}_1' \cap \mathbf{M}_2'' = \emptyset \end{equation}
\item Calculating $\big(\mathbf{M}_2' \cap \mathbf{M}_2''\big)$

\begin{align}\label{eq:m'2capm''2}
      \mathbf{M}_2' \cap \nonumber \mathbf{M}_2'' = &\{(t-1)\theta',\ldots,+\infty\} \cap \\ &\{ts+(t-1)\theta',\ldots,+\infty\} \nonumber \\
      = & \{ts+(t-1)\theta',\ldots,+\infty\}.
\end{align}
\end{itemize}
From (\ref{P(SB)=P'(SB)capP''(SB)}), (\ref{eq:m'1capm''1}), (\ref{eq:m'2capm''1}), (\ref{eq:m'1capm''2}), and (\ref{eq:m'2capm''2}), we have:

\begin{align}\label{eq:P(RB)_set_representation}
      \mathbf{P'}(S_B(x)) \cap \mathbf{P''}(S_B(x))
      = & \{ts+(t-1)\theta',\ldots,+\infty\},
\end{align}
from which the elements of $\mathbf{P}(S_B(x))$ can be selected. As there are $z$ colluding workers, the size of $\mathbf{P}(S_B(x))$ should be $z$
, \ie $|\mathbf{P}(S_B(x))|=z$. On the other hand, since our goal is to reduce the degree of $F_B(x)$ as much as possible, we select the $z$ smallest elements of the set shown in (\ref{eq:P(RB)_set_representation}) to form  
$\mathbf{P}(S_B(x))$: 
\begin{align}
    \mathbf{P}(S_B(x)) = &  \{ts+(t-1)\theta',\ldots,ts+(t-1)\theta'+z-1\} \nonumber \\
\end{align}
This completes the proof of Lemma \ref{lem:P(SB)-z large}.
\hfill $\Box$\\

\begin{lemma}\label{lem:P(SB)-z medium}
If $\theta'-ts-t < z \leq \theta'-ts$ and $s,t \neq 1$, the subsets of all powers of polynomials $S_B(x)$ with non-zero coefficients is defined as the following:
\begin{align}\label{eq:P(RB)_set_representation-z medum}
      \mathbf{P}(S_B(x)) 
       = 
       \{ts+(t-1)\theta'+r,
        \; &0 \leq r \leq z-1,\; \nonumber \\
    & \theta'=t(2s-1)
    \}.
\end{align}
\end{lemma}
{\em Proof:}
For this region, we use $\mathbf{P'}(S_B(x))$ defined in Lemma \ref{lem:P'(SB)-z small} and 
$\mathbf{P''}(S_B(x))$ defined in (\ref{eq:P''(RB)_set_representation}):
\begin{align}\label{P'(SB) set representation-M'--}
   & \mathbf{P'}(S_B(x)) = \mathbf{M}_1' \cup \mathbf{M}_2', \nonumber \\
    & \mathbf{P''}(S_B(x)) = \mathbf{M}_1'' \cup \mathbf{M}_2'', 
\end{align} where,

\begin{align}\label{eq:def-M1'_M1''_M2'_M2''case2}
    & \mathbf{M}_1'= \bigcup\limits_{l'=0}^{t-2} \{\theta' l',\ldots,(l'+1)\theta'-z-t\}, \nonumber \\
    & \mathbf{M}_1'' = \bigcup\limits_{l''=0}^{t-2} \{ts+\theta' l'',\ldots,(l''+1)\theta'-t\}, \nonumber \\
    & \mathbf{M}_2' = \{(t-1)\theta',\ldots,+\infty\}, \nonumber \\
   & \mathbf{M}_2'' = \{ts+(t-1)\theta',\ldots,+\infty\}.
\end{align}
Similar to the proof of Lemma \ref{lem:P(SB)-z large}, we find $\mathbf{P'}(S_B(x)) \cap \mathbf{P''}(S_B(x))$ by calculating
$\big(\mathbf{M}_1' \cap \mathbf{M}_1''\big) \cup \big(\mathbf{M}_2' \cap \mathbf{M}_1''\big) \cup \big(\mathbf{M}_1' \cap \mathbf{M}_2''\big) \cup \big(\mathbf{M}_2' \cap \mathbf{M}_2''\big)$ with the only difference that the definition of $\mathbf{M}_1'$ in (\ref{eq:def-M1'_M1''_M2'_M2''}) is different from the definition of $\mathbf{M}_1'$ in (\ref{eq:def-M1'_M1''_M2'_M2''case2}).

\begin{itemize}
    \item Calculating $\big(\mathbf{M}_1' \cap \mathbf{M}_1''\big)$
    
    We show that each subset of $\mathbf{M}_1'$, \ie $\{\theta' l',\ldots,(l'+1)\theta'-z-t\}$ does not have any overlap with any of the subsets of $\mathbf{M}_1''$, \ie $\{ts+\theta' l'',\ldots,(l''+1)\theta'-t\}, 0\leq l''< t-2$. Similar to the proof of Lemma \ref{lem:P(SB)-z large}, we consider two cases of $l''<l'$ and $l'' \ge l'$.
    
    (i) $l''<l'$: As shown in (\ref{eq:M'1capM''1-l''<l'}), all subsets of $\mathbf{M}_1''$ falls to the left of the subset of $\mathbf{M}_1'$.
    
    (ii) $l'' \ge l'$: In this case, the smallest element of all subsets of $\mathbf{M}_1''$, \ie $\theta' l''+ts$, is greater than the largest element of $\{\theta' l',\ldots,(l'+1)\theta'-z-t\}$. The reason is that:
    \begin{align}\label{eq:M'1capM''1-scenario2-p1}
          l' \leq l''\Rightarrow & \theta' l' \leq \theta' l'', \nonumber \\
          \Rightarrow & \theta' l'+ts < \theta' l''+ts.
   \end{align}
On the other hand we have:
\begin{align}\label{eq:M'1capM''1-scenario2-p2}
& \theta'-ts-t < z \nonumber \\
& \Rightarrow \theta' l'-z < \theta' l'-\theta'+ts+t \nonumber \\
& \Rightarrow (l'+1)\theta'-z-t < \theta' l'+ts.
\end{align}
Therefore, from (\ref{eq:M'1capM''1-scenario2-p1}) and (\ref{eq:M'1capM''1-scenario2-p2}) we have:
\begin{align}
& (l'+1)\theta'-z-t < \theta' l''+ts.
\end{align}
From (i) and (ii) discussed in the above, we conclude that:
\begin{equation}\label{eq:m'1capm''12} \mathbf{M}_1' \cap \mathbf{M}_1''=\emptyset \end{equation}

\item Calculating $\big(\mathbf{M}_2' \cap \mathbf{M}_1''\big)$

The largest element of $\mathbf{M}_1''$, $(t-1)\theta' -t$, is always less than $(t-1)\theta'$, which is the smallest element of $\mathbf{M}_2'$. This results in: \begin{equation}\label{eq:m'2capm''12} \mathbf{M}_2' \cap \mathbf{M}_1'' = \emptyset\end{equation}\\ \\
\item Calculating $\big(\mathbf{M}_1' \cap \mathbf{M}_2''\big)$

The largest element of $\mathbf{M}_1'$, \ie $(t-1)\theta'-z -t$ is always less than $(t-1)\theta'+ts$, which is the smallest element of $\mathbf{M}_2''$. This results in: \begin{equation}\label{eq:m'1capm''22}\mathbf{M}_1' \cap \mathbf{M}_2'' = \emptyset \end{equation}

\item Calculating $\big(\mathbf{M}_2' \cap \mathbf{M}_2''\big)$

\begin{align}\label{eq:m'2capm''22}
      \mathbf{M}_2' \cap \nonumber \mathbf{M}_2'' = &\{(t-1)\theta',\ldots,+\infty\} \cap \\ &\{ts+(t-1)\theta',\ldots,+\infty\} \nonumber \\
      = & \{ts+(t-1)\theta',\ldots,+\infty\}.
\end{align}
\end{itemize}
From (\ref{P(SB)=P'(SB)capP''(SB)}), (\ref{eq:m'1capm''12}), (\ref{eq:m'2capm''12}), (\ref{eq:m'1capm''22}), and (\ref{eq:m'2capm''22}), we have:

\begin{align}\label{eq:P(RB)_set_representation2}
      \mathbf{P'}(S_B(x)) \cap \mathbf{P''}(S_B(x))
      = & \{ts+(t-1)\theta',\ldots,+\infty\}.
\end{align}
$\mathbf{P}(S_B(x))$ is formed by selecting the $z$ smallest elements of the set shown in (\ref{eq:P(RB)_set_representation2}):
\begin{align}
    \mathbf{P}(S_B(x)) = \{ts+(t-1)\theta',\ldots,ts+(t-1)\theta'+z-1\}
\end{align}

This completes the proof.
\hfill $\Box$

\begin{lemma}\label{lem:P(SB)-z small}
If $\frac{\theta'-ts-t+1}{2} <z \leq \theta'-ts-t$ and $s,t \neq 1$, the subsets of all powers of polynomials $S_B(x)$ with non-zero coefficients is defined as the following:
\begin{align}\label{eq:P(RB)_set_representation-zsmall}
    \mathbf{P}(S_B(x)) = & \mathbf{P'}(S_B(x)) \cap \mathbf{P''}(S_B(x)) \nonumber \\
    = & \Big(\bigcup\limits_{l''=0}^{p'-1}\{ts+\theta'l'', \ldots, (l''+1)\theta'-z-t\}\Big) \nonumber \\ & \cup \{ts+p'\theta',\dots, \nonumber \\ & ts+p'\theta'+z-1-p'(\theta'-t-ts-z+1)\}\\
    =&\{ts+\theta'l'+d, d\in\Omega_0^{\theta'-t-ts-z}, l'\in \Omega_0^{p'-1}\} \nonumber \\
    &\cup \{ts+\theta'p'+v, v\in\Omega_0^{z-1-p'(\theta'-ts-t-z+1)}\}\label{eq:psb3}.
\end{align}
\end{lemma}
{\em Proof:}
In this scenario, $\mathbf{P'}(S_B(x))$ and $\mathbf{P''}(S_B(x))$ are equal to the previous case, as shown in (\ref{P'(SB) set representation-M'--}) and (\ref{eq:def-M1'_M1''_M2'_M2''case2}). The difference between this case and the previous case is that $\big(\mathbf{M}_1' \cap \mathbf{M}_1''\big)$ is no longer an empty set. The reason is that as we can see in Fig. \ref{fig:M1'M1''overlap}, each $l^{\text{th}}$ subset of $\mathbf{M}_1'$, \ie $\{\theta' l',\ldots,(l'+1)\theta'-z-t\}, l'=l-1$ has overlap with each $l^{\text{th}}$ subset of $\mathbf{M}_1''$, \ie $\{ts+\theta' l'',\ldots,(l''+1)\theta'-t\}, l''=l-1$: 

\begin{figure}[t]
		\centering
		\includegraphics[width=9.2cm]{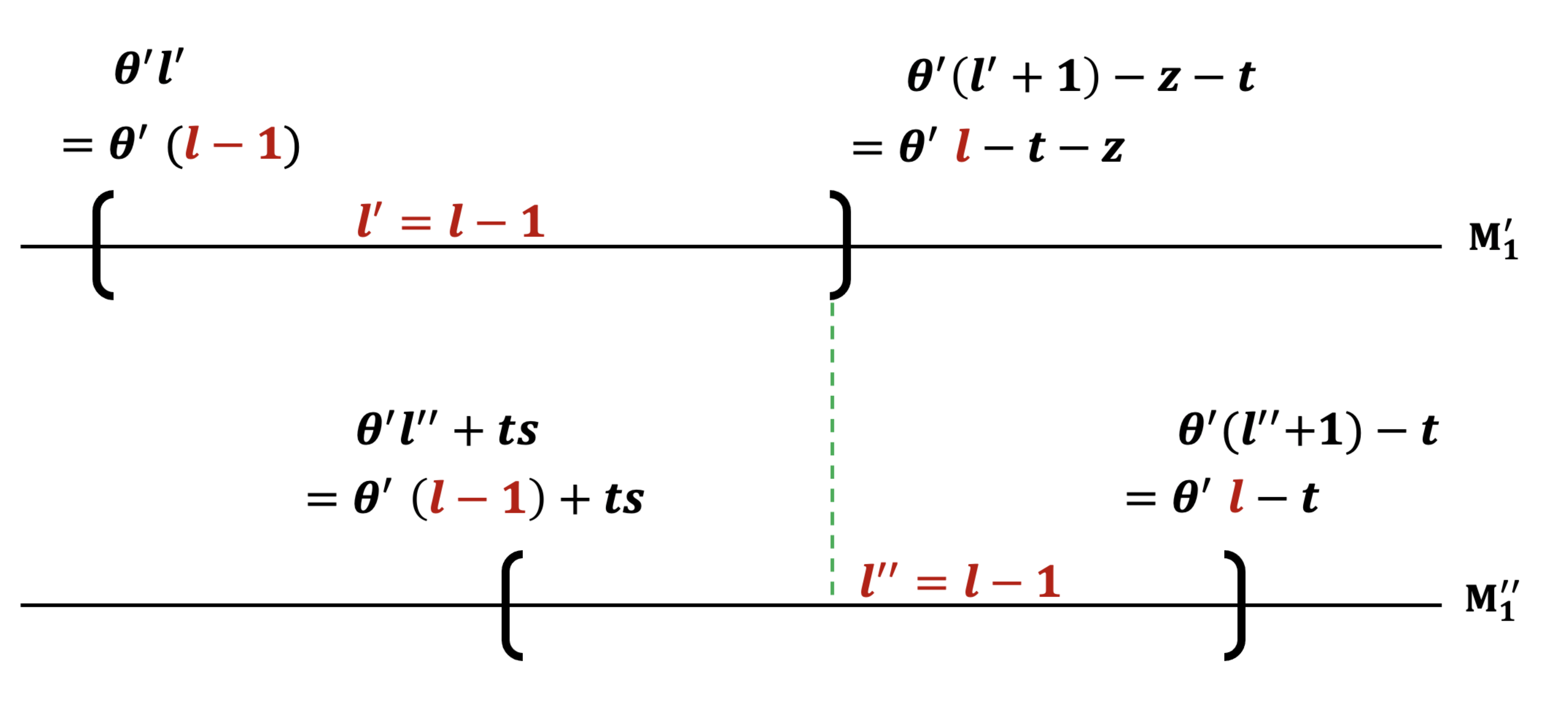}
	\caption{Illustration of the overlap between $\mathbf{M}_1'$ and $\mathbf{M}_1''$ in Lemma \ref{lem:P(SB)-z small}.
	}
\label{fig:M1'M1''overlap}
\vspace{-15pt}
\end{figure}

\begin{align}
     & z \leq \theta'-ts-t \nonumber \\
    & \Rightarrow -z \geq -\theta'+ts+t \nonumber \\ 
    & \Rightarrow \theta' l -t-z \geq \theta' (l-1)+ts \nonumber \\ 
    & \Rightarrow \theta' l -t-z > \theta' (l-1)+ts \nonumber \\
    & \Rightarrow \theta'(l-1) < ts+\theta'(l-1) < l\theta'-z-t < l\theta'-t.
\end{align}
Therefore, we have:
\begin{equation}\label{eq:m'1capm''13}
    \mathbf{M}_1' \cap \mathbf{M}_1'' = \bigcup\limits_{l''=0}^{t-2}\{ts+\theta'l'', \ldots, (l''+1)\theta'-z-t\}
\end{equation}

$(\mathbf{M}_2' \cap \mathbf{M}_1''), (\mathbf{M}_1' \cap \mathbf{M}_2'')$, and $(\mathbf{M}_2' \cap \mathbf{M}_2'')$ can be calculated the same way as they are calculated in the previous case. Therefore, from (\ref{P(SB)=P'(SB)capP''(SB)}), (\ref{eq:m'1capm''13}), (\ref{eq:m'2capm''12}), (\ref{eq:m'1capm''22}), and (\ref{eq:m'2capm''22}), we have:
\begin{align}\label{eq:P(RB)_set_representation when z less than alpha-ts-t}
    &\mathbf{P'}(S_B(x)) \cap \mathbf{P''}(S_B(x)) \nonumber \\
    = &\bigcup\limits_{l''=0}^{t-2}\{ts+\theta'l'', \ldots, (l''+1)\theta'-z-t\} \nonumber \\ &\cup \{ts+(t-1)\theta',\dots,\infty\}.
\end{align}
$\mathbf{P}(S_B(x))$ is formed by selecting the $z$ smallest elements of the set shown in (\ref{eq:P(RB)_set_representation when z less than alpha-ts-t}).
This set consists of $t-1$ finite sets and one infinite set, where each finite set contains $(\theta'-ts-t-z+1)=(ts-2t-z+1)$\footnote{$\theta'$ is defined as $\theta'=t(2s-1)$.} elements. For the case of 
$\frac{\theta'-ts-t+1}{2} < z \leq \theta'-ts-t$, or equivalently $\frac{t(s-2)+1}{2} < z \leq t(s-2)$, $z$ is greater than $ts-2t-z+1$ and thus more than one finite set of (\ref{eq:P(RB)_set_representation when z less than alpha-ts-t}) is required to form $\mathbf{P}(S_B(x))$. Therefore we select $p'+1 \ge 2$ sets, where $p'$ is defined as $p'=
\min\{\floor{\frac{z-1}{ts-2t-z+1}},t-1\}$. With this definition, the first $p$ selected intervals are selected in full, in other words, we select $p'(ts-2t-z+1)$ elements to form the first $p'$ intervals of $\mathbf{P}(S_B(x))$. The remaining $z-p'(ts-2t-z+1)=z-p'(\theta'-t-ts-z+1)$ elements are selected from the $(p'+1)^{\text{st}}$ interval of (\ref{eq:P(RB)_set_representation when z less than alpha-ts-t}). This results in:

\begin{align} 
    \mathbf{P}(S_B(x)) 
    = &\{ts,\dots,\theta'-t-z\} \nonumber \\ & \cup \{ts+\theta',\dots,2\theta'-t-z\} \cup \dots \nonumber \\ & \cup \{ts+p'\theta',\dots, \nonumber \\ & ts+p'\theta'+z-1-p'(\theta'-t-ts-z+1)\}. \nonumber
\end{align}
This completes the proof.
\hfill $\Box$
\begin{lemma}\label{lem:P(SB)-z very small}
If $z \leq \frac{\theta'-ts-t+1}{2}$ and $s,t \neq 1$, the subsets of all powers of polynomial $S_B(x)$ with non-zero coefficients is defined as the following:
\begin{align}\label{eq:P(RB)_set_representation-z very small}
\mathbf{P}(S_B(x)) & = \{ts,\dots,ts+z-1\}\nonumber \\
&=\{ts+v, v\in \Omega_0^{z-1}\}.
\end{align}
\end{lemma}
{\em Proof:}
This case is similar to the previous case, where $\frac{\theta'-ts-t+1}{2} < z \leq \theta'-ts-t$, with the difference that the first subset of (\ref{eq:P(RB)_set_representation when z less than alpha-ts-t}) is sufficient to form $\mathbf{P}(S_B(x))$. The reason is that:
\begin{align}
     & z \leq \frac{\theta'-ts-t+1}{2} \nonumber \\
    & \Rightarrow z \leq \theta'-ts-t-z+1,
    & \Rightarrow z \leq ts-2t-z+1,
\end{align}
and thus the first subset with $ts-2t-z+1$ elements is sufficient to form $z$ elements of $\mathbf{P}(S_B(x))$ as shown in (\ref{eq:P(RB)_set_representation-z very small}). This completes the proof. \hfill $\Box$

\begin{lemma}\label{lem:P(SB)-s=1}
If $s=1$, the set of all powers of polynomial $S_B(x)$ with non-zero coefficients is defined as the following:
\begin{align}\label{eq:finiteP(BA)-s=1}
\mathbf{P}(S_B(x)) = \{t^2,\dots,t^2+z-1\}, \nonumber\\
= \{t^2+r, r\in \Omega_0^{z-1}\}.
\end{align}
\end{lemma}
{\em Proof:}
In this scenario, from lemma \ref{lem:P'(SB)-s=1}, we have $\mathbf{P'}(S_B(x))=\{0,\dots,+\infty\}$, and from Lemma \ref{lem:p''SB-s=1} we have $\mathbf{P''}(S_B(x))=\{t^2,\dots,+\infty\}$. Therefor, in this scenario the intersection of $\mathbf{P'}(S_B(x))$ and $\mathbf{P''}(S_B(x))$ is equal to $\{t^2,\dots,+\infty\}$, and $\mathbf{P}(S_B(x))$ is formed by selecting the $z$ smallest elements of $\{t^2,\dots,+\infty\}$, as shown in (\ref{eq:finiteP(BA)-s=1}). This completes the proof. \hfill $\Box$

\begin{lemma}\label{lem:P(SB)-t=1}
If $t=1$, the set of all powers of polynomial $S_B(x)$ with non-zero coefficients is defined as the following:
\begin{align}\label{eq:finiteP(BA)-t=1}
\mathbf{P}(S_B(x)) = \{s,\dots,s+z-1\}, \nonumber\\
= \{s+r, r\in \Omega_0^{z-1}\}.
\end{align}
\end{lemma}
{\em Proof:}
In this scenario, from lemma \ref{lem:P'(SB)-t=1}, we have $\mathbf{P'}(S_B(x))=\{0,\dots,+\infty\}$, and from Lemma \ref{lem:p''SB-t=1} we have $\mathbf{P''}(S_B(x))=\{s,\dots,+\infty\}$. Therefor, in this scenario the intersection of $\mathbf{P'}(S_B(x))$ and $\mathbf{P''}(S_B(x))$ is equal to $\{s,\dots,+\infty\}$, and $\mathbf{P}(S_B(x))$ is formed by selecting the $z$ smallest elements of $\{s,\dots,+\infty\}$, as shown in (\ref{eq:finiteP(BA)-t=1}). This completes the proof. \hfill $\Box$ 
%

$S_A(x)$ in (\ref{eq:FA1andFA2PolyDotCMPC}) can be directly derived from Lemmas \ref{lem:P(SA)-z large}, \ref{lem:P(SA)-z small}, and \ref{lem:P(SA)-s=1t=1}. Note that (i) when $z\leq \theta'-ts$, we have $p=0$ by definition and thus $ts+\theta' p+u$ in (\ref{eq:FA2PolyDotCMPC}) is equal to $ts+u$ in (\ref{eq:finiteP(SA)-polydot-second-scenario}), (ii) when $s=1$, we have $p=t-1$ and $\theta'=t$ by definition and thus $ts+\theta' p+u$ in (\ref{eq:FA2PolyDotCMPC}) is equal to $t^2+u$ in (\ref{eq:psafors=1}), and (iii) when $t=1$, we have $p=0$ by definition and thus $ts+\theta' p+u$ in (\ref{eq:FA2PolyDotCMPC}) is equal to $s+u$ in (\ref{eq:psafort=1}). 
%
%
Next we explain how to derive (\ref{eq:FBPolyDotCMPC}).

$S_B(x)$ in (\ref{eq:FBPolyDotCMPC}) can be directly derived from Lemmas \ref{lem:P(SB)-z large}, \ref{lem:P(SB)-z medium}, \ref{lem:P(SB)-z small}, \ref{lem:P(SB)-z very small}, \ref{lem:P(SB)-s=1}, and \ref{lem:P(SB)-t=1}. Note that (i) when $z > \theta'-ts, s,t \neq 1$ or $\theta'-ts-t < z \leq \theta'-ts, s,t \neq 1$, $\mathbf{P}(S_B(x))$ in (\ref{eq:P(RB)_set_representation-z large}) and (\ref{eq:P(RB)_set_representation-z medum}) is equal to the powers of $S_B(x)$ in (\ref{eq:FB1PolyDotCMPC}), (ii) when $\frac{\tau+1}{2}=\frac{\theta'-ts-t+1}{2} <z \leq \theta'-ts-t=\tau, s,t \neq 1$, $\mathbf{P}(S_B(x))$ in (\ref{eq:psb3}) is equal to the powers of $S_B(x)$ in (\ref{eq:FB2PolyDotCMPC}), (iii) when $z \leq \frac{\theta'-ts-t+1}{2}, s,t \neq 1$, $\mathbf{P}(S_B(x))$ in (\ref{eq:P(RB)_set_representation-z very small}) is equal to the powers of $S_B(x)$ in (\ref{eq:FB3PolyDotCMPC}), (iv) when $s=1$, we have $\theta' = t$ by definition, and thus $ts+\theta' (t-1)+r$ in (\ref{eq:FB1PolyDotCMPC}) is equal to $t^2+r$ in (\ref{eq:finiteP(BA)-s=1}), and (v) when $t=1$, $ts+\theta' (t-1)+r$ in (\ref{eq:FB1PolyDotCMPC}) is equal to $s+r$ in (\ref{eq:finiteP(BA)-t=1}). 

This completes the derivation of (\ref{eq:FA1andFA2PolyDotCMPC}) and (\ref{eq:FBPolyDotCMPC}).
\hfill $\Box$

\section*{Appendix B: Proof of Theorem \ref{th:N_PolyDot}}
To prove this theorem, we first consider the two cases of $t=1$ and $s=1$ separately and in the rest of this appendix, we consider $s, t \neq 1$.

\begin{lemma}\label{lemma:psit=1}
For $t=1$, $N_{\text{PolyDot-CMPC}}=2s+2z-1=(p+2)ts+\theta'(t-1)+2z-1=\psi_1$.\end{lemma}

{\em Proof:} For $t=1$, $p=0$ by definition. From (\ref{eq:FA2PolyDotCMPC}) and (\ref{eq:FB1PolyDotCMPC}) and by replacing $p$ with $0$, $F_A(x)$ and $F_B(x)$ are calculated as the following:
\begin{align}\label{eq:FA PolyDot-CMPC-t=1}
    F_{A}(x) = & \sum_{j=0}^{s-1} A_{j}x^{j}
    + \sum_{u=0}^{z-1}\bar{A}_{u}x^{s+u},
\end{align}
\begin{align}\label{eq:FB PolyDot-CMPC-t=1}
    F_{B}(x) = & \sum_{k=0}^{s-1} B_{k}x^{s-1-k}
    + \sum_{r=0}^{z-1}\bar{B}_{r}x^{s+r},
\end{align}
which are equal to the secret shares of Entangled-CMPC \cite{8613446}, for $t=1$. Thus, in this case PolyDot-CMPC and Entangled-CMPC are equivalent and as a result we have $N_{\text{PolyDot-CMPC}}=N_{\text{Entangled-CMPC}}=2s+2z-1$ \cite{8613446}, where by replacing $p=0$, we have $2s+2z-1=(p+2)ts+\theta'(t-1)+2z-1=\psi_1$. This completes the proof. \hfill $\Box$

\begin{lemma}\label{lemma:psis=1}
For $s=1$, 
\begin{align}
    N_{\text{PolyDot-CMPC}}
    =\begin{cases}
    2t^2+2z-1=\psi_1 & z>t \\
    t^2+2t+tz-1=\psi_6 & z \leq t
     \end{cases}
\end{align}
\end{lemma}
{\em Proof:}
For $s=1$, $\theta'=t$ and $p=t-1$ by definition. From (\ref{eq:FA2PolyDotCMPC}) and (\ref{eq:FB1PolyDotCMPC}) and by replacing $\theta'$ and $p$ with $t$ and $t-1$, respectively, $F_A(x)$ and $F_B(x)$ are calculated as the following:
\begin{align}\label{eq:FA PolyDot-CMPC-s=1}
    F_{A}(x) = & \sum_{i=0}^{t-1} A_{i}x^{i}
    + \sum_{u=0}^{z-1}\bar{A}_{u}x^{t^2+u},
\end{align}
\begin{align}\label{eq:FB PolyDot-CMPC-s=1}
    F_{B}(x) = & \sum_{l=0}^{t-1} B_{l}x^{tl}
    + \sum_{r=0}^{z-1}\bar{B}_{r}x^{t^2+r},
\end{align}
which are equal to the secret shares of Entangled-CMPC \cite{8613446}, for $s=1$. Thus, in this case PolyDot-CMPC and Entangled-CMPC are equivalent and as a result, we have:
\begin{align}
    &N_{\text{PolyDot-CMPC}}=\nonumber\\
    &N_{\text{Entangled-CMPC}}
    =\begin{cases}
    2t^2+2z-1 & z>t \\
    t^2+2t+tz-1 & z \leq t,
     \end{cases}
\end{align}
where by replacing $p=t-1$ and $\theta'=t$, we have $\psi_1=(p+2)ts+\theta'(t-1)+2z-1=2t^2+2z-1$ and $\psi_6=t^2+2t+tz-1$. This completes the proof. \hfill $\Box$

Now, we consider $s,t \neq 1$. The required number of workers is equal to the number of terms in $H(x)=F_A(x)F_B(x)$ with non-zero coefficients. The set of all powers in polynomial $H(x)$ with non-zero coefficients, shown by $\mathbf{P}({H}(x))$, is equal to:

\begin{align}\label{eq:PHx}
 \mathbf{P}({H}(x)) = \mathbf{D}_1 \cup  \mathbf{D}_2\cup \mathbf{D}_3 \cup \mathbf{D}_4,
 \end{align}
where
\begin{align}
     & \mathbf{D}_1 = \mathbf{P}(C_A(x))+\mathbf{P}(C_B(x))
 \end{align}
\begin{align}
    & \mathbf{D}_2  =\mathbf{P}(C_A(x))+\mathbf{P}(S_B(x))
\end{align}
\begin{align}\label{eq:d3definition}
    & \mathbf{D}_3=\mathbf{P}(S_A(x))+\mathbf{P}(C_B(x))
\end{align}
\begin{align}\label{eq:d4definition}
    & \mathbf{D}_4=\mathbf{P}(S_A(x))+\mathbf{P}(S_B(x))
\end{align}
%
 
Using (\ref{eq:polydot-p(CA)-th}) and (\ref{eq:polydot-p(CB)-th}), $\mathbf{D}_1$ is calculated as:
 \begin{align}\label{eq:d1}
      \mathbf{D}_1 = & \mathbf{P}(C_{A}(x))+\mathbf{P}(C_B(x)) \nonumber \\
     = &  \{i'+tj
     : 0 \leq i' \leq t-1,\; 0 \leq j \leq s-1,\} \nonumber \\ 
     &+ \{tq'+\theta' l' 
     : 0 \leq l' \leq t-1,\; 0 \leq q' \leq s-1\} \nonumber \\
     = &  \{i'+t(j+q')+\theta' l':0 \leq i',l' \leq t-1,\; \nonumber \\
     & 0 \leq j, q' \leq s-1,\}\nonumber \\
     = & \{i'+tj'+\theta' l':0 \leq i',l' \leq t-1,\; 0 \leq j' \leq 2s-2 \}\nonumber \\
     = & \{0,\ldots,t(2s-1)-1\} + \{\theta'l' : 0\leq l'\leq t-1\} \nonumber\\
     = & \{0,\ldots,\theta'-1\} + \{\theta'l': 0\leq l'\leq t-1\} \nonumber\\
     = & \{0,\ldots,t\theta'-1\}.
      \end{align}
In the following, we consider different regions for the value of $z$ and calculate $|\mathbf{P}({H}(x))|$ through calculation of $\mathbf{D}_2$, $\mathbf{D}_3$, and $\mathbf{D}_4$ for each region. In addition, we use the following lemma, which in some cases helps us to calculate $\mathbf{P}({H}(x))$ without requiring to calculate all of the terms $\mathbf{D}_2$, $\mathbf{D}_3$, and $\mathbf{D}_4$.
\begin{lemma}\label{lemma:UpperBoundPHx}
\begin{align}\label{eq:UpperBoundPHx}
    |\mathbf{P}({H}(x))|\leq& \deg(S_A(x))+\max\{\deg(S_B(x)), \deg(C_B(x))\}\nonumber\\&+1.
\end{align}
\end{lemma}
{\em Proof:} 
$|\mathbf{P}({H}(x))|$ which is equal to the number of terms in $H(x)$ with non-zero coefficients is less than or equal to the number of all terms, which is equal to $\deg(H(x))+1$:
\begin{align}\label{eq:phxUpperBound}
    |\mathbf{P}({H}(x))|\leq &\deg(H(x))+1 \nonumber \\ =&\deg((C_A(x)+S_A(x))(C_B(x)+S_B(x)))+1 \nonumber \\
    =&\max\{\deg(C_A(x)),\deg(S_A(x))\} \nonumber \\
    &+\max\{\deg(S_B(x), \deg(C_B(x))\}+1.
\end{align}
From (\ref{eq:polydot-p(CA)-th}), $\deg(C_A(x))=ts-1$. On the other hand, from (\ref{eq:psa12}) and (\ref{eq:finiteP(SA)-polydot-second-scenario}), $\deg(S_A(x))\ge ts$. Therefore, $\max\{\deg(C_A(x)),\deg(S_A(x))\} = \deg(S_A(x))$, which results in (\ref{eq:UpperBoundPHx}). This completes the proof. \hfill $\Box$

%
\begin{lemma}\label{lemma:non-zero-coeff-polydot}
For $z > ts$ or $t = 1$:
\begin{equation}
    |\mathbf{P}({H}(x))|= \psi_1=(p+2)ts+\theta'(t-1)+2z-1
\end{equation}
 \end{lemma}
 {\em Proof:} 
 To prove this lemma, we first calculate $\mathbf{D}_2$ from (\ref{eq:polydot-p(CA)-th}) and (\ref{eq:P(RB)_set_representation-z large}):
 \begin{align}\label{eq:d2}
    \mathbf{D}_2 = & \mathbf{P}(C_{A}(x))+\mathbf{P}(S_B(x)) \nonumber \\
    = & \{0,\ldots,ts-1\} + \{ts+(t-1)\theta',\ldots,ts+(t-1)\theta'+ \nonumber \\
    & z-1\} \nonumber \\
    = & \{ts+(t-1)\theta',\ldots,ts-1+ts+(t-1)\theta'+z-1\} \nonumber \\
    = & \{t\theta'-t(s-1),\ldots,t\theta'+t+z-2\}.
\end{align}
 
From (\ref{eq:d1}) and (\ref{eq:d2}), we can calculate $\mathbf{D}_{1} \cup \mathbf{D}_2$ as:
\begin{align}\label{eq:d12}
    \mathbf{D}_{12} = & \mathbf{D}_1 \cup \mathbf{D}_2 \nonumber \\
    = & \{0,\ldots,t\theta'-1\} \cup \{t\theta'-t(s-1),\ldots,t\theta'+t+z-2\} \nonumber \\
    = & \{0,\ldots,t\theta'+t+z-2\},
\end{align}
where the last equality comes from the fact that $t(s-1)\ge 0$ and thus $(t\theta'-1)+1\ge t\theta'-t(s-1)$. Next, we calculate $\mathbf{D}_4$ and its union with $\mathbf{D}_{12}$.

From (\ref{eq:psa12}) and (\ref{eq:P(RB)_set_representation-z large}), we have:
\begin{align}
     \mathbf{D}_4= &\mathbf{P}(S_A(x))+\mathbf{P}(S_B(x))  \nonumber \\
     =&\{ts+\theta'l+w, l\in\Omega_0^{p-1}, w\in \Omega_0^{t(s-1)-1}\} \nonumber \\
    &\cup \{ts+\theta'p+u, u\in\Omega_0^{z-1-pt(s-1)}\} \nonumber\\
    & + \{ts+(t-1)\theta'+r, \; 0 \leq r \leq z-1\} \nonumber \\
    = & \bigcup\limits_{l=0}^{p-1} \{2ts+(t-1+l)\theta',\ldots,2ts+(t-1+l)\theta'+ \nonumber \\
    & \quad \quad t(s-1)-1+z-1\}
     \nonumber \\
     & \cup \{2ts+(t-1+p)\theta',\ldots,2ts+\theta'p+(t-1)\theta'+z \nonumber \\
     & \quad -1-pt(s-1)+z-1\}\nonumber \\
    = & \bigcup\limits_{l=0}^{p-1} \{2ts+(t-1+l)\theta',\ldots,(t+l)\theta'+ts+z-2\}
     \nonumber \\
     & \cup \{2ts+(t-1+p)\theta',\ldots,(p+2)ts+\theta'(t-1)+ \nonumber \\
     & \quad \quad 2z-2\} \label{eq:d4subsets}\\
     = &\{2ts+(t-1)\theta',\ldots,(p+2)ts+\theta'(t-1)+2z-2\}\label{eq:d42},
 \end{align}
 where the last equality comes from the fact that there is no gap between each two consecutive subsets of (\ref{eq:d4subsets}). The reason is that:
 \begin{align}
    ts < z  
    \Rightarrow &  ts \leq z-1 \nonumber \\
    \Rightarrow & 2ts \leq ts+z-1 \nonumber \\
    \Rightarrow & 2ts+(t+l)\theta' \leq ((t+l)\theta'+ts+z-2)+1.
\end{align}
\begin{figure*}
		\centering
		\includegraphics[width=14cm]{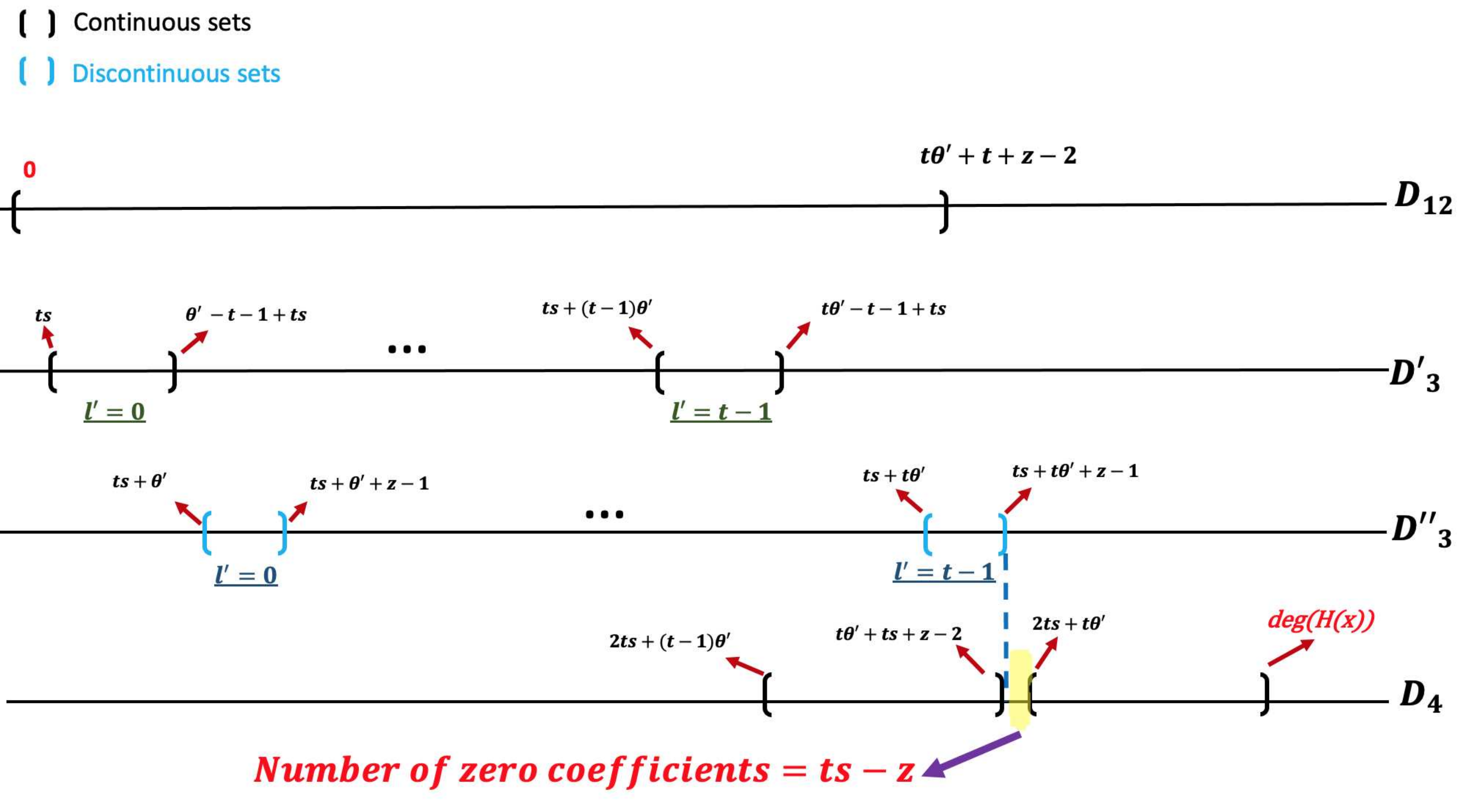}
	\caption{Illustration of $\mathbf{D}_{12} \cup \mathbf{D}_{3} \cup \mathbf{D}_{4}$ for $\theta'-ts<z \leq ts$.
	}
\label{Numberof zerocoeffslemma17-revised}
\vspace{-15pt}
\end{figure*}
Now, we calculate $\mathbf{D}_{12} \cup \mathbf{D}_4$. From (\ref{eq:d12}) and (\ref{eq:d42}), we have:
\begin{align}\label{eq:d124--}
    \mathbf{D}_{1} \cup& \mathbf{D}_2 \cup  \mathbf{D}_4 =\mathbf{D}_{12} \cup \mathbf{D}_4 = \{0,\ldots,t\theta'+t+z-2\}\cup \nonumber \\ & \{2ts+(t-1)\theta',\ldots,(p+2)ts+\theta'(t-1)+2z-2\}\nonumber \\
    &=\{0,\ldots,(p+2)ts+\theta'(t-1)+2z-2\},
\end{align} 
where the last equality comes from the fact that $\mathbf{D}_{12}$ has overlap with $\mathbf{D}_4$ and the upper bound of $\mathbf{D}_4$ is larger than the upper bound of $\mathbf{D}_{12}$. The reason is that:
\begin{align}
0 \leq z-2 
\Rightarrow & 2ts-2ts+t \leq t+z-2 \nonumber \\
\Rightarrow & 2ts-\theta' \leq t+z-2\nonumber \\
\Rightarrow & 2ts + (t-1)\theta' \leq t\theta'+t+z-2,
\end{align}
and
\begin{align}
0 < pts+&z 
\Rightarrow t < pts+t+z \nonumber \\
\Rightarrow & t < (p+2)ts-t(2s-1)+z\nonumber \\
\Rightarrow & t\theta'+t+z-2 < (p+2)ts+\theta'(t-1)+2z-2.
\end{align}
On the other hand, from (\ref{eq:UpperBoundPHx}), (\ref{eq:finite_P(RA)_set_representation-z large}), (\ref{eq:P(RB)_set_representation-z large}), and (\ref{eq:polydot-p(CB)-th}), $|\mathbf{P}({H}(x))|$ is upper bounded by: 
\begin{align}\label{eq:phxUpper}
    |\mathbf{P}({H}(x))|&\leq \deg(S_A(x))+\max\{\deg(S_B(x), \deg(C_B(x))\}\nonumber\\
    &+1\nonumber\\
    &=ts+p\theta'+z-1-p(\theta'-ts)+\nonumber\\
    &\max\{ts+(t-1)\theta'+z-1, t(s-1)+\theta'(t-1)\}\nonumber\\
    &+1\nonumber\\
    &=ts+p\theta'+z-1-p(\theta'-ts)+\nonumber\\
    &ts+(t-1)\theta'+z-1+1,\nonumber\\
    &=(p+2)ts+\theta'(t-1)+2z-1.
\end{align}
From (\ref{eq:PHx}) and (\ref{eq:d124--}), $|\mathbf{P}({H}(x))|$ is lower bounded by:
\begin{align}\label{eq:phxLower}
    |\mathbf{P}({H}(x))|&\ge |\mathbf{D}_1\cup\mathbf{D}_2\cup\mathbf{D}_4|\nonumber\\
    &=(p+2)ts+\theta'(t-1)+2z-1.
\end{align}
From (\ref{eq:phxUpper}) and (\ref{eq:phxLower}), $|\mathbf{P}({H}(x))|=(p+2)ts+\theta'(t-1)+2z-1$. This completes the proof. \hfill $\Box$

\begin{lemma}\label{lemma:non-zero-coeff-polydot-p=1&ts-t<z<ts}
For $\theta'-ts < z \leq ts$ and $s,t \neq 1$:
\begin{equation}
    |\mathbf{P}({H}(x))|= \psi_2=2ts+\theta'(t-1)+3z-1
\end{equation}
\end{lemma}
{\em Proof:}
For $\theta'-ts < z \leq ts$, $\mathbf{D}_1$ and $\mathbf{D}_2$ are calculated as (\ref{eq:d1}) and (\ref{eq:d2}) and thus from (\ref{eq:d12}), $\mathbf{D}_{12}$ is equal to:
\begin{align}\label{eq:d122}
    \mathbf{D}_{12} = \mathbf{D}_1 \cup \mathbf{D}_2=\{0,\ldots,t\theta'+t+z-2\},
\end{align}
Next, we calculate $\mathbf{D}_4$ and $\mathbf{D}_3$. We note that $p$ is equal to 1. The reason is that for this region of $z$, we have:
\begin{align}
\theta'-ts < z \leq ts 
\Rightarrow &\theta'-ts \leq z-1 < ts  \nonumber \\
\Rightarrow &\theta'-ts \leq z-1 < ts+t(s-2) \nonumber \\
\Rightarrow & \theta'-ts \leq z-1 < 2ts-2t\nonumber \\
\Rightarrow & \theta'-ts \leq z-1 < 2\theta'-ts \nonumber \\
\Rightarrow & p = \min\{\floor{\frac{z-1}{\theta'-ts}},t-1\} = 1.
\end{align}
By replacing $p$ with 1 in (\ref{eq:finite_P(RA)_set_representation-z large}) and using (\ref{eq:P(RB)_set_representation-z large}), $\mathbf{D}_4$ is equal to:
\begin{align}
    \mathbf{D}_4 =&\mathbf{P}(S_A(x))+\mathbf{P}(S_B(x))\nonumber \\ 
    =&\{ts,\ldots,\theta'-1\}\cup \{ts+\theta',\ldots,2ts+z-1\}\nonumber\\
    &+\{ts+(t-1)\theta',\ldots,ts+(t-1)\theta'+z-1\}\nonumber\\
    =&\{2ts+(t-1)\theta',\ldots,t\theta'+ts+z-2\}\nonumber\\
    &\cup \{2ts+t\theta',\ldots,3ts+\theta'(t-1)+2z-2\}.
\end{align}
Using (\ref{eq:finite_P(RA)_set_representation-z large}) with $p=1$ and (\ref{eq:polydot-p(CB)-th}), $\mathbf{D}_3$ is equal to:
\begin{align}\label{eq:d32}
    \mathbf{D}_3
    =&\mathbf{P}(S_A(x))+\mathbf{P}(C_B(x))\nonumber \\ 
    =&\{ts,\ldots,\theta'-1\}\cup \{ts+\theta',\ldots,2ts+z-1\}\nonumber\\
    &+\{tq'+\theta'l', 0\leq l' \leq t-1, 0\leq q' \leq s-1\}\nonumber\\
    =&\mathbf{D}'_3 \cup \mathbf{D}''_3,
\end{align}
where $\mathbf{D}'_3$ and $\mathbf{D}''_3$ are defined as follows.
\begin{align}\label{eq:d'3}
    \mathbf{D}'_3=&\{ts,\ldots,\theta'-1\}\nonumber\\
    &+\{tq'+\theta'l', 0\leq l' \leq t-1, 0\leq q' \leq s-1\}\nonumber\\
    =&\bigcup\limits_{l'=0}^{t-1}\bigcup\limits_{q'=0}^{s-1} \{ts+tq'+\theta'l',\ldots,\theta'-1+tq'+\theta'l'\}\nonumber\\
    =&\bigcup\limits_{l'=0}^{t-1}
    \{ts+\theta'l',\ldots,\theta'-1+t(s-1)+\theta'l'\},
\end{align}
where the last equality comes from the fact that there is no gap between each two consecutive subsets of $\bigcup\limits_{q'=0}^{s-1} \{ts+tq'+\theta'l',\ldots,\theta'-1+tq'+\theta'l'\}$. The reason is that:
\begin{align}
s \ge 2 
\Rightarrow &st \ge 2t  \nonumber \\
\Rightarrow &t(2s-1) \ge ts+t \nonumber \\
\Rightarrow & \theta' \ge ts+t\nonumber \\
\Rightarrow & (\theta'-1+tq'+\theta'l')+1 \ge ts+t(q'+1)+\theta'l'.
\end{align}
$\mathbf{D}''_3$ is defined and calculated as:
\begin{align}\label{eq:d''3}
    \mathbf{D}''_3=&\{ts+\theta',\ldots,2ts+z-1\}\nonumber\\
    &+\{tq'+\theta'l', 0\leq l' \leq t-1, 0\leq q' \leq s-1\}\nonumber\\
    =&\bigcup\limits_{l'=0}^{t-1}\bigcup\limits_{q'=0}^{s-1} \{ts+\theta'+tq'+\theta'l',\ldots,\nonumber\\
    &\quad \quad \quad \quad 2ts+z-1+tq'+\theta'l'\}.
\end{align}
To calculate $\mathbf{D}_1 \cup \mathbf{D}_2 \cup \mathbf{D}_3 \cup \mathbf{D}_4$, we first calculate $\mathbf{D}_{12} \cup \mathbf{D}'_3$ using (\ref{eq:d122}) and (\ref{eq:d'3}):
\begin{align}\label{eq:d12Capd'3}
    \mathbf{D}_{12} \cup \mathbf{D}'_3 =& \{0,\ldots,t\theta'+t+z-2\} \nonumber \\ &\cup \bigcup\limits_{l'=0}^{t-1}
    \{ts+\theta'l',\ldots,\theta'-1+t(s-1)+\theta'l'\}\nonumber\\
    =&\mathbf{D}_{12},
\end{align}
where the last equality comes from the fact that the largest element of $\mathbf{D}'_3$, \ie $\theta'-1+t(s-1)+\theta'(t-1)$ is smaller than the largest element of $\mathbf{D}_{12}$, \ie $t\theta'+t+z-2$, as illustrated in Fig. \ref{Numberof zerocoeffslemma17-revised} and shown below:
\begin{align}
z > \theta'-ts 
&\Rightarrow z > ts-t  \nonumber \\
\Rightarrow &z > ts-t-(t-1) \nonumber \\
\Rightarrow &t+z-2 > ts-t-1 \nonumber \\
\Rightarrow &t\theta'+t+z-2 > \theta'-1+t(s-1)+\theta'(t-1).
\end{align}
Next, we calculate $\mathbf{D}_{12} \cup \mathbf{D}_4$ as demonstrated in Fig. \ref{Numberof zerocoeffslemma17-revised}:
\begin{align}\label{eq:d12Cupd4}
    &\mathbf{D}_{12}\cup \mathbf{D}_4\nonumber \\
    =&\{0,\ldots,t\theta'+t+z-2\} \nonumber \\
    & \cup \{2ts+(t-1)\theta',\ldots,t\theta'+ts+z-2\}\nonumber\\
    &\cup \{2ts+t\theta',\ldots,3ts+\theta'(t-1)+2z-2\}\nonumber\\
    =&\{0,\ldots,3ts+\theta'(t-1)+2z-2\} \nonumber \\
    & - \{t\theta'+ts+z-1, \dots, 2ts+t\theta'-1\}.
\end{align}
$z\leq ts$ results in the non-empty set of $\{t\theta'+ts+z-1, \dots, 2ts+t\theta'-1\}$ in the above equation. Now we calculate $\mathbf{D}_{12} \cup \mathbf{D}_4 \cup \mathbf{D}''_3$ using (\ref{eq:d12Cupd4}) and (\ref{eq:d''3}):
\begin{align}\label{eq:d12Capd4Capd''3}
    &\mathbf{D}_{12}\cup \mathbf{D}_4 \cup \mathbf{D}''_3\nonumber \\
    =&(\{0,\ldots,3ts+\theta'(t-1)+2z-2\} \nonumber \\
    & - \{t\theta'+ts+z-1, \dots, 2ts+t\theta'-1\})\nonumber\\
    &\cup \mathbf{D}''_3\nonumber\\
    =&\{0,\ldots,3ts+\theta'(t-1)+2z-2\} \nonumber \\
    & - \{t\theta'+ts+z, \dots, 2ts+t\theta'-1\},
\end{align}
where the last equality comes from the fact that $\mathbf{D}''_3 \subset \{0,\ldots,3ts+\theta'(t-1)+2z-2\}$\footnote{ The reason is that the largest element of $\mathbf{D}''_3$, \ie $t\theta'+ts+z-1$ is smaller than the largest element of $\{0,\ldots,3ts+\theta'(t-1)+2z-2\}$.} and $\mathbf{D}''_3 \cap (\{t\theta'+ts+z-1, \dots, 2ts+t\theta'-1\}) = \{t\theta'+ts+z-1\}$. From (\ref{eq:d122}), (\ref{eq:d32}), (\ref{eq:d12Capd'3}), and (\ref{eq:d12Capd4Capd''3}), we have:
\begin{align}
    \mathbf{D}_1 \cup \mathbf{D}_2 \cup \mathbf{D}_3 \cup \mathbf{D}_4 =&\{0,\ldots,3ts+\theta'(t-1)+2z-2\} \nonumber \\
    &- \{t\theta'+ts+z, \dots, 2ts+t\theta'-1\},
\end{align}
and thus from (\ref{eq:PHx}):
\begin{align}
 |\mathbf{P}({H}(x))|=&(3ts+\theta'(t-1)+2z-2)+1\nonumber\\
 &-(2ts+t\theta'-1-(t\theta'+ts+z)+1)\nonumber\\
 =&2ts+\theta'(t-1)+3z-1.
\end{align}
This completes the proof.
\hfill $\Box$ 

\begin{lemma}\label{lemma:non-zero-coeff-polydot-zless than alpha minus ts1}
For $\theta'-ts-t < z \leq \theta'-ts$ and $s,t \neq 1$:
\begin{equation}
    |\mathbf{P}({H}(x))|= \psi_3= 2ts+\theta'(t-1)+2z-1
\end{equation}
\end{lemma}
{\em Proof:} For $\theta'-ts-t < z \leq \theta'-ts$, $\mathbf{P}(S_B(x))$ is derived from (\ref{eq:P(RB)_set_representation-z medum}), which is equal to $\mathbf{P}(S_B(x))$ used in (\ref{eq:d2}). Therefore, $\mathbf{D}_{2}$ is equal to:
\begin{equation}
    \mathbf{D}_2=\{t\theta'-t(s-1),\ldots,t\theta'+t+z-2\},
\end{equation}
and thus using (\ref{eq:d12}), we have:
\begin{equation}
    \mathbf{D}_1 \cup \mathbf{D}_2 = \{0,\ldots,t\theta'+t+z-2\}.
\end{equation}
From (\ref{eq:finiteP(SA)-polydot-second-scenario}) and (\ref{eq:P(RB)_set_representation-z medum}), $\mathbf{D}_4$ is calculated as:
\begin{align}
    \mathbf{D}_4 =& \mathbf{P}(S_A(x))+\mathbf{P}(S_B(x)) \nonumber\\
    =& \{2ts+\theta'(t-1),\ldots,2ts+\theta'(t-1)+2z-2\}.
\end{align}
Now, from the above two equations, we calculate $\mathbf{D}_1 \cup \mathbf{D}_2 \cup \mathbf{D}_4$:
\begin{align}
    \mathbf{D}_1 \cup \mathbf{D}_2 \cup \mathbf{D}_4 = \{0,\ldots,2ts+\theta'(t-1)+2z-2\}, 
\end{align}
where the equality comes from the fact that:
\begin{align}
z \ge 1 
\Rightarrow &t+z-2+1 \ge t  \nonumber \\
\Rightarrow &(t\theta'+t+z-2)+1 \ge 2ts+\theta'(t-1),
\end{align}
and
\begin{align}
t < t+z 
&\Rightarrow t\theta'+t+z-2 < 2ts+\theta'(t-1)+2z-2.
\end{align}
Therefore, $|\mathbf{P}({H}(x))| \ge |\mathbf{D}_1 \cup \mathbf{D}_2 \cup \mathbf{D}_4|=(2ts+\theta'(t-1)+2z-2)+1$. On the other hand, from (\ref{eq:UpperBoundPHx}), (\ref{eq:finiteP(SA)-polydot-second-scenario}), and (\ref{eq:P(RB)_set_representation-z medum}), we have:
\begin{align}
    |\mathbf{P}({H}(x))|\leq& \deg(S_A(x))+\max\{\deg(S_B(x), \deg(C_B(x))\}\nonumber\\&+1\nonumber\\
    =&(ts+z-1)+\max\{ts+(t-1)\theta'+z-1,\nonumber\\
    &\quad \quad \quad \quad \quad \quad \quad \quad \quad  t(s-1)+\theta'(t-1)\}+1\nonumber\\
    =&2ts+\theta'(t-1)+2z-1.
\end{align}
This results in $|\mathbf{P}({H}(x))|=2ts+\theta'(t-1)+2z-1$, which completes the proof.
\hfill $\Box$ 

\begin{lemma}\label{lemma:non-zero-coeff-polydot-zless than alpha minus ts}
For $\frac{\theta'-ts-t+1}{2}< z \leq \theta'-ts-t$:
\begin{equation}\label{eq:equation_in_lemma30}
    |\mathbf{P}({H}(x))|= \max\{\theta' t+z,(p'+2)ts+p'(z+t-1)+2z-1\}
\end{equation}
\end{lemma}
{\em Proof:}
For $\frac{\theta'-ts-t+1}{2}< z \leq \theta'-ts-t$, $\mathbf{D}_2$ is calculated using (\ref{eq:polydot-p(CA)-th}) and (\ref{eq:P(RB)_set_representation-zsmall}):
\begin{align}\label{eq:d24}
    \mathbf{D}_2 = & \mathbf{P}(C_{A}(x))+\mathbf{P}(S_B(x)) \nonumber \\
    = & \{0,\ldots,ts-1\} + \nonumber \\
    \Big(&\bigcup\limits_{l''=0}^{p'-1}\{ts+\theta'l'', \ldots, (l''+1)\theta'-z-t\}\Big) \nonumber \\ 
    &\cup \{ts+p'\theta',\dots, \nonumber \\ 
    & \quad \quad ts+p'\theta'+z-1-p'(\theta'-t-ts-z+1)\} \nonumber \\
    =\Big(&\bigcup\limits_{l''=0}^{p'-1}\{ts+\theta'l'', \ldots, (l''+1)\theta'-z-t+ts-1\}\Big) \nonumber \\
    &\cup \{ts+p'\theta',\dots, \nonumber \\ 
    & ts+p'\theta'+z-1-p'(\theta'-t-ts-z+1)+ts-1\}\nonumber
    \end{align}
    \begin{align}
    =\Big(&\bigcup\limits_{l''=0}^{p'-1}\{ts+\theta'l'', \ldots, (l''+1)\theta'-z-t+ts-1\}\Big) \nonumber \\
    &\cup \{ts+p'\theta',\dots,2ts+p'(t+ts+z-1)+z-2\}
\end{align}
From (\ref{eq:d1}) and (\ref{eq:d24}), $\mathbf{D}_1 \cup \mathbf{D}_2$ is equal to:
\begin{align}\label{eq:---d124}
    &\mathbf{D}_{12} = \mathbf{D}_1 \cup \mathbf{D}_2 \nonumber \\
    = & \{0,\ldots,t\theta'-1\} \cup \nonumber \\
    \Big(&\bigcup\limits_{l''=0}^{p'-1}\{ts+\theta'l'', \ldots, (l''+1)\theta'-z-t+ts-1\}\Big) \nonumber \\
    &\cup \{ts+p'\theta',\dots,2ts+p'(t+ts+z-1)+z-2\}\nonumber\\
    =& \{0,\dots,\max\{2ts+p'(ts+z+t-1)+z-2 , t\theta'-1\}\},
\end{align}
where the last equality comes from the fact that $\mathbf{D}_1$ has overlap with the last subset of $\mathbf{D}_2$, as shown below:
\begin{align}
    &p' \leq t-1 \nonumber \\
    &\Rightarrow p'\theta' \leq (t-1)\theta' \nonumber \\
    &\Rightarrow p'\theta'+ts \leq t\theta'-ts+t < t\theta'-1 \nonumber \\
    &\Rightarrow p'\theta'+ts < t\theta'-1.
\end{align}
From (\ref{eq:d3definition}), (\ref{eq:finiteP(SA)-polydot-second-scenario}) and (\ref{eq:polydot-p(CB)-th}), $\mathbf{D}_3$ is calculated as:
\begin{align}\label{eq:d34}
    \mathbf{D}_3 = \bigcup\limits_{l'=0}^{t-1}\bigcup\limits_{q'=0}^{s-1} \{ts+tq'+\theta'l',\ldots,ts+z-1+tq'+\theta'l'\}.
\end{align}
From (\ref{eq:d4definition}), (\ref{eq:finiteP(SA)-polydot-second-scenario}), and (\ref{eq:P(RB)_set_representation-zsmall}), $\mathbf{D}_4$ is calculated as:
\begin{align}\label{eq:d441}
    \mathbf{D}_4 = &\Big(\bigcup\limits_{l''=0}^{p'-1}\{2ts+\theta'l'', \ldots, ts-1+(l''+1)\theta'-t\}\Big) \nonumber \\
    &\cup \{2ts+p'\theta',\dots,2ts+p'(t+ts+z-1)+2z-2\}.
\end{align}
To calculate $\mathbf{D}_1 \cup \mathbf{D}_2 \cup \mathbf{D}_3 \cup \mathbf{D}_4$, we consider two cases of (i) $2ts+p'(ts+z+t-1)+z-2 \ge t\theta'-1$ and (ii) $2ts+p'(ts+z+t-1)+z-2 < t\theta'-1$.

(i) $2ts+p'(ts+z+t-1)+z-2 \ge t\theta'-1$: For this case, from (\ref{eq:---d124}), $\mathbf{D}_{12}$ is equal to:
\begin{align}\label{eq:d1242}
    \mathbf{D}_1 \cup \mathbf{D}_2=\{0,\dots,2ts+p'(ts+z+t-1)+z-2\}
\end{align}
From (\ref{eq:d441}) and (\ref{eq:d1242}), we have:
\begin{align}
    \mathbf{D}_1 \cup &\mathbf{D}_2 \cup \mathbf{D}_4 = \{0,\dots,2ts+p'(ts+z+t-1)+z-2\} \nonumber \\
    & \cup \{2ts+p'\theta',\dots,2ts+p'(t+ts+z-1)+2z-2\}\label{eq:Cupd4LastSubset} \\
    & = \{0,\dots,2ts+p'(t+ts+z-1)+2z-2\}\label{eq:d12Cupd441},
\end{align}
where (\ref{eq:Cupd4LastSubset}) and (\ref{eq:d12Cupd441}) come from the fact that each subset of $S_B(x)$ in (\ref{eq:P(RB)_set_representation-zsmall}) is designed to be non-empty:
\begin{align}
&ts+p'\theta' \leq ts+p'\theta'+z-1-p'(\theta'-t-ts-z+1)\nonumber\\
&\Rightarrow 2ts+p'\theta' \leq (2ts+p'(ts+z+t-1)+z-2)+1,
\end{align}
and $2ts+p'(ts+z+t-1)+z-2 < 2ts+p'(ts+z+t-1)+2z-2$. On the other hand,  from the condition considered in (i), the largest element of $\mathbf{D}_3$, \ie $ts+z-1+t(s-1)+\theta'(t-1)=z-1+\theta't$ is less than or equal to $(2ts+p'(ts+z+t-1)+z-2)+z=2ts+p'(t+ts+z-1)+2z-2$, and thus $\mathbf{D}_3 \subset \{0,\dots,2ts+p'(t+ts+z-1)+2z-2\}$:
\begin{align}\label{eq:d1d2d3d4i}
    \mathbf{D}_1 \cup &\mathbf{D}_2 \cup \mathbf{D}_3 \cup \mathbf{D}_4 =\nonumber \\ &\{0,\dots,2ts+p'(t+ts+z-1)+2z-2\},\nonumber \\
    & \text{for  } (2ts+p'(ts+z+t-1)+z-2) \ge t\theta'-1
\end{align}

(ii) $2ts+p'(ts+z+t-1)+z-2 < t\theta'-1$: For this case, from (\ref{eq:---d124}), $\mathbf{D}_{12}$ is equal to:
\begin{align}\label{eq:d1243}
    \mathbf{D}_1 \cup \mathbf{D}_2=\{0,\dots,t\theta'-1\}
\end{align} 
From (\ref{eq:d34}) and (\ref{eq:d1242}), we have:
\begin{align}
    \mathbf{D}_1 \cup \mathbf{D}_2 \cup \mathbf{D}_3 &= \{0,\dots,t\theta'-1\} \nonumber \\
    & \quad \cup \{ts+t(t-1)+\theta'(t-1),\ldots,\nonumber\\
    & \quad \quad ts+z-1+t(s-1)+\theta'(t-1)\} \nonumber \\
    &= \{0,\dots,t\theta'-1\} \cup \{t\theta',\ldots,t\theta'+z-1\} \nonumber \\
    & = \{0,\dots,t\theta'+z-1\}\label{eq:d12Cupd441--},
\end{align}
where the first equality comes from the fact that $\{0,\ldots,t\theta'-1\}$ has overlap with all subsets of $\mathbf{D}_3$ in (\ref{eq:d34}) except for the last subset. On the other hand, from the condition considered in (ii), the largest element of $\mathbf{D}_4$, \ie $2ts+p'(t+ts+z-1)+2z-2$ is less than $t\theta'+z-1$, and thus $\mathbf{D}_4 \subset \{0,\ldots,t\theta'+z-1\}$:
\begin{align}\label{eq:d1d2d3d4ii}
    \mathbf{D}_1 \cup &\mathbf{D}_2 \cup \mathbf{D}_3 \cup \mathbf{D}_4 =\{0,\dots,t\theta'+z-1\} \nonumber\\
    &\text{for  } (2ts+p'(ts+z+t-1)+z-2) < t\theta'-1
\end{align}

From (\ref{eq:d1d2d3d4i}) and (\ref{eq:d1d2d3d4ii}), we have:
\begin{align}
    &|\mathbf{P}({H}(x))|= | \mathbf{D}_1 \cup \mathbf{D}_2 \cup \mathbf{D}_3 \cup \mathbf{D}_4| \nonumber \\
    &= \max\{\theta' t+z,(p'+2)ts+p'(z+t-1)+2z-1\}
\end{align}
This completes the proof.
\hfill $\Box$
\begin{lemma}\label{lemma:psi4-clarification}
For $\frac{\theta'-ts-t+1}{2}<z \leq ts-2t-s+2$ and $s,t \neq 1$:
\begin{align}\label{eq:lemma31FirPart}
    |\mathbf{P}({H}(x))| = t\theta'+z
\end{align}
and for $\max\{st-2t-s+2,\frac{\theta'-ts-t+1}{2}\} < z \leq \theta'-ts-t$:
\begin{align}\label{eq:lemma31SecPart}
    &|\mathbf{P}({H}(x))| = \psi_4= (t+1)ts+(t-1)(z+t-1)+2z-1
\end{align} 
\end{lemma}
{\em Proof:}
To prove this lemma, first, we determine the condition for which $p'=t-1$ and the condition that $p'<t-1$:
\begin{align}\label{eq:p'value determination}
    p'=&\min\{\floor{\frac{z-1}{\theta'-ts-t-z+1}},t-1\}\nonumber\\
    &\begin{cases}
   =t-1 & z>st-2t-s+2\\
   <t-1 & z \leq st-2t-s+2,
\end{cases}
\end{align}
The above equation comes from the following:
\begin{align}
    & z \leq st-2t-s+2\nonumber\\
    \Rightarrow & z-1 < st-2t-s+2 \nonumber \\
    \Rightarrow & t(z-1) < t(s-2)(t-1) \nonumber \\
    \Rightarrow & z-1 < t(s-2)(t-1)-(t-1)(z-1) \nonumber \\
    \Rightarrow & z-1 < (ts-2t-z+1)(t-1) \nonumber \\
    \Rightarrow & \frac{z-1}{\theta'-ts-t-z+1} < t-1 \nonumber \\
    \Rightarrow & \floor{\frac{z-1}{\theta'-ts-t-z+1}} < t-1
\end{align}
Next, we decompose (\ref{eq:equation_in_lemma30}) to determine in which region $|\mathbf{P}({H}(x))| =\psi'_4= t\theta'+z$ and in which region $|\mathbf{P}({H}(x))| =\psi''_4= (t+1)ts+(t-1)(z+t-1)+2z-1$ when $\frac{\theta'-ts-t+1}{2}<z\leq \theta'-ts-t$. For this purpose, we calculate $\psi'_4 - \psi''_4$ as follows:
\begin{align}\label{eq:psi'4-psi''4}
    & \psi'_4-\psi''_4 \nonumber \\
   & = \theta't+z-(p'+2)ts-p'(z+t-1)-2z+1 \nonumber \\
   & = 2st^2-t^2+z-(p'+2)ts-p'(t-1)-z(p'+2)+1 \nonumber \\ 
   & = ts(2t-p'-2)-t(t+p')+p'+1-z(p'+1) \nonumber \\
   & = (p'+1)(ts(\frac{2t-p'-1-1}{p'+1})-t(\frac{p'+1+t-1}{p'+1})+1-z) \nonumber \\
   & = (p'+1)(ts(\frac{2t-2+1}{p'+1}-1)-t(\frac{t-1}{p'+1}+1)+1-z) \nonumber \\
   & = (p'+1)(ts(\frac{2t-2+1}{p'+1})-t(\frac{t-1}{p'+1})-(ts+t)+1-z) \nonumber \\
   & = (p'+1)(2ts(\frac{t-1+1/2}{p'+1})-t(\frac{t-1}{p'+1})-(ts+t)+1-z) \nonumber \\
   & = (p'+1)((\frac{t-1}{p'+1})(2ts-t)+\frac{ts}{p'+1}-(ts+t)+1-z) \nonumber \\
   & = (p'+1)(y-z),
\end{align}
Next, we consider the two cases of (i) $\max\{st-2t-s+2,\frac{\theta'-ts-t+1}{2}\} < z \leq \theta'-ts-t$ and (ii) $\frac{\theta'-ts-t+1}{2}<z \leq ts-2t-s+2$ and calculate $\psi'_4 - \psi''_4$ through comparison of $y$ and $z$.

(i) $\max\{st-2t-s+2,\frac{\theta'-ts-t+1}{2}\} < z \leq \theta'-ts-t$: For this case, from (\ref{eq:p'value determination}), $p'=t-1$ and from (\ref{eq:psi'4-psi''4}), $\psi'_4 - \psi''_4$ is calculated as:
\begin{align}
    \psi'_4 - \psi''_4 & =t(y-z) \nonumber\\
    & = (t-1)(2ts-t)+ts-t(ts+t)+t-tz \nonumber \\
    & = t(-2t-s+2+ts-z)\nonumber\\
    & < 0,
\end{align}
where the last inequality comes from the condition of (i). Therefore, for $\max\{st-2t-s+2,\frac{\theta'-ts-t+1}{2}\} < z \leq \theta'-ts-t$, we have $\max\{\psi'_4, \psi''_4\}=\psi''_4= (t+1)ts+(t-1)(z+t-1)+2z-1$. Since the condition of (i) is a subset of the condition considered in Lemma \ref{lemma:non-zero-coeff-polydot-zless than alpha minus ts}, \ie $\frac{\theta'-ts-t+1}{2} < z \leq \theta'-ts-t$, from (\ref{eq:equation_in_lemma30}), we have $|\mathbf{P}({H}(x))|= \max\{\theta' t+z,(p'+2)ts+p'(z+t-1)+2z-1\} = (t+1)ts+(t-1)(z+t-1)+2z-1$. This proves (\ref{eq:lemma31SecPart}).

(ii) $\frac{\theta'-ts-t+1}{2}<z \leq ts-2t-s+2$: For this case, from (\ref{eq:p'value determination}), $p'<t-1$ and from (\ref{eq:psi'4-psi''4}), $\psi'_4 - \psi''_4$ is calculated as:
\begin{align}
    \psi'_4 - \psi''_4 & =(p'+1)(y-z) \nonumber \\
    & > (p'+1)(\frac{t-1}{t}(2ts-t)+\frac{ts}{t}-(ts+t)+1-z) \nonumber \\
    & = (p'+1)((t-1)(2s-1)+s-(ts+t)+1-z)\nonumber\\
    & = (p'+1)(-s-2t+2+ts-z)\nonumber\\
    & \ge 0,
\end{align}
where the last inequality comes from the condition of (ii). Therefore, for $\frac{\theta'-ts-t+1}{2}<z \leq ts-2t-s+2$, we have $\max\{\psi'_4, \psi''_4\}=\psi'_4= t\theta'+z$. Since the condition of (ii) is a subset of the condition considered in Lemma \ref{lemma:non-zero-coeff-polydot-zless than alpha minus ts}, \ie $\frac{\theta'-ts-t+1}{2} < z \leq \theta'-ts-t$\footnote{This comes from the fact that $0\ge 2-s$ and thus $\theta'-ts-t=ts-2t \ge ts-2t-s+2$.}, from (\ref{eq:equation_in_lemma30}), we have $|\mathbf{P}({H}(x))|= \max\{\theta' t+z,(p'+2)ts+p'(z+t-1)+2z-1\} = \theta' t+z$. This proves (\ref{eq:lemma31FirPart}).

This completes the proof.  \hfill $\Box$

\begin{lemma}\label{lemma:non-zero-coeff-polydot-zless than (alpha minus ts minus t)/2}
For $z \leq \frac{\theta'-ts-t+1}{2}$:
\begin{equation}
    |\mathbf{P}({H}(x))|= t\theta'+z
\end{equation}
\end{lemma}
{\em Proof:}
For $z \leq \frac{\theta'-ts-t+1}{2}$, $\mathbf{P}(S_A(x))$ and $\mathbf{P}(S_B(x))$ are calculated from (\ref{eq:finiteP(SA)-polydot-second-scenario}) and (\ref{eq:P(RB)_set_representation-z very small}). Therefore, using (\ref{eq:polydot-p(CA)-th}) and (\ref{eq:polydot-p(CB)-th}), $\mathbf{D}_2, \mathbf{D}_3$, and $\mathbf{D}_4$ are equal to:
\begin{align}
    \mathbf{D}_2=&\mathbf{P}(C_A(x))+\mathbf{P}(S_B(x))=\{ts,\ldots,2ts+z-2\}\nonumber\\
    \mathbf{D}_3=&\mathbf{P}(S_A(x))+\mathbf{P}(C_B(x))\nonumber \\
    =&\bigcup\limits_{l'=0}^{t-1}\bigcup\limits_{q'=0}^{s-1} \{ts+tq'+\theta'l',\ldots,ts+z-1+tq'+\theta'l'\}\nonumber\\
    \mathbf{D}_3=&\mathbf{P}(S_A(x))+\mathbf{P}(S_B(x))=\{2ts,\ldots,2ts+2z-2\}
\end{align}
From (\ref{eq:d1}) and the above equations, we calculate $\mathbf{D}_1 \cup \mathbf{D}_2 \cup \mathbf{D}_3 \cup \mathbf{D}_4$ as follows:
\begin{align}
    &\mathbf{D}_1 \cup \mathbf{D}_2 \cup \mathbf{D}_3 \cup \mathbf{D}_4 =\{0,\dots,t\theta'-1\} \cup \nonumber \\
    &\{ts,\dots,2ts+z-2\} \cup \nonumber \\
    &\bigcup\limits_{l'=0}^{t-1}\bigcup\limits_{q'=0}^{s-1} \{ts+tq'+\theta'l',\ldots,ts+z-1+tq'+\theta'l'\}\nonumber\\
    &\cup \{2ts,\ldots,2ts+2z-2\} \nonumber
        \end{align}
    \begin{align}
    &=\{0,\dots,t\theta'-1\} \cup \{ts,\dots,2ts+z-2\} \cup \nonumber \\
    &\{ts+t(s-1)+\theta'(t-1),\ldots,\nonumber\\
    &ts+z-1+t(s-1)+\theta'(t-1)\}\nonumber\\
    &\cup \{2ts,\ldots,2ts+2z-2\}\label{eq:d1d2d3d46}\\
    &=\{0,\dots,t\theta'-1\} \cup \{ts,\dots,2ts+2z-2\} \cup \nonumber \\
    &\quad \quad \{\theta't,\ldots,\theta't+z-1\}\nonumber\\
    &=\{0,\ldots,\theta't+z-1\}\cup\{ts,\ldots,2ts+2z-2\}\nonumber\\
    &=\{0,\ldots,t\theta'+z-1\}\label{eq:d1d2d3d48},
\end{align}
where (\ref{eq:d1d2d3d46}) comes from the fact that all subsets of $\mathbf{D}_3$ except for the last one is subsets of $\{0,\ldots,t\theta'-1\}$ and (\ref{eq:d1d2d3d48}) comes from the fact that $2ts+2z-2<t\theta'+z-1$. The reason is that:
\begin{align}
    2ts+2z-2 \leq & 2ts+(\theta'-ts-t+1)-2 \nonumber \\
    = & 2\theta'-ts-1 \nonumber \\
    \leq & t\theta'-ts-1\nonumber \\
    <& t\theta'\nonumber\\
    \leq& t\theta'+z-1.
\end{align}
From (\ref{eq:d1d2d3d48}) we have:
\begin{align}
    &|\mathbf{P}({H}(x))|= | \mathbf{D}_1 \cup \mathbf{D}_2 \cup \mathbf{D}_3 \cup \mathbf{D}_4|= t\theta'+z
\end{align}
This completes the proof.
\hfill $\Box$
\begin{lemma}\label{lemma:psi5-clarification}
For $ z \leq \max\{st-2t-s+2,\frac{\theta'-ts-t+1}{2}\}$ and $s,t \neq 1$:
\begin{align}
    &|\mathbf{P}({H}(x))| = \psi_5= t\theta'+z
\end{align} 
\end{lemma}
{\em Proof:} To prove this lemma we consider two scenarios:

(i) $\frac{\theta'-ts-t+1}{2}<st-2t-s+2$: From Lemma \ref{lemma:psi4-clarification}, for $\frac{\theta'-ts-t+1}{2} < z \leq st-2t-s+2$, we have $|\mathbf{P}({H}(x))| = t\theta'+z$. On the other hand, from Lemma \ref{lemma:non-zero-coeff-polydot-zless than (alpha minus ts minus t)/2}, for $z\leq \frac{\theta'-ts-t+1}{2}$, we have $|\mathbf{P}({H}(x))| = t\theta'+z$. Therefore, we conclude that for $z \leq st-2t-s+2=$, we have $|\mathbf{P}({H}(x))| = t\theta'+z$.

(ii) $st-2t-s+2 \leq \frac{\theta'-ts-t+1}{2}$: 
From Lemma \ref{lemma:non-zero-coeff-polydot-zless than (alpha minus ts minus t)/2}, for $z\leq \frac{\theta'-ts-t+1}{2}$, we have $|\mathbf{P}({H}(x))| = t\theta'+z$.

From (i) and (ii), for $ z \leq \max\{st-2t-s+2,\frac{\theta'-ts-t+1}{2}\}$, $|\mathbf{P}({H}(x))| = t\theta'+z$. This completes the proof. \hfill $\Box$

The required number of workers, $N_{\text{PolyDot-CMPC}}$, is equal to $|\mathbf{P}({H}(x))|$. Therefore, from Lemmas (\ref{lemma:psit=1}), (\ref{lemma:psis=1}), (\ref{lemma:non-zero-coeff-polydot}), (\ref{lemma:non-zero-coeff-polydot-p=1&ts-t<z<ts}), (\ref{lemma:non-zero-coeff-polydot-zless than alpha minus ts1}), (\ref{lemma:psi4-clarification}), and (\ref{lemma:psi5-clarification}), Theorem \ref{th:N_PolyDot} is proved.
\section*{Appendix C: Proof of Lemmas \ref{lemma: regions where N_polydot<N_entangled}, \ref{lemma: regions where N_polydot<N_ssmm}, and \ref{lemma: regions where N_polydot<N_gcsana}}

\subsection{Proof of Lemma \ref{lemma: regions where N_polydot<N_entangled} (PolyDot-CMPC Versus Entangled-CMPC)}
To prove this lemma, we consider different regions for the value of $z$ and compare the required number of workers for PolyDot-CMPC, $N_{\text{PolyDot-CMPC}}$, with Entangled-CMPC, $N_{\text{Entangled-CMPC}}$, in each region. From \cite{8613446}, $N_{\text{Entangled-CMPC}}$ is equal to:
\begin{align}\label{eq:N Entang-CMPC}
N_{\text{Entangled-CMPC}}=\begin{cases}
   2st^2+2z-1,&z>ts-s\\
   st^2+3st-2s+t(z-1)+1,&z \leq ts-s,
\end{cases} 
\end{align}
and we use (\ref{eq:N-PolyDot-DMPC}) for $N_{\text{PolyDot-CMPC}}$ in each region.

(i) $ts<z \text{ or } t=1$: From (\ref{eq:N-PolyDot-DMPC}), $N_{\text{PolyDot-CMPC}} = \psi_1 = (p+2)ts+\theta'(t-1)+2z-1$ and from (\ref{eq:N Entang-CMPC}), $N_{\text{Entangled-CMPC}}=2st^2+2z-1$, thus we have:
\begin{align}\label{eq:compare psi-1 with Entang}
 & N_{\text{PolyDot-CMPC}} - N_{\text{Entangled-CMPC}} \nonumber \\
 = & (p+2)ts+\theta'(t-1)+2z-1-(2st^2+2z-1) \nonumber \\
 = & pts+2ts+(2ts-t)(t-1)+2z-1-2st^2-2z+1 \nonumber \\
 = & t(ps-t+1).
 \end{align} 
From the above equation, if $p<\frac{t-1}{s}$ and $t\neq 1$, we have $N_{\text{PolyDot-CMPC}}<N_{\text{Entangled-CMPC}}$, otherwise, $N_{\text{PolyDot-CMPC}} \ge N_{\text{Entangled-CMPC}}$\footnote{Note that for $t=1$, $N_{\text{PolyDot-CMPC}} = N_{\text{Entangled-CMPC}}$.}. This along with the condition of (i), provides condition 1 for $N_{\text{PolyDot-CMPC}}<N_{\text{Entangled-CMPC}}$ in Lemma \ref{lemma: regions where N_polydot<N_entangled}.

(ii) $ts-t < z \leq ts$ and $s,t \neq 1$: From (\ref{eq:N-PolyDot-DMPC}), $N_{\text{PolyDot-CMPC}} = \psi_2=2ts+\theta'(t-1)+3z-1$ and from (\ref{eq:N Entang-CMPC}), $N_{\text{Entangled-CMPC}}=2st^2+2z-1$ for $z>ts-s$ and $N_{\text{Entangled-CMPC}}=st^2+3st-2s+t(z-1)+1$ for $z\leq ts-s$, thus we have:

(a) $z>ts-s$ and $t-1>s$: For this case, we have:
\begin{align}
 & N_{\text{PolyDot-CMPC}} - N_{\text{Entangled-CMPC}} \nonumber \\
 = & 2ts+\theta'(t-1)+3z-1-(2st^2+2z-1) \nonumber \\
 = & 2ts+(2ts-t)(t-1)+3z-1-2st^2-2z+1 \nonumber \\
 = & z-t(t-1)\nonumber\\
 < & z-ts\label{eq:temppppps}\\
 \leq & 0,
\end{align}
where (\ref{eq:temppppps}) comes from the condition of (a), $t-1>s$ and the last inequality comes from the condition of (ii), $z\leq ts$. Therefore, for the combination of conditions (ii) and (a), \ie $ts-s<z\leq ts$ and $t-1>s$, we have $ N_{\text{PolyDot-CMPC}} < N_{\text{Entangled-CMPC}}$. This provides condition 2 for $N_{\text{PolyDot-CMPC}}<N_{\text{Entangled-CMPC}}$ in Lemma \ref{lemma: regions where N_polydot<N_entangled}.

(b) $z>ts-s$ and $s= t-1$: For this case, we have:
\begin{align}\label{eq:compare psi-2 with Entang-(ii-a)}
 & N_{\text{PolyDot-CMPC}} - N_{\text{Entangled-CMPC}} \nonumber \\
 = & 2ts+\theta'(t-1)+3z-1-(2st^2+2z-1) \nonumber \\
 = & 2ts+(2ts-t)(t-1)+3z-1-2st^2-2z+1 \nonumber \\
 = & z-(t^2-t)\nonumber\\
 \leq & 0,
\end{align} 
where the last inequality comes from the condition of (ii), $z\leq ts=t(t-1)$. From the above equation, for $z<t^2-t$, we have $N_{\text{PolyDot-CMPC}}<N_{\text{Entangled-CMPC}}$, otherwise, $N_{\text{PolyDot-CMPC}}=N_{\text{Entangled-CMPC}}$. By replacing $s$ with $t-1$ and combining the conditions of (ii), (b), and $z<t^2-t$, \ie $t^2-2t+1<z<t^2-t, s=t-1$, condition 3 for $N_{\text{PolyDot-CMPC}}<N_{\text{Entangled-CMPC}}$ in Lemma \ref{lemma: regions where N_polydot<N_entangled} is derived.

(c) $z>ts-s$ and $s> t-1$: For this case, we have:
\begin{align}
 & N_{\text{PolyDot-CMPC}} - N_{\text{Entangled-CMPC}} \nonumber \\
 = & 2ts+\theta'(t-1)+3z-1-(2st^2+2z-1) \nonumber \\
 = & 2ts+(2ts-t)(t-1)+3z-1-2st^2-2z+1 \nonumber \\
 = & z-t(t-1)\nonumber\\
 \ge & z-t(s-1)\label{eq:tempp54}\\
 > & 0,
\end{align} 
where (\ref{eq:tempp54}) comes from the condition of (c), $s>t-1$ and the last inequality comes from the condition of (ii), $z>ts-t$.

(d) $z\leq ts-s, t>3$: For this case, we have:
\begin{align}\label{eq:compare psi-2 with Entang-(ii-d)}
 & N_{\text{PolyDot-CMPC}} - N_{\text{Entangled-CMPC}} \nonumber \\
 = & 2ts+\theta'(t-1)+3z-1\nonumber\\
 &\quad \quad -(st^2+3st-2s+t(z-1)+1) \nonumber \\
 = & 2ts+(2ts-t)(t-1)+3z-1\nonumber\\
 &\quad \quad -st^2-3st+2s-tz+t-1 \nonumber \\
 = & st^2-t^2+2t-3st+2s-2-z(t-3) \nonumber \\
 = & st^2-3st-t^2+3t-t+3+2s-5-z(t-3) \nonumber \\
 = & st(t-3)-t(t-3)-(t-3)+2s-5-z(t-3) \nonumber \\
 = &(t-3)(st-t-1+\frac{2s-5}{t-3})-(t-3)z.
\end{align} 
From the above equation, if $z>(st-t-1+\frac{2s-5}{t-3})$, we have $ N_{\text{PolyDot-CMPC}} < N_{\text{Entangled-CMPC}}$\footnote{Note that in this case $t\ge 3$.}, otherwise $ N_{\text{PolyDot-CMPC}} \ge N_{\text{Entangled-CMPC}}$. By combining the conditions of (ii), (d), and $z>(st-t-1+\frac{2s-5}{t-3})$, \ie $ts-t-\min\{0,1-\frac{2s-5}{t-3}\}<z\leq ts-s, t>3$, condition 4 for $N_{\text{PolyDot-CMPC}}<N_{\text{Entangled-CMPC}}$ in Lemma \ref{lemma: regions where N_polydot<N_entangled} is derived.

(e) $z\leq ts-s, t=3$: For this case, we have:
\begin{align}\label{eq:compare psi-2 with Entang-(ii-e)}
 & N_{\text{PolyDot-CMPC}} - N_{\text{Entangled-CMPC}} \nonumber \\
 = & 2ts+\theta'(t-1)+3z-1\nonumber\\
 &\quad \quad -(st^2+3st-2s+t(z-1)+1) \nonumber \\
 = & 2ts+(2ts-t)(t-1)+3z-1\nonumber\\
 &\quad \quad -st^2-3st+2s-tz+t-1 \nonumber \\
 =&2s-5.
\end{align} 
From the above equation, if $s=2$, $N_{\text{PolyDot-CMPC}} < N_{\text{Entangled-CMPC}}$, otherwise $N_{\text{PolyDot-CMPC}} > N_{\text{Entangled-CMPC}}$. By combining the conditions of (ii), (e), and $s=2$, \ie $s=2, t=3,z=4$, condition 5 for $N_{\text{PolyDot-CMPC}}<N_{\text{Entangled-CMPC}}$ in Lemma \ref{lemma: regions where N_polydot<N_entangled} is derived.

(f) $z\leq ts-s, t=2$: This condition is not possible, because $s\ge 2$ and thus $2s-2 \ge s$. Therefore, there is no overlap between the condition of (ii), $z>ts-t=2s-2$ and the condition of (f), $z\leq ts-s=s$. 

(iii) $ts-2t < z \leq ts-t$ and $s,t \neq 1$: From (\ref{eq:N-PolyDot-DMPC}), $N_{\text{PolyDot-CMPC}} = \psi_3=2ts+\theta'(t-1)+2z-1$ and from (\ref{eq:N Entang-CMPC}), $N_{\text{Entangled-CMPC}}=2st^2+2z-1$ for $z>ts-s$ and $N_{\text{Entangled-CMPC}}=st^2+3st-2s+t(z-1)+1$ for $z\leq ts-s$, thus we have:

(a) $t\ge s$: For this case, we have:
\begin{align}\label{eq:compare psi-3 with Entang-iii-a}
 & N_{\text{PolyDot-CMPC}} - N_{\text{Entangled-CMPC}} \nonumber \\
 = & 2ts+\theta'(t-1)+2z-1 \nonumber\\
 &\quad \quad - (st^2+3st-2s+t(z-1)+1) \nonumber \\
 = & 2ts+(2ts-t)(t-1)+2z-1 \nonumber\\
 &\quad \quad -st^2-3st+2s-tz+t-1 \nonumber \\
 = & st^2-2st-st-t^2+2t+2s-2-z(t-2) \nonumber \\
 = & st(t-2)-t(t-2)-s(t-2)-2-z(t-2) \nonumber \\
 = & (t-2)(st-t-s-\frac{2}{t-2})-z(t-2).
\end{align} 
From the above equation, if $t=2$, $N_{\text{PolyDot-CMPC}} < N_{\text{Entangled-CMPC}}$. By replacing $t=2$ in the conditions of (iii) and (a), \ie $t=2, s=2, z=1,2$, condition 6 for $N_{\text{PolyDot-CMPC}}<N_{\text{Entangled-CMPC}}$ in Lemma \ref{lemma: regions where N_polydot<N_entangled} is derived. In addition, if $t>2$ and $z>st-t-s-\frac{2}{t-2}$, $N_{\text{PolyDot-CMPC}} < N_{\text{Entangled-CMPC}}$, otherwise, $N_{\text{PolyDot-CMPC}} \ge N_{\text{Entangled-CMPC}}$. By combining the conditions of (iii), (a), and $t>2, z>st-t-s-\frac{2}{t-2}$, \ie $\max\{st-t-s-\frac{2}{t-2}, ts-2t\} < z \leq ts-t, t>2, t\ge s$, condition 7 for $N_{\text{PolyDot-CMPC}}<N_{\text{Entangled-CMPC}}$ in Lemma \ref{lemma: regions where N_polydot<N_entangled} is derived.

(b) $2t \ge s>t, z>ts-s$: For this case, we have:
\begin{align}\label{eq:compare psi-3 with Entang-iii-b}
 & N_{\text{PolyDot-CMPC}} - N_{\text{Entangled-CMPC}} \nonumber \\
 = & 2ts+\theta'(t-1)+2z-1-(2st^2+2z-1) \nonumber \\
 = & 2ts+(2ts-t)(t-1)+2z-1-2st^2-2z+1 \nonumber \\
 = & -t(t-1)\nonumber\\
 <&0,
\end{align}
From the above equation, for this case, $N_{\text{PolyDot-CMPC}} < N_{\text{Entangled-CMPC}}$. By combining the conditions of (iii) and (b), \ie $t<s\leq 2t, ts-s<z\leq ts-t$, condition 8 for $N_{\text{PolyDot-CMPC}}<N_{\text{Entangled-CMPC}}$ in Lemma \ref{lemma: regions where N_polydot<N_entangled} is derived.

(c) $2t \ge s>t, z\leq ts-s$: For this case, we have:
\begin{align}\label{eq:compare psi-3 with Entang-iii-c}
 & N_{\text{PolyDot-CMPC}} - N_{\text{Entangled-CMPC}} \nonumber \\
 = & 2ts+\theta'(t-1)+2z-1 - (st^2+3st-2s+t(z-1)+1) \nonumber \\
 = & 2ts+(2ts-t)(t-1)+2z-1\nonumber\\
 &\quad \quad -st^2-3st+2s-tz+t-1 \nonumber \\
 = & st(t-2)-t(t-2)-s(t-2)-2-z(t-2) \nonumber \\
 = & (t-2)(st-t-s-\frac{2}{t-2})-z(t-2).
\end{align} 
From the above equation, if $t=2$, $N_{\text{PolyDot-CMPC}} < N_{\text{Entangled-CMPC}}$. By replacing $t=2$ in the conditions of (iii) and (c), \ie $t=2, 3\leq s \leq 4, 2(s-2)<z\leq 2(s-1)$, condition 9 for $N_{\text{PolyDot-CMPC}}<N_{\text{Entangled-CMPC}}$ in Lemma \ref{lemma: regions where N_polydot<N_entangled} is derived. In addition, if $t>2$ and $z>st-t-s-\frac{2}{t-2}$, $N_{\text{PolyDot-CMPC}} < N_{\text{Entangled-CMPC}}$, otherwise, $N_{\text{PolyDot-CMPC}} \ge N_{\text{Entangled-CMPC}}$. By combining the conditions of (iii), (c), and $t>2, z>st-t-s-\frac{2}{t-2}$, \ie $st-2t < z \leq ts-s, t>2, t< s\leq 2t$, condition 10 for $N_{\text{PolyDot-CMPC}}<N_{\text{Entangled-CMPC}}$ in Lemma \ref{lemma: regions where N_polydot<N_entangled} is derived.

(d) $s>2t$: For this case, we have:
\begin{align}
 & N_{\text{PolyDot-CMPC}} - N_{\text{Entangled-CMPC}} \nonumber \\
 = & 2ts+\theta'(t-1)+2z-1-(2st^2+2z-1) \nonumber \\
 = & 2ts+(2ts-t)(t-1)+2z-1-2st^2-2z+1 \nonumber \\
 = & -t(t-1)\nonumber\\
 <&0,
\end{align}
From the above equation, for this case, $N_{\text{PolyDot-CMPC}} < N_{\text{Entangled-CMPC}}$. By combining the conditions of (iii) and (d), \ie $s>2t, ts-2t<z\leq ts-t$, condition 11 for $N_{\text{PolyDot-CMPC}}<N_{\text{Entangled-CMPC}}$ in Lemma \ref{lemma: regions where N_polydot<N_entangled} is derived.

(iv) $\max\{ts-2t-s+2, \frac{ts-2t+1}{2}\} < z \leq st-2t$ and $s,t \ne 1$: From (\ref{eq:N-PolyDot-DMPC}), $N_{\text{PolyDot-CMPC}} = \psi_4=(t+1)ts+(t-1)(z+t-1)+2z-1$ and from (\ref{eq:N Entang-CMPC}), $N_{\text{Entangled-CMPC}}=2st^2+2z-1$ for $z>ts-s$ and $N_{\text{Entangled-CMPC}}=st^2+3st-2s+t(z-1)+1$ for $z\leq ts-s$, thus we have:

(a) $2t\ge s$: For this case, we have:
\begin{align}\label{eq:compare psi-4 with Entang-iv-a}
 & N_{\text{PolyDot-CMPC}} - N_{\text{Entangled-CMPC}} \nonumber \\
 = & (t+1)ts+(t-1)(z+t-1)+2z-1 \nonumber \\ -&(st^2+3st-2s+t(z-1)+1) \nonumber \\
 = & z-(2ts-t^2+t-2s+1).
\end{align} 
From the above equation, if $z<(2ts-t^2+t-2s+1)$, we have $N_{\text{PolyDot-CMPC}} < N_{\text{Entangled-CMPC}}$, otherwise, $N_{\text{PolyDot-CMPC}} \ge N_{\text{Entangled-CMPC}}$. By combining the conditions of (iv), (a), and $z<(2ts-t^2+t-2s+1)$, \ie $2t\ge s, \max\{ts-2t-s+2, \frac{ts-2t+1}{2}\} < z \leq \min\{st-2t, 2ts-t^2+t-2s+1\}$, condition 12 for $N_{\text{PolyDot-CMPC}} < N_{\text{Entangled-CMPC}}$ in Lemma \ref{lemma: regions where N_polydot<N_entangled} is derived.

(b) $2t<s, ts-s<z\leq st-2t, t\neq 2$: For this case, $\max\{ts-2t-s+2, \frac{ts-2t+1}{2}\}=ts-2t-s+2<ts-s$. The reason is summarized as follows:
\begin{align}\label{eq:temmp}
    & s>2t \nonumber\\
    \Rightarrow & s(t-2)>2t\nonumber\\
    \Rightarrow & s(t-2)+3>2t\nonumber\\
    \Rightarrow & ts-2t-2s+4>1\nonumber\\
    \Rightarrow & ts-2t-s+2>\frac{ts-2t+1}{2}
\end{align}
For this case, we have:
\begin{align}
 & N_{\text{PolyDot-CMPC}} - N_{\text{Entangled-CMPC}} \nonumber \\
 = & (t+1)ts+(t-1)(z+t-1)+2z-1 \nonumber \\ 
 -&(2st^2+2z-1) \nonumber \\
 = & (t-1)(z-1+t-ts)\nonumber\\
 <&0,
\end{align} 
where the last inequality comes from the condition of (b), $z\leq st-2t$, as $st-2t<st-t+1$ and thus $z< st-t+1$. By combining the conditions of (iv) and (b) \ie $s>2t, ts-s<z\leq ts-2t, t\neq 2$, condition 13 for $N_{\text{PolyDot-CMPC}}<N_{\text{Entangled-CMPC}}$ in Lemma \ref{lemma: regions where N_polydot<N_entangled} is derived.

(c) $2t<s, ts-s<z\leq st-2t, t= 2$: By replacing $t=2$ in conditions of (iv) and (c), we have $4<s<z<2s-4$. Therefore, for this case, we have:
\begin{align}
 & N_{\text{PolyDot-CMPC}} - N_{\text{Entangled-CMPC}} \nonumber \\
 = & (t+1)ts+(t-1)(z+t-1)+2z-1 \nonumber \\ 
 -&(2st^2+2z-1) \nonumber \\
 = & (t-1)(z-1+t-ts)\nonumber\\
 = & z-1+2-2s\nonumber\\
 <&-3 < 0.
\end{align}
The condition of this case, \ie $4<s<z<2s-4, t=2$, provides condition 14 for $N_{\text{PolyDot-CMPC}}<N_{\text{Entangled-CMPC}}$ in Lemma \ref{lemma: regions where N_polydot<N_entangled}.

(d) $2t<s, ts-2t-s+2<z\leq ts-s$: For this case, we have:
\begin{align}
 & N_{\text{PolyDot-CMPC}} - N_{\text{Entangled-CMPC}} \nonumber \\
 = & (t+1)ts+(t-1)(z+t-1)+2z-1 \nonumber \\ -&(st^2+3st-2s+t(z-1)+1) \nonumber \\
 = & z-(2ts-t^2+t-2s+1).
\end{align} 
From the above equation, if $z<2ts-t^2+t-2s+1$, we have $ N_{\text{PolyDot-CMPC}} < N_{\text{Entangled-CMPC}}$, otherwise, $N_{\text{PolyDot-CMPC}} \ge N_{\text{Entangled-CMPC}}$.
On the other hand, $\max\{ts-2t-s+2, \frac{ts-2t+1}{2}\}=ts-2t-s+2<ts-s$, which is derived from (\ref{eq:temmp}) for $t\neq 2$. For $t=2$, $\max\{ts-2t-s+2, \frac{ts-2t+1}{2}\}=\max\{s-2, s-2+1/2\}=s-1.5$, however, we consider $s-2=ts-2t-s+2$ as $s$ and $z$ are integers and $z>s-1.5$ is equivalent to $z>s-2$. Therefore, by combining the conditions of (iv) and (d), \ie 
$ts-2t-s+2<z<ts-s, 2t<s$,
condition 15 for $N_{\text{PolyDot-CMPC}}<N_{\text{Entangled-CMPC}}$ in Lemma \ref{lemma: regions where N_polydot<N_entangled} is derived. The reason for this combination is that:
\begin{align}
    & 2ts-t^2+t-2s+1 \nonumber \\
    &= 2ts-2s-t(t-1)+1 \nonumber \\
    &= ts-s+s(t-1)-t(t-1)+1 \nonumber \\
    &= ts-s+(t-1)(s-t)+1 \nonumber \\
    &> ts-s+(t-1)(2t-t)+1 \text{\;since $s>2t$} \nonumber \\
    &= ts-s+t(t-1)+1 \nonumber \\
   &>ts-s \nonumber \\
   \Rightarrow & \min\{ts-s,2ts-t^2+t-2s+1\} = ts-s.
\end{align}
%

(v) $z \leq \max\{ts-2t-s+2, \frac{ts-2t+1}{2}\}$ and $s,t \neq 1$: For this case, we have, $z\leq ts-s$. The reason is that $ts-s>ts-s-2t+2$ and $ts-s>\frac{ts-2t+1}{2}$\footnote{This can be directly derived from the fact that $s(t-2)\ge 0 >-2t+1$.}, therefore, from (\ref{eq:N Entang-CMPC}), 
$N_{\text{Entangled-CMPC}}=st^2+3st-2s+t(z-1)+1$ and from (\ref{eq:N-PolyDot-DMPC}), $N_{\text{PolyDot-CMPC}} = \psi_5=\theta' t+z$, thus we have:
%
\begin{align}
 & N_{\text{PolyDot-CMPC}} - N_{\text{Entangled-CMPC}} \nonumber \\
 = & \theta't+z-(st^2+3st-2s+t(z-1)+1) \nonumber \\
 = & 2st^2-t^2+z-st^2-3st+2s-tz+t-1 \nonumber \\
 = & (t-1)(st-2s-t-\frac{1}{t-1})-z(t-1).
\end{align}
From the above equation, if $z>st-2s-t-\frac{1}{t-1}$, we have $N_{\text{PolyDot-CMPC}} < N_{\text{Entangled-CMPC}}$, otherwise, $N_{\text{PolyDot-CMPC}} \ge N_{\text{Entangled-CMPC}}$. By combining (v), (a), and $z>st-2s-t-\frac{1}{t-1}$, \ie $st-2s-t-\frac{1}{t-1}<z\leq \max\{ts-2t-s+2, \frac{ts-2t+1}{2}\}$, condition 16 for $N_{\text{PolyDot-CMPC}}<N_{\text{Entangled-CMPC}}$ in Lemma \ref{lemma: regions where N_polydot<N_entangled} is derived.

(vi) $s=1 \text{ and } t \ge z$: From (\ref{eq:N-PolyDot-DMPC}), $N_{\text{PolyDot-CMPC}} = \psi_6=t^2+2t+tz-1$ and from (\ref{eq:N Entang-CMPC}), $N_{\text{Entangled-CMPC}}=2t^2+2z-1$ for $z>t-1$ and $N_{\text{Entangled-CMPC}}=t^2+3t-2+t(z-1)+1$ for $z\leq t-1$, thus we have:

(a) $z=t$: For this case, we have:
\begin{align}
 & N_{\text{PolyDot-CMPC}} - N_{\text{Entangled-CMPC}} \nonumber \\
 = & t^2+2t+tz-1-(2t^2+2z-1) \nonumber \\
 = & (z-t)(t-2) \nonumber \\
 =&0.
\end{align}
From the above equation, for this condition, $N_{\text{PolyDot-CMPC}} = N_{\text{Entangled-CMPC}}$.

(b) $z\leq t-1$: For this case, we have:
\begin{align}
 & N_{\text{PolyDot-CMPC}} - N_{\text{Entangled-CMPC}} \nonumber \\
 = & t^2+2t+tz-1-(t^2+3t-2+t(z-1)+1) \nonumber \\
 =&0.
\end{align}
From the above equation, for this condition, $N_{\text{PolyDot-CMPC}} = N_{\text{Entangled-CMPC}}$.

\subsection{Proof of Lemma \ref{lemma: regions where N_polydot<N_ssmm} (PolyDot-CMPC Versus SSMM)}
To prove this lemma, we consider different regions for the value of $z$ and compare the required number of workers for PolyDot-CMPC, $N_{\text{PolyDot-CMPC}}$, with SSMM, $N_{\text{SSMM}}$, in each region. From \cite{Zhu2021ImprovedCF}, $N_{\text{SSMM}} = (t+1)(ts+z)-1$ and we use (\ref{eq:N-PolyDot-DMPC}) for $N_{\text{PolyDot-CMPC}}$ in each region.

(i) $ts<z \text{ or } t=1$: From (\ref{eq:N-PolyDot-DMPC}), $N_{\text{PolyDot-CMPC}} = \psi_1 = (p+2)ts+\theta'(t-1)+2z-1$ and thus we have:
\begin{align}\label{eq:compare psi-1 with N-SSMM}
    & N_{\text{PolyDot-CMPC}} - N_{\text{SSMM}} \nonumber \\
    = & (p+2)ts+\theta'(t-1)+2z-1-(t+1)(ts+z)+1 \nonumber \\
    = & pts+2ts+2t^2s-2ts-t^2+t+2z-t^2s-ts-(t+1)z \nonumber \\
    = & pts+(t-1)ts-t(t-1)-(t-1)z. 
\end{align}
From the above equation, if $z>\frac{pts}{t-1}+ts-t$ and $t\neq 1$, we have $N_{\text{PolyDot-CMPC}} < N_{\text{SSMM}}$, otherwise $N_{\text{PolyDot-CMPC}} \geq N_{\text{SSMM}}$\footnote{Note that for $t=1$, $N_{\text{PolyDot-CMPC}} = N_{\text{SSMM}}$.}. Therefore, from the condition of (i), we have $N_{\text{PolyDot-CMPC}} < N_{\text{SSMM}}$ only if $z > \max\{ts,ts-t+\frac{pts}{t-1}\}, t\neq 1$. This provides one of the conditions that $N_{\text{PolyDot-CMPC}} < N_{\text{SSMM}}$ in Lemma \ref{lemma: regions where N_polydot<N_ssmm}.

(ii) $ts-t < z \leq ts $: From (\ref{eq:N-PolyDot-DMPC}), $N_{\text{PolyDot-CMPC}} = \psi_2=2ts+\theta'(t-1)+3z-1$ and thus we have:
\begin{align}\label{eq:compare psi-2 with N-SSMM}
 & N_{\text{PolyDot-CMPC}} - N_{\text{SSMM}} \nonumber \\
 = & 2ts+\theta'(t-1)+3z-1 - (t+1)(ts+z)+1 \nonumber \\
 = & 2ts+2t^2s-2ts-t^2+t+3z-1-t^2s-ts-(t+1)z+1 \nonumber \\
 = & st^2-st-t^2+t-(t-2)z \nonumber \\
 = & st(t-1)-t(t-1)-(t-2)z \nonumber \\
 = & (t-1)(st-t)-(t-2)z.
\end{align}
From the above equation, if $z>\frac{(st-t)(t-1)}{t-2}$, we have $N_{\text{PolyDot-CMPC}}<N_{\text{SSMM}}$ otherwise,  $N_{\text{PolyDot-CMPC}} \geq N_{\text{SSMM}}$. Therefore, from the condition of (ii), we have $N_{\text{PolyDot-CMPC}}<N_{\text{SSMM}}$ only if $\frac{t-1}{t-2}(st-t)<z \leq ts$. This provides the other condition that $N_{\text{PolyDot-CMPC}}<N_{\text{SSMM}}$ in Lemma \ref{lemma: regions where N_polydot<N_ssmm}.

(iii) $ts-2t < z \leq ts-t$: From (\ref{eq:N-PolyDot-DMPC}), $N_{\text{PolyDot-CMPC}} = \psi_3=2ts+\theta'(t-1)+2z-1$ and thus we have:
\begin{align}\label{eq:compare psi-3 with N-SSMM}
 & N_{\text{PolyDot-CMPC}} - N_{\text{SSMM}} \nonumber \\
 &= 2ts+\theta'(t-1)+2z-1 - (t+1)(ts+z)+1 \nonumber \\
 &= 2ts+2t^2s-2ts-t^2+t-st^2-st-(t-1)z \nonumber \\
 &=-t^2+t+st^2-st-(t-1)z \nonumber \\
 &= (ts-t)(t-1)-(t-1)z.
\end{align}
From the above equation and the condition of (iii), $N_{\text{PolyDot-CMPC}} \geq N_{\text{SSMM}}$ for $ts-2t <z \leq ts-t$. 

(iv) $\max\{ts-2t-s+2, \frac{ts-2t+1}{2}\} < z \leq st-2t$: From (\ref{eq:N-PolyDot-DMPC}), $N_{\text{PolyDot-CMPC}} = \psi_4=(t+1)ts+(t-1)(z+t-1)+2z-1$ and thus we have:
\begin{align}\label{eq:compare psi-4 with N-SSMM}
 & N_{\text{PolyDot-CMPC}} - N_{\text{SSMM}} \nonumber \\
 = & (t+1)ts+(t-1)(z+t-1)+2z-1 -(t+1)(ts+z)+1
\nonumber \\
= &  (t+1)ts+(t+1)z+(t-1)^2-(t+1)(ts+z) \nonumber \\
= & (t-1)^2 > 0.
\end{align}
From the above equation, $N_{\text{PolyDot-CMPC}} > N_{\text{SSMM}}$ for $\max\{ts-2t-s+2, \frac{ts-2t+1}{2}\} < z \leq st-2t$.

(v) $z \leq \max\{ts-2t-s+2, \frac{ts-2t+1}{2}\}$: From (\ref{eq:N-PolyDot-DMPC}), $N_{\text{PolyDot-CMPC}} = \psi_5=\theta' t+z$ and thus we have:
\begin{align}
 & N_{\text{PolyDot-CMPC}} - N_{\text{SSMM}} \nonumber \\
 &= \theta't+z - (t+1)(ts+z)+1 \nonumber \\
 &= 2t^2s-t^2+z-t^2s-ts-(t+1)z+1 \nonumber \\
 &= t^2s-t^2-ts+1-tz \nonumber \\
 &= t(ts-t-s+\frac{1}{t}-z)\nonumber\\
 & \ge t(\max\{ts-2t-s+2, \frac{ts-2t+1}{2}\}-z) \label{eq:compare1v1}\\
 & \geq 0 \label{eq:compare psi-5 with N-SSMM},
 \end{align}
where, (\ref{eq:compare1v1}) comes from:
\begin{align}
  & ts-t-s+\frac{1}{t} - (ts-2t-s+2) \nonumber \\
  = & ts-t-s+\frac{1}{t}-ts+2t+s-2\nonumber \\
  = & t+\frac{1}{t}-2>0,
\end{align}
and
\begin{align}
  & ts-t-s+\frac{1}{t} - (\frac{ts-2t+1}{2}) \nonumber \\
  = & \frac{s(t-2)+2/t-1}{2} \ge 0,
\end{align}
and (\ref{eq:compare psi-5 with N-SSMM}) comes from the condition of the (v), \ie $z \leq ts-2t-s+1$. Therefore, $N_{\text{PolyDot-CMPC}} \ge N_{\text{SSMM}}$ for $z \leq ts-2t-s+1$.

(vi) $s=1 \text{ and } t \ge z$: From (\ref{eq:N-PolyDot-DMPC}), $N_{\text{PolyDot-CMPC}} = \psi_6=t^2+2t+tz-1$ and thus we have:
\begin{align}
 & N_{\text{PolyDot-CMPC}} - N_{\text{SSMM}} \nonumber \\
 &= t^2+2t+tz-1 - (t+1)(ts+z)+1 \nonumber \\
 &= t^2+2t-t^2s-ts-z \nonumber \\
 &= t^2+2t-t^2-t-z \nonumber \\
 &= t-z\nonumber\\
 & \geq 0 \label{eq:compare psi-6 with N-SSMM},
 \end{align}
From (i), (ii), (iii), (iv), (v) and (vi), the only conditions that $N_{\text{PolyDot-CMPC}} < N_{\text{SSMM}}$, are $z > \max\{ts,ts-t+\frac{pts}{t-1}\}, t\neq 1$ and $\frac{t-1}{t-2}(st-t)<z \leq ts$. In all other conditions, we have $N_{\text{PolyDot-CMPC}} \ge N_{\text{SSMM}}$.
This completes the proof. \hfill $\Box$

\subsection{Proof of Lemma \ref{lemma: regions where N_polydot<N_gcsana} (PolyDot-CMPC Versus GCSA-NA)}\label{subsec:polydot-cmpc vs gcsa-na}
To prove this lemma, we consider different regions for the value of $z$ and compare the required number of workers for PolyDot-CMPC, $N_{\text{PolyDot-CMPC}}$, with GCSA-NA, $N_{\text{GCSA-NA}}$, in each region. From \cite{9333639}, $N_{\text{GCSA-NA}}$ for one matrix multiplication (the number of batch is one) is equal to $N_{\text{GCSA-NA}} = 2st^2+2z-1$ and we use (\ref{eq:N-PolyDot-DMPC}) for $N_{\text{PolyDot-CMPC}}$ in each region.

(i) $ts<z \text{ or } t=1$: From (\ref{eq:N-PolyDot-DMPC}), $N_{\text{PolyDot-CMPC}} = \psi_1 = (p+2)ts+\theta'(t-1)+2z-1$ and thus we have:
\begin{align}\label{eq:compare psi-1 with GCSA-NA}
 & N_{\text{PolyDot-CMPC}} - N_{\text{GCSA-NA}} \nonumber \\
 = & (p+2)ts+\theta'(t-1)+2z-1-(2st^2+2z-1) \nonumber \\
 = & pts+2ts+(2ts-t)(t-1)+2z-1-2st^2-2z+1 \nonumber \\
 = & t(ps-t+1).
 \end{align} 
From the above equation, if $p<\frac{t-1}{s}$ and $t\neq 1$, we have $N_{\text{PolyDot-CMPC}}<N_{\text{GCSA-NA}}$, otherwise, $N_{\text{PolyDot-CMPC}} \ge N_{\text{GCSA-NA}}$\footnote{Note that for $t=1$, $N_{\text{PolyDot-CMPC}} = N_{\text{GCSA-NA}}$.}. This along with the condition of (i), provides one of the conditions that $N_{\text{PolyDot-CMPC}}<N_{\text{GCSA-NA}}$ in Lemma \ref{lemma: regions where N_polydot<N_gcsana}.

(ii) $ts-t < z \leq ts$: From (\ref{eq:N-PolyDot-DMPC}), $N_{\text{PolyDot-CMPC}} = \psi_2=2ts+\theta'(t-1)+3z-1$ and thus we have:
\begin{align}\label{eq:compare psi-2 with GCSA-NA}
 & N_{\text{PolyDot-CMPC}} - N_{\text{GCSA-NA}} \nonumber \\
 = & 2ts+\theta'(t-1)+3z-1-(2st^2+2z-1) \nonumber \\
 = & 2ts+(2ts-t)(t-1)+3z-1-2st^2-2z+1 \nonumber \\
 = & z-(t^2-t).
\end{align} 
From the above equation, if $z<t(t-1)$, we have $N_{\text{polyDot-CMPC}}<N_{\text{GCSA-NA}}$, otherwise, $N_{\text{polyDot-CMPC}}\ge N_{\text{GCSA-NA}}$. From the condition of (ii), $ts-t<z\leq ts$. Therefore, $N_{\text{polyDot-CMPC}}<N_{\text{GCSA-NA}}$ only if $ts-t<z \leq \min\{ts, t(t-1)-1\}$, which also requires that $s<t$. This is another condition that $N_{\text{PolyDot-CMPC}}<N_{\text{GCSA-NA}}$in Lemma \ref{lemma: regions where N_polydot<N_gcsana}.

(iii) $ts-2t < z \leq ts-t$: From (\ref{eq:N-PolyDot-DMPC}), $N_{\text{PolyDot-CMPC}} = \psi_3=2ts+\theta'(t-1)+2z-1$ and thus we have:
\begin{align}\label{eq:compare psi-3 with GCSA-NA}
 & N_{\text{PolyDot-CMPC}} - N_{\text{GCSA-NA}} \nonumber \\
 = & 2ts+\theta'(t-1)+2z-1-(2st^2+2z-1) \nonumber \\
 = & 2ts+(2ts-t)(t-1)+2z-1-2st^2-2z+1 \nonumber \\
 = & t(1-t)\nonumber\\
 <&0.
\end{align}
From the above equation, for $ts-2t < z \leq ts-t$, we have $N_{\text{PolyDot-CMPC}} < N_{\text{GCSA-NA}}$. This provides part of the third condition that $N_{\text{PolyDot-CMPC}}<N_{\text{GCSA-NA}}$ in Lemma \ref{lemma: regions where N_polydot<N_gcsana}.

(iv) $\max\{ts-2t-s+2, \frac{ts-2t+1}{2}\} < z \leq st-2t$: From (\ref{eq:N-PolyDot-DMPC}), $N_{\text{PolyDot-CMPC}} = \psi_4=(t+1)ts+(t-1)(z+t-1)+2z-1$ and thus we have:
\begin{align}\label{eq:compare psi-4 with GCSA-NA}
 & N_{\text{PolyDot-CMPC}} - N_{\text{GCSA-NA}} \nonumber \\
 = & (t+1)ts+(t-1)(z+t-1)+2z-1-(2st^2+2z-1) \nonumber \\
 = & t^2s+ts+(t-1)(z+t-1)-2st^2 \nonumber \\
 = & (t-1)(z-(st-t+1)).
\end{align} 
From the above equation, if $z<st-t+1$, we have $N_{\text{polyDot-CMPC}}<N_{\text{GCSA-NA}}$. This condition is satisfied for the condition of (iv), $\max\{ts-2t-s+2, \frac{ts-2t+1}{2}\} < z \leq st-2t$, as $st-t-t<st-t+1$. Therefore, for $\max\{ts-2t-s+2, \frac{ts-2t+1}{2}\} < z \leq st-2t$, we have $N_{\text{PolyDot-CMPC}} < N_{\text{GCSA-NA}}$. This provides part of the third condition that $N_{\text{PolyDot-CMPC}}<N_{\text{GCSA-NA}}$ in Lemma \ref{lemma: regions where N_polydot<N_gcsana}.

(v) $z \leq \max\{ts-2t-s+2, \frac{ts-2t+1}{2}\}$: From (\ref{eq:N-PolyDot-DMPC}), $N_{\text{PolyDot-CMPC}} = \psi_5=\theta' t+z$ and thus we have:
\begin{align}\label{eq:compare psi-5 with GCSA-NA}
 & N_{\text{PolyDot-CMPC}} - N_{\text{GCSA-NA}} \nonumber \\
 = & \theta't+z-(2st^2+2z-1) \nonumber \\
 = & 2st^2-t^2+z-2st^2-2z+1 \nonumber \\
 = & -t^2-z+1\nonumber\\
 <&0.
\end{align} 
From the above equation, for $z \leq \max\{ts-2t-s+2, \frac{ts-2t+1}{2}\}$, we have $N_{\text{PolyDot-CMPC}} < N_{\text{GCSA-NA}}$. This provides part of the third condition that $N_{\text{PolyDot-CMPC}}<N_{\text{GCSA-NA}}$ in Lemma \ref{lemma: regions where N_polydot<N_gcsana}.

(vi) $s=1 \text{ and } t \ge z$: From (\ref{eq:N-PolyDot-DMPC}), $N_{\text{PolyDot-CMPC}} = \psi_6=t^2+2t+tz-1$ and thus we have:
\begin{align}\label{eq:compare psi-6 with GCSA-NA}
 & N_{\text{PolyDot-CMPC}} - N_{\text{GCSA-NA}} \nonumber \\
 = & t^2+2t+tz-1-(2st^2+2z-1) \nonumber \\
 = & 2t+tz-t^2-2z \nonumber \\
 = & (2-t)(t-z)\nonumber\\
 \leq &0.
\end{align}
From the above equation and the condition of (vi), if $s=1, t>z$ and $t\neq 2$, we have $N_{\text{polyDot-CMPC}}<N_{\text{GCSA-NA}}$. This provides the last condition that $N_{\text{PolyDot-CMPC}}<N_{\text{GCSA-NA}}$ in Lemma \ref{lemma: regions where N_polydot<N_gcsana}, and completes the proof. \hfill $\Box$

\end{document}